\documentclass[aps,prb,twocolumn,superscriptaddress,floatfix,10pt,letterpaper,showpacs]{revtex4-1}
\usepackage{amsmath,amssymb,amsfonts,stmaryrd,wasysym,graphicx,multirow,color,textcomp}
\usepackage{url}\usepackage{diagrams}
\usepackage{multirow}
\usepackage[colorlinks=true,linkcolor=blue,citecolor=blue,urlcolor=blue]{hyperref}

\newcommand{\TI}{\hyperlink{TI}{TI} }
\newcommand{\FTI}{\hyperlink{FTI}{FTI} }
\newcommand{\TR}{\hyperlink{TR}{TR} }
\newcommand{\AFTR}{\hyperlink{AFTR}{AFTR} }

\newcommand{\ETCR}{\hyperlink{ETCR}{ETCR} }

\newcommand{\CFT}{\hyperlink{CFT}{CFT} }
\newcommand{\PH}{\hyperlink{PH}{PH} }
\newcommand{\FQH}{\hyperlink{FQH}{FQH} }
\newcommand{\OPE}{\hyperlink{OPE}{OPE} }

\begin{document}

\title{Paired parton quantum Hall states: a coupled wire construction}
\author{Alexander Sirota}\affiliation{Department of Physics, University of Virginia, Charlottesville, VA22904, USA}
\author{Sharmistha Sahoo}\affiliation{Department of Physics, University of Virginia, Charlottesville, VA22904, USA}\affiliation{Stewart Blusson Quantum Matter Institute, University of British Columbia, Vancouver, British Columbia, Canada V6T 1Z4}
\author{Gil Young Cho}\affiliation{Department of Physics, Korea Advanced Institute of Science and Technology, Daejeon 305-701, Korea}\affiliation{Department of Physics, Pohang University of Science and Technology (POSTECH), Pohang 37673, Republic of Korea}
\author{Jeffrey C. Y. Teo}\email{jteo@virginia.edu}\affiliation{Department of Physics, University of Virginia, Charlottesville, VA22904, USA}

\begin{abstract}
The Pfaffian fractional quantum Hall (FQH) states are incompressible non-Abelian topological fluids present in a half-filled electron Landau level, where there is a balanced population of electrons and holes. They give rise to half-integral quantum Hall plateaus that divide critical transitions between integer quantum Hall (IQH) states. On the other hand, there are Abelian FQH states, such as the Laughlin state, that can be understood using partons, which are fermionic divisions of the electron. In this paper, we propose a new family of incompressible paired parton FQH states at filling $\nu=1/6$ (modulo 1) that emerge from critical transitions between IQH states and Abelian FQH states at filling $\nu=1/3$ (modulo 1). These paired parton states are originated from a half-filled parton Landau level, where there is an equal amount of partons and holes. They generically support Ising-like anyonic quasiparticle excitations and carry non-Abelian Pfaffian topological orders (TO) for partons. We prove the principle existence of these paired parton states using exactly solvable interacting arrays of electronic wires under a magnetic field. Moreover, we establish a new notion of particle-hole (PH) symmetry for partons and relate the PH symmetric parton Pfaffian TO with the gapped symmetric surface TO of a fractional topological insulator in three dimension.
\end{abstract}

\maketitle

\section{Introduction}
Topological phases of matter~\cite{WenRMP2017} have been playing a pivotal role in the development of modern condensed matter physics. Several fundamental concepts such as fractionalization~\cite{Laughlin83}, topological order~\cite{Wentopologicalorder90}, and anyon~\cite{ArovasSchriefferWilczek84,Wilczekbook} have emerged from the studies of topological states and made striking experimental success. For example, the celebrated Laughlin state~\cite{Laughlin83} and more general fractional quantum Hall ({\color{blue}\hypertarget{FQH}{FQH}}) states~\cite{FQHE_Review} are inarguably the most well-understood strongly-correlated electronic phases beyond one dimensional systems, and have been a `motif' for many interesting quantum phases. On the other hand, another experimentally-observed `topological' state, namely topological band insulator~\cite{HasanKane10,QiZhangreview11,RMP}, embodies a remarkable interplay between symmetries and topologies of electronic wavefunctions. The topological insulator has made a broad impact on a number of disciplines of physics, from experimental and theoretical condensed matter physics to theoretical high-energy physics, and is still being actively investigated. In particular, physical properties of fractional topological insulators~\cite{MaciejkoQiKarchZhang10,SwingleBarkeshliMcGreevySenthil11,LevinBurnellKochStern11,maciejko2012models,ye2016composite,maciejko2015fractionalized,stern2016fractional,YeChengFradkin17} ({\color{blue}\hypertarget{FTI}{FTI}}), which are fractional analogues of topological insulators, are largely unexplored. 

Conventional topological insulators ({\color{blue}\hypertarget{TI}{TI}}) in three dimensions are intricately related to the quantum Hall effect in two dimensions. The massless surface Dirac fermions on the boundary of a \TI inspired a new dual low-energy description~\cite{Son15,WangSenthil15,MetlitskiVishwanath16} of the compressible composite Fermi liquid at a half-filled Landau level~\cite{KalmeyerZhang92,HalperinLeeRead93} at the critical transition between adjacent integer quantum Hall plateaus. Under the duality, the role of time-reversal ({\color{blue}\hypertarget{TR}{TR}}) symmetry of the former is replaced by particle-hole symmetry~\cite{Girvin84,Son15,BarkeshliMulliganFisher15,WangSenthil16,BalramJain17,NguyenGolkarRobertsSon18} of the latter. As a consequence, a thin \TI slab with finite thickness and \TR breaking massive top and bottom surfaces can be regarded as a quasi-2D system that is topologically equivalent to an incompressible quantum anomalous Hall state~\cite{liu2016quantum} which violates the particle-hole symmetry. This is because the two systems share identical 2D Chern invariant~\cite{TKNN} in the bulk as well as chiral charge and energy transport along their edges. Moreover, the symmetry-preserving many-body interacting gapped surface states~\cite{WangPotterSenthilgapTI13,MetlitskiKaneFisher13b,ChenFidkowskiVishwanath14,BondersonNayakQi13} of a \TI exhibit similar topological orders to the incompressible Pfaffian quantum Hall state~\cite{MooreRead} at half-filling. For instance, the T-Pfaffian \TI surface state~\cite{ChenFidkowskiVishwanath14} has an identical anyon structure to the particle-hole symmetric Pfaffian quantum Hall state~\cite{Son15}.

Similar correspondences hold between 3D \FTI and 2D \FQH states. The \FTI considered in this article are electronic states where the electrons are divided into emergent charge $e/3$ fermionic components, referred to as partons. More precisely, we use the parton ansatz~\cite{Jain89,WenEdgeStatesQHE92,MaciejkoQiKarchZhang10} \begin{align}\Psi_{\mathrm{el}}=\pi^1\pi^2\pi^3,\label{FTIparton}\end{align} where $\pi^a$, for $a=1,2,3$, are the fermionic partons. The partons are coupled to a dynamical $\mathbb{Z}_3$ gauge field~\cite{maciejko2012models}. Each $\pi^a$ carries a unit $\mathbb{Z}_3$ gauge charge so that the electronic quasiparticle $\Psi_{\mathrm{el}}$ is $\mathbb{Z}_3$ neutral. Moreover, each parton species fills a topological band insulator spectrum. From the construction, the \FTI surface must have the ``fractional parity anomaly" $\sigma_{xy}=\frac{e^2}{6h}$, which is the same Hall conductivity as of a \FQH state with filling $\nu=1/6$. In previous works~\cite{SahooSirotaChoTeo17,ChoTeoFradkin17}, we showed that the \FTI can host two types of charge-conserving and $\mathbb{Z}_3$-symmetric surface states with a finite excitation energy gap. The first is a ferromagnetic surface state that breaks \TR symmetry. The second is an anomalous symmetry-preserving surface state, referred to as the fractionalized T-Pfaffian state (denoted by $\mathcal{T}$-$\mathrm{Pf}^\ast$ in our previous work). This state is topologically-ordered and supports further fractionalized surface quasiparticle excitations, such as the charge $e/12$ Ising anyon.

\begin{figure}[htbp]
\centering\includegraphics[width=0.45\textwidth]{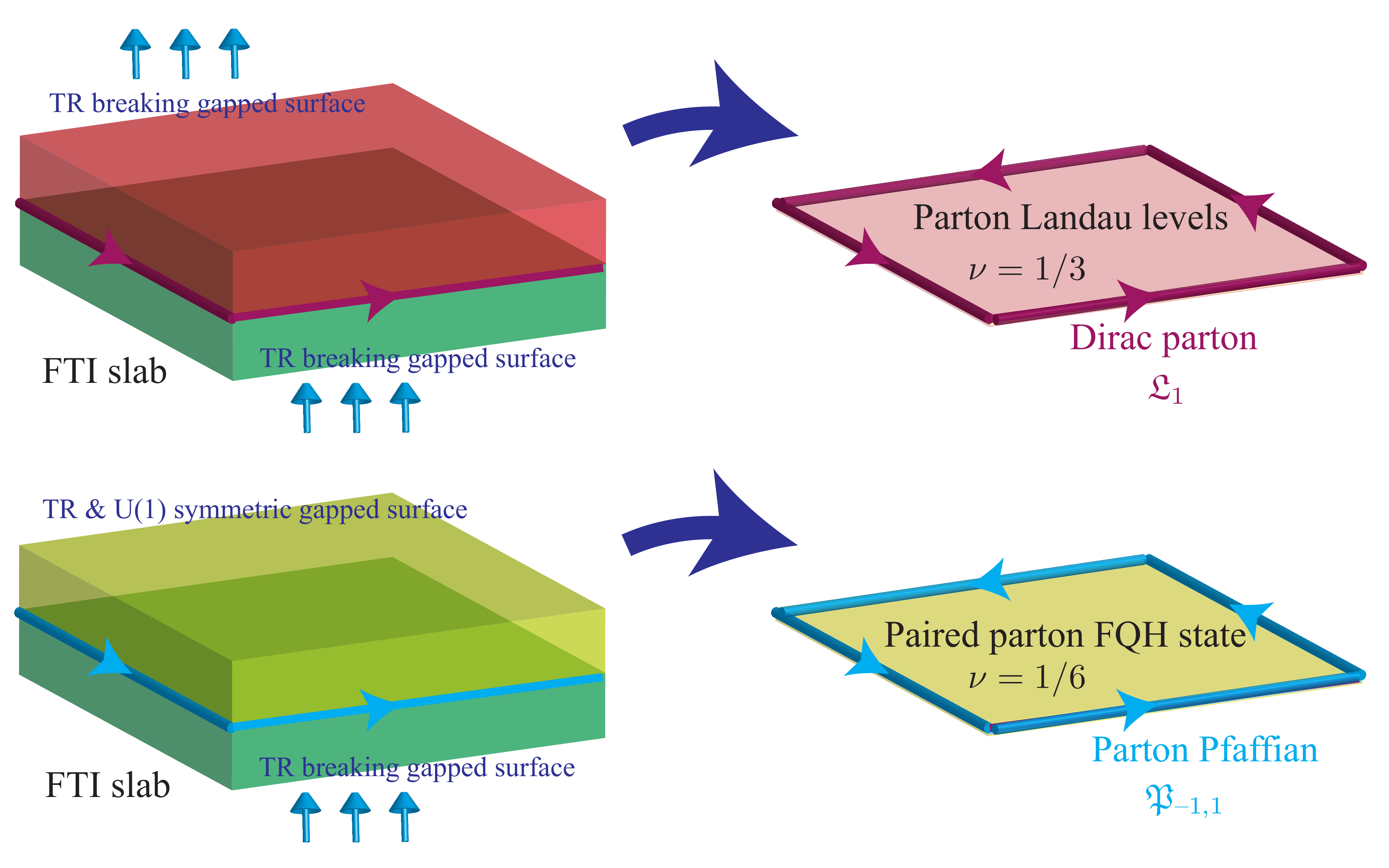}
\caption{Correspondence between thin FTI slabs with gapped surfaces and parton fractional quantum Hall states.}\label{fig:FTIslab}
\end{figure}

The three-dimensional \FTI in a slab geometry with a finite thickness can be topologically regarded as a quasi-two-dimensional system (see figure~\ref{fig:FTIslab}). When the two horizontal dimensions are infinite or much longer than the vertical one, the third axis reduces to a finite set of local variables in long length scale. If the top and bottom surfaces are gapped, then the \FTI slab is topologically equivalent to a 2D \FQH state. While there is currently no confirmation of 3D \FTI materials, the thin slab correspondence allows theoretical proposals and predictions of new \FQH states, which may arise in existing 2D quantum Hall materials or heterostructures given the right conditions. 

\begin{table}[htbp]
\begin{tabular}{l|ll}
Partons FQH & \multirow{2}{*}{$\mathfrak{L}_q$} & \multirow{2}{*}{$\mathfrak{P}_{p,q}$}\\
states & & \\\hline
filling fraction & $\nu=1/3$ & $\nu=1/6$ \\
central charge & $c=1+2q$ & $c=1+q+p/2$ \\
known examples & Laughlin state $\mathfrak{L}_0$ & Q-Pfaffian state $\mathfrak{P}_{1,0}$\\
FTI-inspired & \multirow{2}{*}{Dirac parton $\mathfrak{L}_1$} & Particle-hole symmetric\\
examples & & parton Pfaffian $\mathfrak{P}_{-1,1}$
\end{tabular}
\caption{Summary of the series of Abelian parton FQH states $\mathfrak{L}_q$ and paired parton FQH states $\mathfrak{P}_{p,q}$.}\label{tab:summary}
\end{table}

In this article, we propose two new classes of parton \FQH states inspired by \FTI slabs (see table~\ref{tab:summary} for a summary). Each \FQH state is characterized by its filling fraction $\nu$ and chiral central charge $c$ that respectively specify the electric and thermal Hall conductance~\cite{KaneFisher97,Cappelli01,Kitaev06} \begin{align}\sigma_{xy}=\frac{dI_{\mathrm{charge}}}{dV}=\nu\frac{e^2}{h},\quad\kappa_{xy}=\frac{dI_{\mathrm{energy}}}{dT}=c\frac{\pi^2k_B^2}{3h}T,\label{sigmakappaxy}\end{align} where $dV$ and $dT$ are the transverse potential and temperature differences across sample. The first class is a series of Abelian \FQH states $\mathfrak{L}_q$ with filling fraction $\nu=1/3$ and chiral central charge $c=1+2q$, where $q$ is an integer. For example, $\mathfrak{L}_{q=0}$ is the Laughlin \FQH state $U(1)_3$ where partons are confined. $\mathfrak{L}_{q=1}$ represents a deconfined parton \FQH phase $U(3)_1/\mathbb{Z}_3=U(1)_3\otimes SU(3)_1$, where each of the three partons $\pi^a$ completely fills a Landau level and an electrically neutral $SU(3)_1$ sector emerges. This $(2+1)$D phase is topologically equivalent to a thin \FTI slab with \TR breaking top and bottom surfaces~\cite{SahooSirotaChoTeo17}. The second class is a series of paired parton \FQH states $\mathfrak{P}_{p,q}$ with filling $\nu=1/6$ and chiral central charge $c=1+q+p/2$, where $p$ and $q$ are integers. They generically carry non-Abelian topological orders and support Ising anyons with charge $e/12$. For example, $\mathfrak{P}_{p=1,q=0}$ is identical to one of the $Q$-Pfaffian states in ref.~\onlinecite{MooreRead} where partons are confined. On the other hand, $\mathfrak{P}_{p=-1,q=1}$ represents a \FQH phase that extend the Pfaffian topological order by deconfined partons. We refer to this phase as the parton Pfaffian state and denote it by $\mathrm{Pf}^\ast$. It is topologically equivalent to a thin \FTI slab where one of the two surfaces are \TR preserving and the other is \TR breaking~\cite{SahooSirotaChoTeo17}.

One of the most important issues in condensed matter theory is to design and find microscopic electronic Hamiltonians for the strongly-interacting topological states. The parton ansatz \eqref{FTIparton} that artificially divides an electron into fractional components is an adhoc theoretical construction for the purpose of providing an interpretation of a topological state. It is only valid when the emergence of partons is supported by a microscopic Hamiltonian of many-body interacting electrons. Instead of relying on the parton ansatz as a premise, in this article, we will use the exactly solvable model approach where Hamiltonians of interacting electrons are first designed and subsequently shown to carry gapped parton excitations in the $(2+1)$D bulk and gapless parton $(1+1)$D conformal field theories ({\color{blue}\hypertarget{CFT}{CFT}}) along the boundary. These model Hamiltonians of parton \FQH states will be constructed by the coupled wire method. It involves a highly anisotropic description where the low-energy electronic degrees of freedom are confined along an array of continuous one-dimensional wires and many-body interactions take the form of inter- and intra-wire electron backscatterings. The technique was derived from sliding Lutthinger liquids~\cite{OHernLubenskyToner99,EmeryFradkinKivelsonLubensky00,VishwanathCarpentier01,SondhiYang01,MukhopadhyayKaneLubensky01} and was first employed by Kane, Mukhopadhyay and Lubensky~\cite{KaneMukhopadhyayLubensky02} in the study of Laughlin~\cite{Laughlin83} and Haldane-Halperin hierarchy~\cite{Haldane83,Halperin84} \FQH states. The method was later applied to more general \FQH states~\cite{TeoKaneCouplewires,KlinovajaLoss14,MengStanoKlinovajaLoss14,SagiOregSternHalperin15,KaneSternHalperin17} such as the Moore-Read Pfaffian~\cite{MooreRead,ReadMoore} and Read-Rezayi~\cite{ReadRezayi} \FQH states. In particular, our article can be regarded as a parton generalization of Kane, Stern and Halperin's work~\cite{KaneSternHalperin17}, which provided a coupled-wire description of a sequence of ``conventional" Pfaffian \FQH states at filling $\nu=1/2$.

Apart from being a tool to build exactly solvable models, the coupled wire construction has two more valuable features. First, the topological order of the $(2+1)$D bulk can be inferred from the low-energy \CFT that describes the system's $(1+1)$D boundary. The coupled wire Hamiltonian introduces an excitation energy gap in the bulk but leaves behind gapless degrees of freedom along the edge. The gapless boundary modes of the Abelian parton states $\mathfrak{L}_q$ and the paired parton states $\mathfrak{P}_{p,q}$ are effectively described by the $(1+1)$D \hyperlink{CFT}{CFT}s \begin{align}\begin{split}\mathfrak{L}_q&=U(1)_3\otimes[SU(3)_1]^q,\\\mathfrak{P}_{p,q}&=U(1)_{24}\otimes_{\mathrm{el}}U(1)^q_{12}\otimes_{\mathrm{el}}\mathrm{Ising}^p.\end{split}\label{LqPpqdef}\end{align} Second, the series of paired parton states $\mathfrak{P}_{p,q}$ exhibits a new notion of parton particle-hole ({\color{blue}\hypertarget{PH}{PH}}) symmetry, which applies to the half-filled parton Landau levels. These paired parton \FQH states can be related to one another under the \PH conjugation (to be defined in details in the main text) \begin{align}\mathrm{PH}_\pi\left(\mathfrak{P}_{p,q}\right)\equiv\mathfrak{L}_1\boxtimes\overline{\mathfrak{P}_{p,q}}=\mathfrak{P}_{-p-2,2-q}\end{align} that ``subtracts" $\mathfrak{P}_{p,q}$ from the parton Landau level $\mathfrak{L}_1=U(3)_1/\mathbb{Z}_3$. In particular, the parton Pfaffian state $\mathfrak{P}_{-1,1}=\mathrm{Pf}^\ast$ is \PH symmetric. It is not a coincidence that the parton Pfaffian state contains the fractional $\mathcal{T}$-$\mathrm{Pf}^\ast$ topological order that describes the symmetric gapped surface state of the \hyperlink{FTI}{FTI}. A parallel analogy can be drawn at the half-filled electronic Landau level, where the \PH symmetric Pfaffian state~\cite{Son15} has identical topological order to the T-Pfaffian state~\cite{ChenFidkowskiVishwanath14} of the symmetric gapped surface of a conventional \hyperlink{TI}{TI}. The exact correspondence between the symmetric \FTI surface and the \PH symmetric half-filled parton Landau levels is out of the scope of this article and we leave this implication for future studies.

This article will be presented in the following order. In section~\ref{sec:partonCFTs}, we describe (1+1)D chiral \hyperlink{CFT}{CFT}s, which will serve as the building blocks for the (2+1)D parton \FQH states and will emerge at the system edge. They also describe gapless modes at the domain walls between gapped regions on the surface of \hyperlink{FTI}{FTI}s. We will introduce the Dirac parton \CFT in section~\ref{sec:DiracParton}, the parton Pfaffian \CFT in section~\ref{sec:PartonPfaffian}, and their relationship through ``gluing and splitting" in section~\ref{sec:gluingsplitting}. In section~\ref{sec:fillingonethird} and \ref{sec:fillingonesixth}, we present the models for the series of Abelian parton \FQH states $\mathfrak{L}_q$ at filling one-third and the non-Abelian paired parton \FQH state $\mathfrak{P}_{p,q}$ at filling one-sixth by coupling electron wires in the presence of magnetic field. In section~\ref{sec:partonTPfaffian}, we present a model for the surface topological order of the \FTI that is closely-related to parton Pfaffian quantum Hall state at filling one-sixth. In section~\ref{sec:particlehole}, we introduce the emergent notion of particle-hole conjugation and symmetry in the context of partons and demonstrate the \PH action on the paired parton states $\mathfrak{P}_{p,q}$ with respect to each of the Abelian states $\mathfrak{L}_{q_0}$ that we found in the section~\ref{sec:CWM}. Finally, in section~\ref{sec:conclusion}, we summarize our results and discuss open implications. For completeness, we provide in appendix~\ref{app:KacMoody} a review of the relevant Kac-Moody algebras.

\section{Parton conformal field theories}\label{sec:partonCFTs}

\begin{figure}[htbp]
\includegraphics[width=0.4\textwidth]{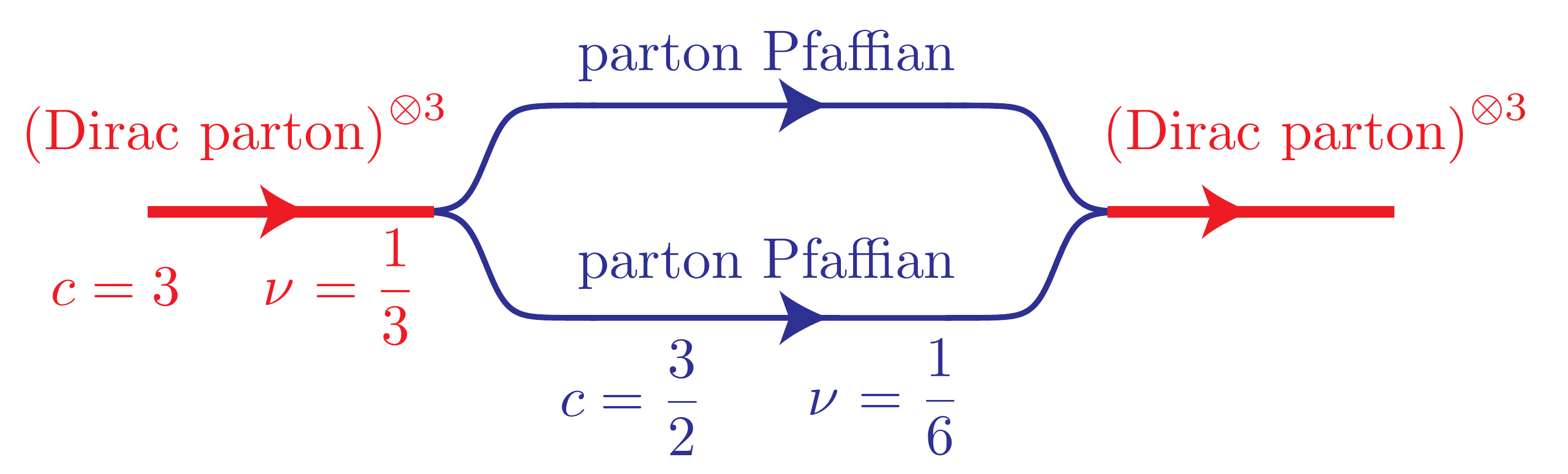}
\caption{Gluing and splitting between a pair of parton Pfaffian CFTs and a Dirac parton triplet. The Dirac parton and parton Pfaffian CFTs are defined in \eqref{DiracPartondef} and \eqref{Pfdef}. The chiral central charge $c$ and ``filling fraction" $\nu$ indicate the differential thermal and electric conductance $\kappa=c(\pi^2k_B^2/3h)T$ and $\sigma=\nu(e^2/h)$ respectively.}\label{fig:glueingsplitting}
\end{figure}

In this section, we present the relevant conformal field theories~\cite{bigyellowbook} (\hyperlink{CFT}{CFT}s) that describe the low-energy degrees of freedom along the $(1+1)$D edges of the parton fractional quantum Hall (\hyperlink{FQH}{FQH}) states. In particular, in section~\ref{sec:DiracParton} and \ref{sec:PartonPfaffian}, we focus on the Dirac parton triplet and the parton Pfaffian \hyperlink{CFT}{CFT}s that live on the edge of the parton Landau levels $\mathfrak{L}_1$ and the particle-hole symmetric parton Pfaffian \FQH state $\mathfrak{P}_{-1,1}$ respectively. In addition to the local electron, these \hyperlink{CFT}{CFT}s carry charge $e/3$ fermionic partons along with other fractional quasiparticles as primary fields. We will characterize the charge and energy transport in these \hyperlink{CFT}{CFT}s, and show that the parton Pfaffian carries half the degrees of freedom in the Dirac parton triplet (see figure~\ref{fig:glueingsplitting}). This means a pair of parton Pfaffians can be glued into a Dirac parton triplet by a condensation Hamiltonian consists of many-body electron backscattering (see \eqref{gluingH} in section~\ref{sec:gluingsplitting}). On the other hand, a Dirac parton triplet can be split into a pair of parton Pfaffians by a fractional basis transformation (see figure~\ref{fig:splitting} and eq.\eqref{splittingflowchart} in section~\ref{sec:gluingsplitting}). The $(2+1)$D parton \FQH states that support these fractional boundary modes will be constructed in the next section, and their topological orders can be inferred from the edge \hyperlink{CFT}{CFT}s through the bulk-boundary correspondence~\cite{MooreRead}.

\subsection{The Dirac parton triplet}\label{sec:DiracParton}
We begin with the triplet of chiral Dirac parton channels. It appears along the 1D interface separating two time-reversal symmetry breaking domains with opposite orientations on the surface of the fractional topological insulator (\hyperlink{FTI}{FTI}). It also appears along the 1D boundary of the 2D parton \FQH state $\mathfrak{L}_1$ where each of the three species of deconfined partons completely occupies a Landau level.

In low energy, the triplet is described by the orbifold~\cite{bigyellowbook} \CFT $U(3)_1/\mathbb{Z}_3$ (to be defined below). The $U(3)_1$ theory consists of three copies of chiral Dirac parton channels, where the fermionic partons can be represented by normal-ordered bosonized vertex operators $\pi^a\sim e^{i\tilde\phi_a}$, for $a=1,2,3$. In low-energy, the bosonized variables are described by the $U(3)_1$ Lagrangian density \begin{align}\mathcal{L}_{U(3)_1}=\frac{1}{2\pi}\sum_{a=1}^3\partial_{\mathsf{t}}\tilde\phi_a\partial_{\mathsf{x}}\tilde\phi_a\label{U(3)Lagrangian}\end{align} up to non-universal kinetic velocity terms. They follow the equal-time commutation relation ({\color{blue}\hypertarget{ETCR}{ETCR}}) \begin{align}\left[\partial_{\mathsf{x}}\tilde\phi_a(\mathsf{x}),\tilde\phi_b(\mathsf{x}')\right]=2\pi i\delta_{ab}\delta(\mathsf{x}-\mathsf{x}'),\label{ETCRpartons}\end{align} which is equivalent to the time-ordered correlation \begin{align}\langle\tilde\phi_a(\mathsf{z})\tilde\phi_b(\mathsf{w})\rangle=-\delta_{ab}\log(\mathsf{z}-\mathsf{w})+\frac{i\pi}{2}S_{ab},\label{TOC}\end{align} where $\mathsf{z},\mathsf{w}\sim\tau+i\mathsf{x}$ are complex space-time parameters (or $\mathsf{z},\mathsf{w}\sim e^{\tau+i\mathsf{x}}$ in a radially-ordered complex space-time geometry where $\mathsf{x}$ is periodic and the Euclidean time $\tau$ parametrizes the radial direction). The non-singular second term containing \begin{align}S=(S_{ab})_{3\times3}=\begin{pmatrix}0&1&-1\\-1&0&1\\1&-1&0\end{pmatrix}\label{KleinS}\end{align} is put in \eqref{TOC} to ensure anticommuting correlations between distinct fermions, $\langle e^{i\tilde\phi_a}e^{i\tilde\phi_b}\rangle=-\langle e^{i\tilde\phi_b}e^{i\tilde\phi_a}\rangle$. The particular form of $S_{ab}$ is chosen to respect the cyclic threefold rotation of parton labels $a,b=1,2,3$.

Each parton carries the electric charge $e^\ast=e/3$, where $e$ is the charge of the electron. The diagonal combination $\Psi_{\mathrm{el}}\sim\pi^1\pi^2\pi^3\sim e^{i(\tilde\phi_1+\tilde\phi_2+\tilde\phi_3)}$ is a local electronic quasiparticle with charge $e$. In addition to $\Psi_{\mathrm{el}}$, there are local spin 1 bosons generated by the electrically neutral and $\mathbb{Z}_3$ neutral combinations $E^{ab}=e^{i(\tilde\phi_a-\tilde\phi_b)}\sim\pi^a(\pi^b)^\dagger$, for $a\neq b$. They form the roots of a $SU(3)$ affine Kac-Moody current Lie algebra at level 1. It will become clear in section~\ref{sec:fillingonethird} that these operators are integral combinations of electrons. 

The $\mathbb{Z}_3$ symmetry rotates the phases of the partons $\pi^a\to e^{2\pi i/3}\pi^a$, and leaves the electronic quasiparticle $\Psi_\mathrm{el}$ invariant. The orbifold construction allows additional twist fields that corresponds to twisted boundary conditions according to the $\mathbb{Z}_3$ symmetry when the $(1+1)$D system is compactified on a torus where both the spatial and temporal directions are periodic. Physically, the orbifold construction addresses the issue of electron locality. If $\pi^a$ were local fermions, the \CFT would not support any additional twist fields that carry non-trivial monodromy with $\pi^a$. However, since the partons $\pi^a$ are fractional quasiparticles that carry a non-trivial $\mathbb{Z}_3$ gauge charge, they are non-local with respect to the $\mathbb{Z}_3$ gauge fluxes.

The orbifold theory includes twist field that carries a $\mathbb{Z}_3$ flux component, such as the Laughlin quasiparticle $\lambda\sim e^{i(\tilde\phi_1+\tilde\phi_2+\tilde\phi_3)/3}$. It carries the electric charge $e^\ast=e/3$ and spin $h_\lambda=1/6$, and generate the $U(1)_3$ charge sector. The bosonic twist fields $\zeta^{ab}=e^{-i(\tilde\phi_a+\tilde\phi_b)}\sim(\pi^a)^\dagger(\pi^b)^\dagger$, for $a\neq b$, serve as representatives for the pure $\mathbb{Z}_3$ charge. This is because they obey the fractional mutual statistics with the Laughlin quasiparticle \begin{align}\langle\lambda(\mathsf{z})\zeta^{ab}(\mathsf{w})\rangle=\frac{1}{(\mathsf{z}-\mathsf{w})^{2/3}}\end{align} The fractional exponent corresponds to a branch cut in the correlation function and gives rise to the non-trivial $e^{2\pi i/3}$ monodromy after a braiding cycle, $\mathsf{z}(s)=\mathsf{w}+|r|e^{is}$, where $s$ runs from 0 to $2\pi$. Electrically neutral combinations of the Laughlin quasiparticle and $\mathbb{Z}_3$ charges, such as $\lambda^2\zeta^{23}\sim e^{i(2\tilde\phi_1-\tilde\phi_2-\tilde\phi_3)/3}$, generate the primary fields for an emergent $SU(3)_1$ sector (see \eqref{SU3fundrep} below). 

The charged $U(1)_3$ and the electrically neutral $SU(3)_1$ sectors fully decouple from each other, and their mutual operator product expansions ({\color{blue}\hypertarget{OPE}{OPE}}s) are non-singular. The decomposition \begin{align}\mathfrak{L}_1&=(\mbox{Dirac parton})^{\otimes3}\nonumber\\&=U(3)_1/\mathbb{Z}_3=U(1)_3\otimes SU(3)_1.\label{DiracPartondef}\end{align} can be summarized by the following fractional basis transformation of bosonized variables. \begin{align}\begin{pmatrix}\phi^0_\pi\\\phi^1_\pi\\\phi^2_\pi\end{pmatrix}=\frac{1}{3}\begin{pmatrix}1&1&1\\2&-1&-1\\1&1&-2\end{pmatrix}\begin{pmatrix}\tilde\phi_1\\\tilde\phi_2\\\tilde\phi_3\end{pmatrix},\label{CWCtrans}\end{align} where the above $3\times3$ transformation matrix is the inverse of \begin{align}A=\begin{pmatrix}1&1&0\\1&-1&1\\1&0&-1\end{pmatrix},\end{align} whose columns are the simple roots of $U(1)\times SU(3)$. In \eqref{CWCtrans}, $\tilde\phi_a$ ($\phi^I$) are referred to as bosonized variables in the Cartan-Weyl (resp.~Chevalley) basis. The Lagrangian density \eqref{U(3)Lagrangian} turns into \begin{align}\mathcal{L}_{U(3)_1/\mathbb{Z}_3}=\frac{1}{2\pi}K_{IJ}\partial_{\mathsf{x}}\phi^I_\pi\partial_{\mathsf{t}}\phi^J_\pi\label{CSLagrangian}\end{align} under the basis transformation. The $K$-matrix in the Chevalley basis is given by \begin{align}K=A^TA=\begin{pmatrix}3&0&0\\0&2&-1\\0&-1&2\end{pmatrix},\label{partonKmatrix}\end{align} and it governs the \ETCR of the Chevalley bosonized variables \begin{align}\left[\partial_{\mathsf{x}}\phi^I_\pi(\mathsf{x}),\phi^J_\pi(\mathsf{x}')\right]=2\pi i(K^{-1})^{IJ}\delta(\mathsf{x}-\mathsf{x}').\label{PartonChevalleyETCR}\end{align} 

The (1,1)-entry of \eqref{partonKmatrix} recovers the $K$-matrix of the Laughlin $\nu=1/3$ \FQH state. The lower $2\times2$ block of \eqref{partonKmatrix} is identical to the Cartan matrix of $SU(3)$. Primary fields are in general represented by vertex operators $e^{im_I\phi^I_\pi}=e^{im_I(A^{-1})^{Ia}\tilde\phi_a}$, where ${\bf m}=(m_0,m_1,m_2)$ are vectors with integral entries. In particular, vectors that fall inside the image of $K$, i.e. ${\bf m}=K{\bf n}$ for some integral vector ${\bf n}$, correspond to local fields that have trivial monodromy with all primary fields and account for all integral electronic combinations. For example, $\Psi_{\mathrm{el}}\sim e^{i(\tilde\phi_1+\tilde\phi_2+\tilde\phi_3)}=e^{i3\phi^0_\pi}$ is the smallest local electronic quasiparticle in the Laughlin charge sector. The neutral combinations $\pi^a(\pi^b)^\dagger\sim E^{ab}=e^{i(\tilde\phi_a-\tilde\phi_b)}=e^{\pm i(2\phi^1_\pi-\phi^2_\pi)}$, $e^{\pm i(-\phi^1_\pi+2\phi^2_\pi)}$ or $e^{\pm i(\phi^1_\pi+\phi^2_\pi)}$, for $a\neq b$, are local fields that form the 6 roots of the $SU(3)$ affine Kac-Moody current Lie algebra~\cite{bigyellowbook} at level 1. The $U(1)_3\times SU(3)_1$ currents obey the singular \OPE \begin{align}i\partial\tilde\phi_a(\mathsf{z})i\partial\tilde\phi_b(\mathsf{w})&=\frac{\delta_{ab}}{(\mathsf{z}-\mathsf{w})^2}+\ldots,\nonumber\\i\partial\tilde\phi_a(\mathsf{z})E^{bc}(\mathsf{w})&=\frac{\delta_a^b-\delta_a^c}{\mathsf{z}-\mathsf{w}}E^{bc}(\mathsf{w})+\ldots,\label{SU3currentOPE}\\E^{ab}(\mathsf{z})E^{cd}(\mathsf{w})&=\frac{\delta^{ad}\delta^{bc}}{(\mathsf{z}-\mathsf{w})^2}+\frac{if^{(ab)(cd)(ef)}}{\mathsf{z}-\mathsf{w}}E^{ef}(\mathsf{w})+\ldots\nonumber\end{align} where $f^{(ab)(cd)(ef)}=\varepsilon^{abd}\delta^{bc}\delta^{ae}\delta^{df}-\varepsilon^{abc}\delta^{ad}\delta^{bf}\delta^{ce}$ is the structure factor of $SU(3)$ where $a\neq b$, $c\neq d$ and $e\neq f$. The derivations of the current algebra \eqref{SU3currentOPE} as well as \hyperlink{OPE}{OPE}s between vertex operators in general are available in appendix~\ref{app:partonalgebra}.

The Laughlin quasiparticle $\lambda\sim e^{i\phi^0_\pi}=e^{i(\tilde\phi_1+\tilde\phi_2+\tilde\phi_3)/3}$ is the smallest non-trivial primary field in the $U(1)_3$ charge sector. It decouples from the $SU(3)_1$ neutral sector and the \OPE between $\lambda$ and any of the $SU(3)_1$ roots $E^{ab}$ is non-singular. The non-trivial primary fields of $SU(3)_1$ are given by the two fundamental representations of the Lie algebra. Linear combinations of primary fields in the super-selection sectors \begin{gather}\mathcal{E}=\mathrm{span}\left\{\mathcal{E}^1,\mathcal{E}^2,\mathcal{E}^3\right\},\quad\overline{\mathcal{E}}=\mathrm{span}\left\{(\mathcal{E}^1)^\dagger,(\mathcal{E}^2)^\dagger,(\mathcal{E}^3)^\dagger\right\},\nonumber\\\mathcal{E}^a=\lambda^\dagger\pi^a\sim e^{i[\tilde\phi_a-(\tilde\phi_1+\tilde\phi_2+\tilde\phi_3)/3]}\label{SU3fundrep}\end{gather}
rotate according to the \hyperlink{OPE}{OPE}s \begin{align}&E^{ab}(\mathsf{z})\mathcal{E}^c(\mathsf{w})=-iS_{ab}\frac{\delta^{bc}\delta^a_d}{\mathsf{z}-\mathsf{w}}\mathcal{E}^d(\mathsf{w})+\ldots\label{SU3OPE}\\&E^{ab}(\mathsf{z})\mathcal{E}^c(\mathsf{w})^\dagger=iS_{ab}\frac{\delta^{ac}\delta^b_d}{\mathsf{z}-\mathsf{w}}\mathcal{E}^d(\mathsf{w})^\dagger+\ldots\nonumber\end{align}
and form a conjugate pair of three dimensional irreducible representations of $SU(3)$. For instance, the matrices $L^{ab}=(L^{abc}_d)_{3\times3}$ and $\overline{L}^{ab}=(\overline{L}^{abc}_d)_{3\times3}$, for $L^{abc}_d=\delta^{bc}\delta^a_d$ and $\overline{L}^{abc}_d=\delta^{ac}\delta^b_d$, are the raising and lowering matrices for the off-diagonal Gell-Mann matrices $L^{ab}+\overline{L}^{ab}$ and $i(L^{ab}-\overline{L}^{ab})$, for $a\neq b$. All primary fields in $\mathcal{E}$ and $\overline{\mathcal{E}}$ are electrically neutral and carry spin $h=1/3$. Since the root operators $E^{ab}$ are local electronic, the \hyperlink{OPE}{OPE}s in \eqref{SU3OPE} show that any two primary fields within the same sector $\mathcal{E}$ or $\overline{\mathcal{E}}$ are equivalent up to local electrons. They obey the fusion rules \begin{align}\mathcal{E}\times\mathcal{E}=\overline{\mathcal{E}},\quad\overline{\mathcal{E}}\times\overline{\mathcal{E}}=\mathcal{E},\quad \mathcal{E}\times\overline{\mathcal{E}}=1,\end{align} which abbreviate the \hyperlink{OPE}{OPE}s \begin{align}&\mathcal{E}^a(\mathsf{z})\mathcal{E}^b(\mathsf{w})=\frac{-i\varepsilon^{abc}}{(\mathsf{z}-\mathsf{w})^{1/3}}\mathcal{E}^c(\mathsf{w})^\dagger+\ldots,\nonumber\\&\mathcal{E}^a(\mathsf{z})^\dagger\mathcal{E}^b(\mathsf{w})^\dagger=\frac{-i\varepsilon^{abc}}{(\mathsf{z}-\mathsf{w})^{1/3}}\mathcal{E}^c(\mathsf{w})+\ldots,\nonumber\\&\mathcal{E}^a(\mathsf{z})\mathcal{E}^b(\mathsf{w})^\dagger=\frac{\delta^{ab}}{(\mathsf{z}-\mathsf{w})^{2/3}}+\ldots.\end{align}

The Dirac parton triplet carries electric and energy transport. The external electromagntic gauge field is coupled to the theory through the $U(1)_{EM}$ transformation $\tilde\phi_a\to\tilde\phi_a+\Lambda/3$ [or equivalently $\phi^I_\pi\to\phi^I_\pi+\Lambda(K^{-1})^{IJ}t_J$, where the charge vector is ${\bf t}^T=(t_0,t_1,t_2)=(1,0,0)$], when the electronic quasiparticle transforms by $\Psi_{\mathrm{el}}=\pi^1\pi^2\pi^3\to e^{i\Lambda}\Psi_{\mathrm{el}}$. In particular, the differential conductance of the parton channel is given by \begin{align}\sigma=\frac{dI}{dV}=\frac{e^2}{h}{\bf t}^TK^{-1}{\bf t}=\frac{1}{3}\frac{e^2}{h},\end{align} which associates to the filling fraction $\nu=1/3$ of a \FQH state that supports a boundary Dirac parton triplet \hyperlink{CFT}{CFT}. As all three bosons propagate in the same direction, the channel carries a chiral central charge $c=c_R-c_L=3$, which dictates the differential thermal conductance~\cite{KaneFisher97,Cappelli01,Kitaev06} \begin{align}\kappa=\frac{dI_{E}}{dT}=c\left(\frac{\pi^2k_B^2}{3h}\right)T\end{align} in low temperature.

\subsection{The parton Pfaffian}\label{sec:PartonPfaffian}
The parton Pfaffian \hyperlink{CFT}{CFT}, denoted by $\mathrm{Pf}^\ast$, carries exactly half of the degrees of freedom of the Dirac parton triplet. It consists of an electrically charged $U(1)_{24}$ sector and a neutral $U(1)_{12}\times\overline{\mathrm{Ising}}$ sector. \begin{align}\mathfrak{P}_{-1,1}=\mathrm{Pf}^\ast=U(1)_{24}\otimes_{\mathrm{el}}U(1)_{12}\otimes_{\mathrm{el}}\overline{\mathrm{Ising}}.\label{Pfdef}\end{align} Each sector is generated by a central ``almost local" primary field that has non-trivial monodromy only with Ising anyons or $\pi$-fluxes. \begin{align}\begin{split}&U(1)_{24}\ni\mbox{charge $e$ spin 3 boson }e^{i\Phi_\rho}\\&U(1)_{12}\ni\mbox{neutral spin $3/2$ Dirac fermion }e^{i\Phi_d}\\&\overline{\mathrm{Ising}}\ni\mbox{Majorana fermion }\psi.\end{split}\end{align} The electrically charged $U(1)_{24}$ and neutral $U(1)_{12}$ sectors are Abelian and can be described by a two-component bosonized theory \begin{align}\mathcal{L}&=\frac{1}{2\pi}\left(K_\rho\partial_{\mathsf{t}}\phi_\rho\partial_{\mathsf{x}}\phi_\rho+K_d\partial_{\mathsf{t}}\phi_d\partial_{\mathsf{x}}\phi_d\right)\label{CSPfAbCFT}\end{align} up to non-universal velocity terms, where the $K$-matrices \begin{gather}K_\rho=24,\quad K_d=12\end{gather} define the levels of the two $U(1)$ sectors. The $\overline{\mathrm{Ising}}$ sector is generated by a Majorana fermions $\psi$, which is described by the Lagrangian density \begin{align}\mathcal{L}=i\psi\left(\partial_t+v\partial_x\right)\psi.\label{CSPfMF}\end{align} The Majorana fermion propagates in the opposite direction to the two $U(1)$ sectors.

The primary field content of the Abelian part of \eqref{Pfdef} can be represented by vertex operators $e^{i(a\phi_\rho+b\phi_d)}$, where $a,b$ are integers. The charged $U(1)_{24}$ sector supports primary fields $e^{ia\phi_\rho}$, denoted by $[a]_\rho$, that carry charge $q_a=a/12$ (in units of the electric charge $e$) and spin $h_a=a^2/48$. In particular, $[12]_\rho$ represents the ``almost local" charge $e$ boson $e^{i\Phi_\rho}$, where $\Phi_\rho=12\phi_\rho$. The external $U(1)_{\mathrm{EM}}$ that transforms $e^{i\Phi_\rho}\to e^{i\Phi_\rho}e^{i\Lambda}$ shifts $\phi_\rho\to\phi_\rho+\Lambda/12=\phi_\rho+\Lambda K_\rho^{-1}t_\rho$, where the charge vector is $t_\rho=2$. This set the filling fraction \begin{align}\nu=t_\rho K_\rho^{-1}t_\rho=\frac{1}{6}\end{align} that determines the differential conductance $\sigma=\nu(e^2/h)$ of the charge sector. The primary fields satisfy the fusion rule \begin{align}[a]_\rho\times[a']_\rho=[a+a']_\rho,\label{Abfusion}\end{align} which is an abbreviation of the operator product expansion $e^{ia\phi_\rho(\mathsf{z})}e^{ia'\phi_\rho(\mathsf{w})}=e^{i(a+a')\phi_\rho(\mathsf{w})}(\mathsf{z}-\mathsf{w})^{aa'/24}+\ldots$. 

The neutral $U(1)_{12}$ sector supports similar primary fields. The vertex operators $e^{ib\phi_d}$ are denoted by $[b]_d$. They carry spin $h_b=b^2/24$ and follow similar fusion rules in \eqref{Abfusion}. $[6]_d$ represents the ``almost local" neutral spin-$3/2$ Dirac fermion $e^{i\Phi_d}$, where $\Phi_d=6\phi_d$. Both the charged $U(1)_{24}$ and neutral $U(1)_{12}$ sectors are $\mathbb{Z}_2$ graded with respect to their central elements $[12]_\rho$ and $[6]_d$ respectively. Each primary field is assigned a parity according its monodromy with the central element. The monodromy can be determined by the time-ordered correlations \begin{align}\begin{split}&\left\langle e^{ia\phi_\rho(\mathsf{z})}e^{i\Phi_\rho(\mathsf{w})}\right\rangle\sim(\mathsf{z}-\mathsf{w})^{a/2},\\&\left\langle e^{ib\phi_d(\mathsf{z})}e^{i\Phi_d(\mathsf{w})}\right\rangle\sim(\mathsf{z}-\mathsf{w})^{b/2},\end{split}\end{align} which carry branch cuts (i.e.~odd monodromy) when $a,b$ are odd. 
Odd primary fields are also referred to as $\pi$-fluxes.

The $\overline{\mathrm{Ising}}$ \CFT supports two non-trivial primary fields. The first is the Majorana fermion $\psi$. It carries spin $h_\psi=-1/2$ and follows the fusion rule $[\psi]\times[\psi]=1$, which is an abbreviation for the operator product expansion $\psi(\overline{\mathsf{z}})\psi(\overline{\mathsf{w}})=1/(\overline{\mathsf{z}}-\overline{\mathsf{w}})+\ldots$. The second is the Ising twist field $\sigma$. It carries spin $h=-1/16$ and following the non-Abelian fusion rules \begin{align}[\sigma]\times[\psi]=[\sigma],\quad[\sigma]\times[\sigma]=1+[\psi].\end{align} They corresponds to the \hyperlink{OPE}{OPE}s $\sigma(\overline{\mathsf{z}})\psi(\overline{\mathsf{w}})=\sigma(\overline{\mathsf{w}})/(\overline{\mathsf{z}}-\overline{\mathsf{w}})^{1/2}+\ldots$ and $\sigma(\overline{\mathsf{z}})\sigma(\overline{\mathsf{w}})=1/(\overline{\mathsf{z}}-\overline{\mathsf{w}})^{1/8}+\psi(\overline{\mathsf{w}})(\overline{\mathsf{z}}-\overline{\mathsf{w}})^3/8+\ldots$. Like the other two Abelian sectors, the $\overline{\mathrm{Ising}}$ \CFT is also $\mathbb{Z}_2$-graded with respect to the fermion $\psi$. The 1 and $[\psi]$ primary fields are even, and the Ising twist field $[\sigma]$ is odd as it has odd monodromy with $\psi$.

We associate the three primary field combinations \begin{align}\begin{split}&\Psi_{\mathrm{el}_1}=[12]_\rho\otimes[\psi]\sim e^{i\Phi_\rho}\psi\\&\Psi_{\mathrm{el}_2}=[12]_\rho\otimes[6]_d\sim e^{i\Phi_\rho}e^{i\Phi_d}\\&\Psi_{\mathrm{el}_3}=[12]_\rho\otimes[-6]_d\sim e^{i\Phi_\rho}e^{-i\Phi_d}\end{split}\label{Pflocalel}\end{align} to be electronic. This means that they are treated as combinations of integral products of electronic operators. All of them are charge $e$ Dirac fermions. They carry spin $h_1=5/2$ and $h_2=h_3=9/2$ respectively. In general, a primary field in the parton Pfaffian \CFT is local electronic if it can be expressed as combinations of $\Psi_{\mathrm{el}_1}$, $\Psi_{\mathrm{el}_2}$ and $\Psi_{\mathrm{el}_3}$. For example, $\Psi_{\mathrm{el}_1}^\dagger\Psi_{\mathrm{el}_2}=[6]_d\otimes[\psi]$ is a spin 1 neutral boson, and is an integral combination of electrons. The locality of the electron forbids the presence of any twist field that exhibit non-trivial monodromy with any of the $\Psi_{\mathrm{el}}$'s. This includes all the $\pi$ fluxes to the electron, such as $[1]_\rho$, $[1]_d$ and $[\sigma]$. The ``confinement" of $\pi$ fluxes is indicated by the ``electronic" tensor product $\otimes_{\mathrm{el}}$ in \eqref{Pfdef}. A general physical primary field of the parton Pfaffian theory takes a tensor product form, and belong in one of the following types \begin{align}\openone_{a,b}&=[a]_\rho\otimes[b]_d\nonumber\\&\sim e^{i(a\phi_\rho+b\phi_d)},\quad\mbox{for $a$ and $b$ even},\nonumber\\\Psi_{a,b}&=[a]_\rho\otimes[b]_d\otimes[\psi]\nonumber\\&\sim e^{i(a\phi_\rho+b\phi_d)}\psi,\quad\mbox{for $a$ and $b$ even},\label{Pfprimaryfieldcontent}\\\Sigma_{a,b}&=[a]_\rho\otimes[b]_d\otimes[\sigma]\nonumber\\&\sim e^{i(a\phi_\rho+b\phi_d)}\sigma,\quad\mbox{for $a$ and $b$ odd}.\nonumber\end{align} In other words, the three sector components must be either all even or all odd according to the $\mathbb{Z}_2$ grading. All other combinations are not allowed by electron locality.

The electric charges, spins and fusion rules of these primary fields can be deduced from that of the three components of $\mathrm{Pf}^\ast$ in \eqref{Pfdef}. $X_{a,b}$ carries the electric charge $q_{X_{a,b}}=a/12$ (in unit of the electric charge $e$), for $X=\openone,\Psi,\Sigma$. Their spins are given by \begin{align}h_{\openone_{a,b}}&=\frac{a^2}{48}+\frac{b^2}{24},\nonumber\\h_{\Psi_{a,b}}&=\frac{a^2}{48}+\frac{b^2}{24}-\frac{1}{2},\nonumber\\h_{\Sigma_{a,b}}&=\frac{a^2}{48}+\frac{b^2}{24}-\frac{1}{16}.\label{Pfspins}\end{align} They follow the fusion rule \begin{gather}\openone_{a,b}\times\openone_{a',b'}=\Psi_{a,b}\times\Psi_{a',b'}=\openone_{a+a',b+b'}\nonumber\\\openone_{a,b}\times\Psi_{a',b'}=\Psi_{a+a',b+b'}\nonumber\\\openone_{a,b}\times\Sigma_{a',b'}=\Psi_{a,b}\times\Sigma_{a',b'}=\Sigma_{a',b'}\nonumber\\\Sigma_{a,b}\times\Sigma_{a',b'}=\openone_{a+a',b+b'}+\Psi_{a+a',b+b'}.\label{Pffusion}\end{gather} Within \eqref{Pfprimaryfieldcontent}, $\openone_{4,\pm2}$ and $\Psi_{4,\pm4}$ are charge $e/3$ spin $1/2$ fermions and serve as the deconfined partons that decompose the electronic quasiparticles \begin{align}\Psi_{\mathrm{el}_1}&=\Psi_{4,\pm4}\times\openone_{4,\mp2}\times\openone_{4,\mp2}=\Psi_{4,\pm4}\times\Psi_{4,\pm4}\times\Psi_{4,\pm4},\nonumber\\\Psi_{\mathrm{el}_2}&=\Psi_{4,4}\times\Psi_{4,4}\times\openone_{4,-2}=\openone_{4,2}\times\openone_{4,2}\times\openone_{4,2},\label{Pfaffianpartons}\\\Psi_{\mathrm{el}_3}&=\Psi_{4,-4}\times\Psi_{4,-4}\times\openone_{4,2}=\openone_{4,-2}\times\openone_{4,-2}\times\openone_{4,-2}.\nonumber\end{align} 

By summing the contributions from the three components, the overall electric and thermal conductance $\sigma=\nu e^2/h$ and $\kappa=c(\pi^2k_B^2/3h)T$ are specified by \begin{align}\nu=\frac{1}{6}+0+0=\frac{1}{6},\quad c=1+1-\frac{1}{2}=\frac{3}{2},\end{align} each of which is half of that of the parton Dirac triplet $U(3)_1/\mathbb{Z}_3$ described in the previous subsection. This suggests the decomposition or conformal embedding \begin{align}U(3)_1/\mathbb{Z}_3\sim\mathrm{Pf}^\ast\otimes\mathrm{Pf}^\ast,\end{align} which will be discussed in the following subsection. 

Before doing so, it is worth noticing that within the parton Pfaffian theory, there is a subset of primary fields \begin{align}I_a&=[a]_\rho\otimes[a]_d\sim e^{ia(\phi_\rho+\phi_d)},\quad\mbox{for $a$ even},\nonumber\\\psi_a&=[a]_\rho\otimes[a]_d\otimes[\psi]\sim e^{ia(\phi_\rho+\phi_d)}\psi,\quad\mbox{for $a$ even},\label{TPfcontent}\\\sigma_a&=[a]_\rho\otimes[a]_d\otimes[\sigma]\sim e^{ia(\phi_\rho+\phi_d)}\sigma,\quad\mbox{for $a$ odd},\nonumber\end{align} that are bosonic, fermionic, or semionic combinations with spins \begin{align}h_{I_a}&=\frac{(a/2)^2}{4},\quad\mbox{for $a$ even},\nonumber\\h_{\psi_a}&=\frac{(a/2)^2}{4}-\frac{1}{2},\quad\mbox{for $a$ even},\nonumber\\h_{\sigma_a}&=\frac{a^2-1}{16},\quad\mbox{for $a$ odd},\label{TPfspins}\end{align} and are closed under fusion \begin{gather}I_a\times I_{a'}=\psi_a\times\psi_{a'}=I_{a+a'},\nonumber\\I_a\times\psi_{a'}=\psi_{a+a'},\quad\psi_a\times\sigma_{a'}=\sigma_{a+a'}\nonumber\\\sigma_a\times\sigma_{a'}=I_{a+a'}+\sigma_{a+a'}.\label{TPffusion}\end{gather} This sub-collection generates a theory, referred to as the parton $T$-Pfaffian state and denoted by $\mathcal{T}$-$\mathrm{Pf}^\ast$, that describes the gapped time-reversal symmetric surface state of a fractional topological insulator~\cite{SahooSirotaChoTeo17,ChoTeoFradkin17}. The $\mathcal{T}$-$\mathrm{Pf}^\ast$ state has a similar topological anyon content as the conventional T-Pfaffian~\cite{ChenFidkowskiVishwanath14} surface state of a single-body topological insulator. Except the charge assignment of each anyon is $1/3$ of the conventional one. Also, there is a threefold increase of periodicity. In particular, the spin-$1/2$ quasiparticle $\psi_4$ in the $\mathcal{T}$-$\mathrm{Pf}^\ast$ takes the role of the fermionic parton and carries electric charge $e/3$. $\psi_{12}$ is the electronic quasiparticle and $1_{24}$ is the charge $2e$ bosonic Cooper pair.

\subsection{Gluing and splitting}\label{sec:gluingsplitting}

A pair of parton Pfaffian channels can be glued into a parton Dirac channel through an anyon condensation~\cite{BaisSlingerlandCondensation} process \begin{align}U(3)_1/\mathbb{Z}_3=\mathrm{Pf}^\ast\boxtimes\mathrm{Pf}^\ast,\end{align} which can be carried out explicitly by a sine-Gordon Hamiltonian. We begin with a pair of parton Pfaffian theories $\mathrm{Pf}^\ast_A\otimes\mathrm{Pf}^\ast_B$, each described by \eqref{Pfdef}. First, we focus on the two neutral Ising sectors $\overline{\mathrm{Ising}}_A\otimes\overline{\mathrm{Ising}}_B$. By condensing the bosonic fermion pair $[\psi]_A\otimes[\psi]_B$, they becomes \begin{align}\overline{SO(2)_1}=\overline{\mathrm{Ising}}_A\boxtimes\overline{\mathrm{Ising}}_B\label{SO2Isingsq}\end{align} where the product notation $\boxtimes$ here stands for a tensor product and the condensation of the fermion pair. We define the electrically neutral spin $h=-1/2$ Dirac fermion \begin{align}d_0=\frac{1}{\sqrt{2}}\left(\psi_A+i\psi_B\right)\sim e^{i2\phi_0}\label{gluingMajdecomp}\end{align} The neutral fermion $d_0$ is fractional, however it can be made local electronic by combining with the charge $e$ boson from either one of the charge sectors $U(1)^A_{24}$ or $U(1)^B_{24}$. The bosonized variable $\phi_0$ is described by the Lagrangian density \begin{align}\mathcal{L}_0=\frac{1}{2\pi}K_0\partial_{\mathsf{t}}\phi_0\partial_{\mathsf{x}}\phi_0\label{gluingL0}\end{align} up to non-universal velocity terms, where $K_0=-4$. The vertex operators $s_\pm=e^{\pm i\phi_0}$ are spin $h=-1/8$ spinor fields that originated from the pair of Ising twist fields \begin{align}[\sigma]_A\otimes[\sigma]_B=s_++s_-.\label{Isingsqspinor}\end{align} Similar to the Ising twist fields, these spinor fields $s_\pm$ have non-trivial monodromy with the electronic quasiparticles $\Psi_{\mathrm{el}_j}^{A,B}$ in \eqref{Pflocalel} and are confined. However, they can be combined with the $\pi$-fluxes in the Abelian sectors $U(1)_{24}^{A,B}$ and $U(1)_{12}^{A,B}$ to form deconfined excitations. The spin $-2$ boson $d_0^2\sim e^{i4\phi_0}$ condenses because $\psi\times\psi=1$. This fixes the fusion rules \begin{gather}d_0\times d_0=1,\quad d_0\times s_\pm=s_\mp,\nonumber\\s_\pm\times s_\pm=d_0,\quad s_\pm\times s_\mp=1.\label{SO2fusion}\end{gather}

Next, we take the rest of the Abelian components $\left[U(1)_{24}^A\times U(1)_{12}^A\right]\times\left[U(1)_{24}^B\times U(1)_{12}^B\right]$ into account. The parton Pfaffian pair is now described by the five-component bosonized Lagrangian density \begin{align}\mathcal{L}_{\mathrm{Pf}^\ast\otimes\mathrm{Pf}^\ast}&=\mathcal{L}_0+\mathcal{L}_d^A+\mathcal{L}_d^B+\mathcal{L}_\rho^A+\mathcal{L}_\rho^B,\label{gluingL1}\\\mathcal{L}_d^{A,B}&=\frac{1}{2\pi}K_d\partial_{\mathsf{t}}\phi_d^{A,B}\partial_{\mathsf{x}}\phi_d^{A,B},\nonumber\\\mathcal{L}_\rho^{A,B}&=\frac{1}{2\pi}K_\rho\partial_{\mathsf{t}}\phi_\rho^{A,B}\partial_{\mathsf{x}}\phi_\rho^{A,B}\nonumber\end{align} up to non-universal velocity terms, where $K_\rho=24$ and $K_d=12$ for both the $A$ and $B$ sectors. The first three bosonzied variables represent the neutral degrees of freedom and are invariant under $U(1)_{\mathrm{EM}}$. The final two represent the charged sectors and transforms under $\phi_\rho^{A,B}\to\phi_\rho^{A,B}+\Lambda/12$ when electron operators change by $\Psi_{\mathrm{el}}\to e^{i\Lambda}\Psi_{\mathrm{el}}$.

We define a new basis within the neutral sectors by the transformation \begin{align}\begin{pmatrix}\phi^{n0}_L\\\phi^{n1}_R\\\phi^{n2}_R\end{pmatrix}=2\begin{pmatrix}1&-1&1\\-1&2&-1\\-1&1&-2\end{pmatrix}\begin{pmatrix}\phi_0\\\phi_d^A\\\phi_d^B\end{pmatrix},\label{Pf2condensedtransformation}\end{align} where the $3\times3$ matrix is unimodular and has the inverse \begin{align}B=\begin{pmatrix}1&-1&1\\-1&2&-1\\-1&1&-2\end{pmatrix}^{-1}=\begin{pmatrix}3&1&1\\1&1&0\\-1&0&-1\end{pmatrix}.\label{Bmatirx}\end{align} The new bosonized variables are described by the Lagrangian density \begin{align}\mathcal{L}_n&=\mathcal{L}_0+\mathcal{L}_d^A+\mathcal{L}_d^B=\frac{1}{2\pi}(K_n)_{IJ}\partial_{\mathsf{t}}\phi^{nI}\partial_{\mathsf{x}}\phi^{nJ},\end{align} where the $K$-matrix (suppressing all zeros) is \begin{align}K_n=\begin{pmatrix}-3&&\\&2&-1\\&-1&2\end{pmatrix}=B^T\begin{pmatrix}-1&&\\&3&\\&&3\end{pmatrix}B.\label{gluingKn}\end{align} 

Lastly, we introduce the gluing sine-Gordon potential \begin{align}\mathcal{H}_{\mathrm{gluing}}&=u\cos\left(12\phi^{n0}_L-24\phi_\rho^A+24\phi_\rho^B\right)\nonumber\\&=u\cos\left[24\left(\phi_0-\phi_d^A+\phi_d^B-\phi_\rho^A+\phi_\rho^B\right)\right].\label{gluingH}\end{align} It is straightforward to check that the angle variable $\Theta=\phi_0-\phi_d^A+\phi_d^B-\phi_\rho^A+\phi_\rho^B$ satisfies the ``Haldane's nullity condition"~\cite{Haldane95} $[\Theta(\mathsf{x}),\Theta(\mathsf{x}')]=0$. The factor of 24 makes sure $\mathcal{H}_{\mathrm{gluing}}$ is constructed by integral combinations of local electronic operators. The potential therefore introduces a finite excitation energy gap to a counter-propagating pair of boson modes and removes them from low-energy. It is charge preserving as $\Theta$ is invariant under $U(1)_{\mathrm{EM}}$. The gapping potential pins a finite ground state expectation value for $\langle\Theta(\mathsf{x})\rangle$ and condenses the bosonic combination \begin{align}1\propto e^{i\langle\Theta\rangle}&\propto\left\langle e^{i\phi_0}e^{-i(\phi^A_\rho+\phi^A_d)}e^{i(\phi^B_\rho+\phi^B_d)}\right\rangle\nonumber\\&\sim\left\langle s_+{I^A_1}^\dagger I^B_1\right\rangle=\left\langle{\sigma^A_1}^\dagger\sigma^B_1\right\rangle.\end{align} This is because the Ising anyon $\sigma_1$ (defined in \eqref{TPfcontent}) is the product $\sigma I_1$ for both the $A$ and $B$ sectors, and from \eqref{Isingsqspinor}, the even spinor $s_+$ is originated from the product $\sigma^A\sigma^B$. Together with the bosonic fermion pair $\psi_0^A\psi_0^B$ that was condensed previously in \eqref{SO2Isingsq}, they generate all the electrically neutral bosonic pairs in $\mathrm{Pf}^\ast_A\otimes\mathrm{Pf}^\ast_B$ \begin{align}\mathcal{B}=\left\{\begin{array}{*{20}c}{I_{4m}^A}^\dagger I_{4m}^B,{\psi_{4m}^A}^\dagger\psi_{4m}^B,{\psi_{4m+2}^A}^\dagger I_{4m+2}^B,\\{I_{4m+2}^A}^\dagger\psi_{4m+2}^B,{\sigma^A_{2m+1}}^\dagger\sigma^B_{2m+1}\end{array}\right\}_{m\in\mathbb{Z}}.\label{condensedbosonsPF2}\end{align} Bosons in $\mathcal{B}$ have trivial mutual monodromy, and the sine-Gordon potential \eqref{gluingH} condenses all bosons in $\mathcal{B}$. For example, the Ising pairs take non-vanishing ground state expectation values thanks to $\mathcal{H}_{\mathrm{gluing}}$ \begin{align}1&\propto e^{i(2m+1)\langle\Theta\rangle}\nonumber\\&\propto\left\langle e^{i(2m+1)\phi_0}e^{-i(2m+1)(\phi^A_\rho+\phi^A_d)}e^{i(2m+1)(\phi^B_\rho+\phi^B_d)}\right\rangle\nonumber\\&\sim\left\langle s_\pm{I^A_{2m+1}}^\dagger I^B_{2m+1}\right\rangle=\left\langle{\sigma^A_{2m+1}}^\dagger\sigma^B_{2m+1}\right\rangle,
\end{align} where the parity of the spinor field is fixed by the fusion rule \eqref{SO2fusion}, $e^{i(2m+1)\phi_0}=s_+$ (or $s_-$) if $m$ is even (resp.~odd). We denote the {\em relative} tensor product $\mathrm{Pf}^\ast_A\boxtimes_\mathcal{B}\mathrm{Pf}^\ast_B$ to be the remaining low-energy \CFT that is unaffected by the condensation. 

From the $K$-matrix \eqref{gluingKn}, it is clear that the two electrically neutral modes $\phi^{n1}$ and $\phi^{n2}$ are completely decoupled from the rest and therefore are unaffected by the sine-Gordon potential $\mathcal{H}_{\mathrm{gluing}}$. It is also straightforward to check that the sum $\phi_\rho=2(\phi_\rho^A+\phi_\rho^B)$ commutes with $\Theta$ and thus decouples from $\mathcal{H}_{\mathrm{gluing}}$ as well. Here we normalizes $\phi_\rho$ with the factor of 2 because the odd vertices, such as $e^{i(\phi_\rho^A+\phi_\rho^B)}$, have non-trivial monodromy with the electronic quasiparticles $\psi_{12}^A$ and $\psi_{12}^B$ and are not allowed by electron locality. Grouping together, the bosonized variables $(\phi^0_\pi,\phi^1_\pi,\phi^2_\pi)=(\phi_\rho,\phi^{n1},\phi^{n2})$ generate the relative tensor product $\mathrm{Pf}^\ast_A\boxtimes_\mathcal{B}\mathrm{Pf}^\ast_B$. They obey the same equal-time commutation relation in \eqref{PartonChevalleyETCR}, and therefore are described by the same $U(1)_3\times SU(3)_1$ parton Lagrangian density \eqref{CSLagrangian}. 

Moreover, the bosonized variables $(\phi_\rho,\phi^{n1},\phi^{n2})$ respect the same charge assignment and electron locality as the Dirac parton triplet. $\phi_\rho$ is the only bosonized variable that transform under $U(1)_{\mathrm{EM}}$. The quasiparticle $e^{i\phi_\rho}=e^{i2(\phi_\rho^A+\phi_\rho^B)}$ carries charge $e/3$ and represents the Laughlin quasiparticle. We now show the triple $e^{i3\phi_\rho}=e^{i6(\phi_\rho^A+\phi_\rho^B)}$ is an integral electronic combination. First, it can be changed into $I_{-6}^A\psi_6^B\sim e^{i6(-\phi_\rho^A-\phi_d^A+\phi_\rho^B+\phi_d^B)}\psi^B$ by absorbing the electronic combination ${\Psi_{\mathrm{el}_2}^A}^\dagger\Psi_{\mathrm{el}_1}^B{\Psi_{\mathrm{el}_3}^B}^\dagger=e^{-i(12\phi_\rho^A+6\phi_d^A)}e^{i6\phi_d}\psi^B_0$. Second, we observe that $I_{-6}^A\psi_6^B={I_6^A}^\dagger\psi_6^B$ is one of the boson in \eqref{condensedbosonsPF2} that condensed to the ground state. This proves $e^{i3\phi_\rho}$ represents an electronic quasiparticle after the condensation. Furthermore, in the neutral sector, the simple $SU(3)$ roots $e^{i(2\phi^{n1}-\phi^{n2})}$ and $e^{i(-\phi^{n1}+2\phi^{n2})}$ are also local electronic. Using the basis transformation \eqref{Pf2condensedtransformation}, the roots are $e^{-i(2\phi_0-6\phi_d^A)}$ and $e^{-i(2\phi_0+6\phi_d^B)}$, which are respectively identical to the electronic operators $\psi^A_0e^{i6\phi_d^A}=\Psi_{\mathrm{el}_1}^A{\Psi_{\mathrm{el}_3}^A}^\dagger$ and $\psi^B_0e^{-i6\phi_d^B}={\Psi_{\mathrm{el}_1}^B}^\dagger\Psi_{\mathrm{el}_3}^B$ up to the condensate $\psi^A_0\psi^B_0$. 

This concludes the gluing procedure \begin{align}\mathrm{Pf}^\ast\otimes\mathrm{Pf}^\ast\xrightarrow{\mathcal{H}_{\mathrm{gluing}}}&\mathrm{Pf}^\ast\boxtimes_{\mathcal{B}}\mathrm{Pf}^\ast=U(3)_1/\mathbb{Z}_3.\end{align} Similar gluing procedure was also presented by us in an earlier work~\cite{SahooSirotaChoTeo17}.

\begin{figure}[htbp]
\centering\includegraphics[width=0.48\textwidth]{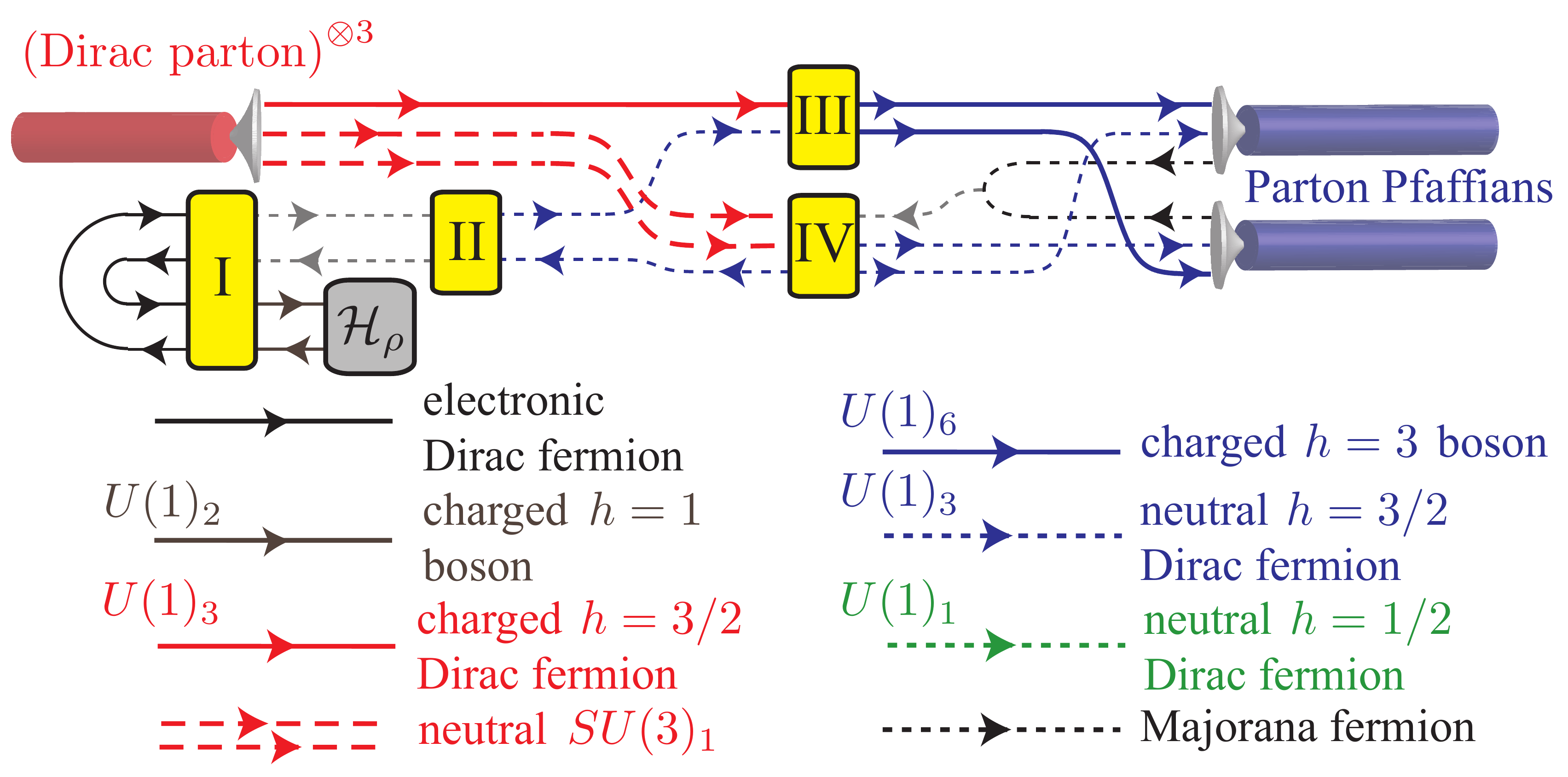}
\caption{Splitting of a chiral Dirac triplet channel into a pair of parton Pfaffian channels. The basis transformations (I), (II), (III) and (IV) are defined in \eqref{splittingbasistrans1}, \eqref{splittingbasistrans2}, \eqref{splittingsd} and \eqref{splittingB} respectively. The backscattering Hamiltonian is defined in \eqref{splittingHrho}.}\label{fig:splitting}
\end{figure}

Next, we present the reversal -- the splitting of the chiral Dirac parton triplet $U(3)_1/\mathbb{Z}_3$ into a pair of decoupled parton Pfaffian \hyperlink{CFT}{CFT}s. The fractionalization is facilitated by an extension of the Dirac partons that includes an additional counter-propagating pair of neutral spin-$3/2$ Dirac fermions, $U(1)_3\times\overline{U(1)_3}$. We first show that the extension can be supported in a purely 1D setting, such as an edge reconstruction, and does not require additional 2D topological order. In other words, the non-chiral $U(1)_3\times\overline{U(1)_3}$ \CFT can be realized in a 1D electronic wire without being holographically supported on the boundary of a 2D topological state. To achieve this, we start with two electronic wires that contains two counter-propagating pairs of Dirac electrons $c^\sigma_a\sim e^{i\tilde\Phi^\sigma_a}$ at Fermi level, where $a=0,1$ is the wire index and $\sigma=R,L=+,-$ labels the propagating direction. They are represented by the solid black lines in figure~\ref{fig:splitting}.  We will introduce a many-body interacting potential that leaves behind a pair of neutral Dirac modes in low-energy.

The interaction is based on the following fractional basis transformation of the bosonized variables \begin{align}\begin{pmatrix}\tilde\Phi^R_\rho\\\tilde\Phi^R_n\\\tilde\Phi^L_\rho\\\tilde\Phi^L_n\end{pmatrix}=\begin{pmatrix}3/2 & 1 & -1/2 & -1 \\ -1/2 & 1 & -1/2 & 0 \\ -1/2 & -1 & 3/2 & 1 \\ -1/2 & 0 & -1/2 & 1 \end{pmatrix}
\begin{pmatrix}\tilde\Phi^R_0\\\tilde\Phi^R_1\\\tilde\Phi^L_0\\\tilde\Phi^L_1\end{pmatrix}.
\label{splittingbasistrans1}\end{align} This transformation will also be useful in the coupled wire construction in section~\ref{sec:fillingonesixth} (c.f.~eq.\eqref{Pfbasistrans1}). We notice that the new bosonized variables contain half-integral combinations of the original ones. Consequently, the vertex operators $e^{i\tilde\Phi^\sigma_\rho},e^{i\tilde\Phi^\sigma_n}$ are non-local and cannot be expressed as integral combination of electronic ones. On the other hand, the sum and differences $\tilde\Phi^R_\rho\pm\tilde\Phi^L_\rho$, $\tilde\Phi^R_n\pm\tilde\Phi^L_n$ and $\tilde\Phi^\sigma_\rho\pm\tilde\Phi^{\sigma'}_n$ are integral. The variables $\tilde\Phi^{R,L}_\rho$ represent electrically charged sector and shift by $\tilde\Phi^\sigma_\rho\to\tilde\Phi^\sigma_\rho+\Lambda$ under a $U(1)_{\mathrm{EM}}$ transformation that change the phase of an electron $c\to e^{i\Lambda}c$. The other two variables $\tilde\Phi^{R,L}_n$ represent electrically neutral modes and are invariant under $U(1)_{\mathrm{EM}}$. The bosonized variables are described by the Lagrangian density \begin{align}\mathcal{L}&=\frac{1}{2\pi}\sum_{\sigma=+,-}\sum_{a=0,1}\sigma\partial_{\mathsf{t}}\tilde\Phi^\sigma_a\partial_{\mathsf{x}}\tilde\Phi^\sigma_a\nonumber\\&=\frac{1}{2\pi}\sum_{\sigma=+,-}\sigma\left(\frac{1}{2}\partial_{\mathsf{t}}\tilde\Phi^\sigma_\rho\partial_{\mathsf{x}}\tilde\Phi^\sigma_\rho+\partial_{\mathsf{t}}\tilde\Phi^\sigma_n\partial_{\mathsf{x}}\tilde\Phi^\sigma_n\right)\label{splttingL1}\end{align} up to non-universal velocity terms.

We introduce the sine-Gordon potential \begin{align}\mathcal{H}_\rho&=u\cos\left(\tilde\Phi^R_\rho-\tilde\Phi^L_\rho\right)\nonumber\\&=u\cos\left(2\tilde\Phi^R_0+2\tilde\Phi^R_1-2\tilde\Phi^L_0-2\tilde\Phi^L_1\right),\label{splittingHrho}\end{align} which can be constructed from electron backscattering and preserves charge $U(1)_{\mathrm{EM}}$. This gives a finite excitation energy gap to the charged sector $\rho$ and removes it from low-energy, and leaves behind a counter-propagating pair of spin-$1/2$ neutral Dirac fermions $c^{i\tilde\Phi^{R,L}_n}$. Next, we perform another fractional basis transformation \begin{align}\begin{pmatrix}\phi^R_D\\\phi^L_D\end{pmatrix}=\frac{1}{3}\begin{pmatrix}2&-1\\-1&2\end{pmatrix}\begin{pmatrix}\tilde\Phi^R_n\\\tilde\Phi^L_n\end{pmatrix}.\label{splittingbasistrans2}\end{align} The new variables are described by the Lagrangian density \begin{align}\mathcal{L}_D&=\frac{1}{2\pi}\sum_{\sigma=+,-}\sigma\partial_{\mathsf{t}}\tilde\Phi^\sigma_n\partial_{\mathsf{x}}\tilde\Phi^\sigma_n\nonumber\\&=\frac{1}{2\pi}\sum_{\sigma=+,-}\sigma K_D\partial_{\mathsf{t}}\phi^\sigma_D\partial_{\mathsf{x}}\phi^\sigma_D,\label{splittingLD}\end{align} where $K_D=3$. The composite vertices $e^{i3\phi^R_D}=e^{i(2\tilde\Phi^R_n-\tilde\Phi^L_n)}$ and $e^{i3\phi^L_D}=e^{i(-\tilde\Phi^R_n+2\tilde\Phi^L_n)}$ are two decoupled counter-propagating spin $\pm3/2$ neutral Dirac fermions, and generate the non-chiral $U(1)_3\times\overline{U(1)_3}$ \hyperlink{CFT}{CFT}. They are represented by the dashed blue lines in figure~\ref{fig:splitting}.

Before proceeding, we notice that these neutral fermions are not integral combinations of the original electrons. Therefore, to be precise, the bosonic doubles $e^{i6\phi^\sigma_D}$ should be regarded as the local fundamental constituents instead. This changes the compactification radius and the level of the $U(1)$ \hyperlink{CFT}{CFT}. In additional to the original primary fields $e^{ia_\sigma\phi^\sigma_D}$, this also allows $\pi$-fluxes to the neutral Dirac fermions, which are represented by half-vertex twist fields $e^{ia_\sigma\phi^\sigma_D/2}$. We rescale $\phi^\sigma_d=\phi^\sigma_D/2$, which turns \eqref{splittingLD} into \begin{align}\mathcal{L}_D=\frac{1}{2\pi}\sum_{\sigma=+,-}\sigma K_d\partial_{\mathsf{t}}\phi^\sigma_d\partial_{\mathsf{x}}\phi^\sigma_d,\end{align} where $K_d=12$ and matches the $\mathcal{L}_d$ in \eqref{CSPfAbCFT} and \eqref{gluingL1}. The rescaled bosonized variables $\phi^{R,L}_d$ generate the non-chiral $U(1)_{12}\times\overline{U(1)_{12}}$ \hyperlink{CFT}{CFT}. The inclusion of the $\pi$-fluxes allows additional characters that correspond to anti-periodic boundary conditions of the Dirac fermion in a closed $(1+1)$D system. This process is known as $\mathbb{Z}_2$-orbifolding~\cite{bigyellowbook} in the \CFT context, and in this case, it extends $U(1)_3$ to $U(1)_3/\mathbb{Z}_2=U(1)_{12}$. Nevertheless, to simplify the notations, we suppress the rescaling/orbifolding and return back to the previous normalization $\mathcal{L}_D$ in \eqref{splittingLD} until the subtlety of electron locality arises again.

We now combine this counter-propagating pair of neutral Dirac fermions with the Dirac parton triplet. To avoid confusion, we differentiate the auxiliary and the parton sectors by $D$ and $\pi$. We denote the auxiliary neutral Dirac fermion sectors by $U(1)^D_3\times\overline{U(1)^D_3}$ and its bosonized variables by $\phi^{R,L}_D$. We denote the the Dirac partons by $U(1)^\pi_3\times SU(3)_1^\pi$, which we recall is identical to $U(3)_1^\pi/\mathbb{Z}_3$ through a basis transformation \eqref{CWCtrans}, and its (right-moving) bosonzied variables by $\phi^0_\pi,\phi^1_\pi,\phi^2_\pi$. The total Lagrangian density is the combination of \eqref{CSLagrangian} and \eqref{splittingLD}. We first perform a basis transformation between $U(1)^\pi_3\times U(1)^D_3$ \begin{align}\begin{array}{*{20}l}4\phi^A_\rho=\phi^0_\pi+\phi^R_D\\4\phi^B_\rho=\phi^0_\pi-\phi^R_D\end{array}.\label{splittingsd}\end{align} Recall $\phi^0_\pi$ generates the charged Laughlin sector, and $\lambda\sim e^{i\phi^0_\pi}$ is the Laughlin $q=e/3$ quasiparticle. The $A$ and $B$ bosons here correspond to two decoupled charged sectors $U(1)^A_{24}\times U(1)^B_{24}$ \begin{align}\mathcal{L}_\pi^0+\mathcal{L}_D^R&=\frac{1}{2\pi}\left(3\partial_{\mathsf{t}}\phi^0_\pi\partial_{\mathsf{x}}\phi^0_\pi+3\partial_{\mathsf{t}}\phi^R_D\partial_{\mathsf{x}}\phi^R_D\right)=\mathcal{L}^A_\rho+\mathcal{L}^B_\rho,\nonumber\\\mathcal{L}_\rho^{A,B}&=\frac{1}{2\pi}K_\rho\partial_{\mathsf{t}}\phi^{A,B}_\rho\partial_{\mathsf{x}}\phi^{A,B}_\rho,\end{align} where $K_\rho=24$ which is identical to that of \eqref{CSPfAbCFT} and \eqref{gluingL1}. The bosons are shifted by $\phi^{A,B}_\rho\to\phi^{A,B}_\rho+\Lambda/12$ under a $U(1)_{\mathrm{EM}}$ transformation that adds the phase $c\to e^{i\Lambda}c$ to electrons. Each of the charged sectors carries the differential electric conductance $\sigma^{A,B}=(1/6)e^2/h$, and the two add up to the conductance of the Laughlin channel.

There are three modes $\phi^L_D,\phi^1_\pi,\phi^2_\pi$ remaining and they correspond to the electrically neutral sectors $\overline{U(1)^D_3}\times SU(3)_1^\pi$. We rotate to a new basis \begin{align}\begin{pmatrix}\phi_0\\\phi^A_d\\\phi^B_d\end{pmatrix}=\frac{1}{2}B\begin{pmatrix}\phi^L_D\\\phi^1_\pi\\\phi^2_\pi\end{pmatrix}\label{splittingB}\end{align} using the unimodular $B$ transformation defined in \eqref{Bmatirx}. This turns the neutral sectors into $\overline{SO(2)_1}\times U(1)_{12}^A\times U(1)_{12}^B$ \begin{align}\mathcal{L}^L_D+\mathcal{L}_{SU(3)_1^\pi}&=-\frac{1}{2\pi}3\partial_{\mathsf{t}}\phi^L_D\partial_{\mathsf{x}}\phi^L_D\nonumber\\&\;\;\;+\frac{1}{2\pi}\sum_{i,j=1,2}K_{ij}^{SU(3)}\partial_{\mathsf{t}}\phi^i_\pi\partial_{\mathsf{x}}\phi^j_\pi\nonumber\\&=\mathcal{L}_{\overline{SO(2)_1}}+\mathcal{L}^A_d+\mathcal{L}^B_d,\\\mathcal{L}_{\overline{SO(2)_1}}&=\frac{1}{2\pi}K_0\partial_{\mathsf{t}}\phi_0\partial_{\mathsf{x}}\phi_0,\nonumber\\\mathcal{L}^{A,B}_d&=\frac{1}{2\pi}K_d\partial_{\mathsf{t}}\phi_d^{A,B}\partial_{\mathsf{x}}\phi_d^{A,B},\nonumber\end{align} where $K_0=-4$ and $K_d=12$, both of which match that of the previous Lagrangian densities \eqref{gluingL0} and \eqref{CSPfAbCFT} respectively, and $K^{SU(3)}$ is the $2\times2$ Cartan matrix of $SU(3)$ defined in the lower $2\times2$ block of $\mathcal{L}_n$ in \eqref{gluingKn}.

Lastly, we decompose the spin $1/2$ Dirac fermion $d_0\sim e^{i2\phi_0}=(\psi_A+i\psi_B)/\sqrt{2}$ into Majorana components (see eq.\eqref{gluingMajdecomp}). They generate two decoupled Ising sectors $\overline{\mathrm{Ising}^A}\times\overline{\mathrm{Ising}^B}$. This completes the splitting process, which can be summarized by the following flow chart. \begin{gather}\begin{diagram}(\mbox{Dirac Parton})^{\otimes3}=U(1)^\pi_3\times SU(3)^\pi_1\\\dTo^{\mathrm{extend}}\\\left[U(1)^\pi_3\times U(1)^D_3\right]\times\left[\overline{U(1)^D_3}\times SU(3)^\pi_1\right]\\\dTo^{\eqref{splittingsd}}\quad\quad\quad\quad\quad\quad\dTo^{\eqref{splittingB}}\\\Big[U(1)^A_{24}\times U(1)^B_{24}\Big]\times\left[\overline{SO(2)_1}\times U(1)^A_{12}\times U(1)^B_{12}\right]\\\dTo^{\mbox{\small Majorana decomposition}}\\\left[\mathrm{Pf}^\ast_A=U(1)^A_{24}\times U(1)^A_{12}\times\overline{\mathrm{Ising}^A}\right]\\\times\left[\mathrm{Pf}^\ast_B=U(1)^B_{24}\times U(1)^B_{12}\times\overline{\mathrm{Ising}^B}\right]\end{diagram}\label{splittingflowchart}\end{gather}

We conclude this subsection by addressing electron locality in the splitting process. First, we show that \begin{align}\openone_{12,\pm6}^{A,B}=e^{i12\phi_\rho^{A,B}}e^{\pm i6\phi_d^{A,B}},\quad\Psi_{12,0}^{A,B}=e^{i12\phi_\rho^{A,B}}\psi_{A,B}\label{elecQPsplitting}\end{align} are electronic quasiparticles in both sectors (c.f.~eq.\eqref{Pflocalel}). Without loss of generality, we illustrate this on the $A$ sector. Written in terms of the basis of the original Dirac partons and the auxiliary electrons, \begin{align}\openone_{12,\pm6}^A&=e^{i3\phi^0_\pi+i3\phi^R_D}e^{\pm i3\phi^L_D\pm i3\phi^1_\pi}\nonumber\\&=e^{i3\phi^0_\pi}e^{i3(\phi^R_D\pm\phi^L_D)}e^{\pm i3\phi^1_\pi}\\\Psi_{12,0}^A&=e^{i3\phi^0_\pi+i3\phi^R_D}\cos\left(3\phi^L_D+\phi^1_\pi+\phi^2_\pi\right)\nonumber\\&\sim e^{i3\phi^0_\pi}e^{i3(\phi^R_D+\phi^L_D)}e^{i(\phi^1_\pi+\phi^2_\pi)}\nonumber\\&\;\;\;+e^{i3\phi^0_\pi}e^{i3(\phi^R_D-\phi^L_D)}e^{-i(\phi^1_\pi+\phi^2_\pi)}.\end{align} The first piece $e^{i3\phi^0_\pi}$ is the combination of three Laughlin quasiparticles, and is identical to the electronic spin $3/2$ Dirac fermion $\lambda^3\sim\pi^1\pi^2\pi^3$. From the basis transformations \eqref{splittingbasistrans1} and \eqref{splittingbasistrans2}, it is straightforward to check that the second pieces $e^{i3(\phi^R_D+\phi^L_D)}=e^{i(\tilde\Phi^R_0-\tilde\Phi^R_1-\tilde\Phi^L_0+\tilde\Phi^L_1)}$ and $e^{i3(\phi^R_D-\phi^L_D)}=e^{i(3\tilde\Phi^R_0-3\tilde\Phi^L_1)}$ are integral combination of the auxiliary electrons. Using the basis transformation \eqref{CWCtrans}, the third and last pieces are $e^{i3\phi^1_\pi}=e^{i(2\tilde\phi_1-\tilde\phi_2-\tilde\phi_3)}$ and $e^{i(\phi^1_\pi+\phi^2_\pi)}=e^{i(\tilde\phi_1-\tilde\phi_3)}$, which are identical to the neutral local electronic combinations $E^{12}E^{13}=(\pi^1)^2(\pi^2)^\dagger(\pi^3)^\dagger$ and $E^{13}=\pi^1(\pi^3)^\dagger$ respectively. The electron locality of $\pi^1\pi^2\pi^3$ and $\pi^a(\pi^b)^\dagger$ will be shown later in section~\ref{sec:fillingonethird} when the Dirac parton triplet is constructed explicitly from electronic wires. In particular, $\pi^1\pi^2\pi^3\sim e^{i\Phi^R_\rho}$, $\pi^1(\pi^2)^\dagger\sim e^{i\Phi^R_{n1}}$ and $\pi^2(\pi^3)^\dagger\sim e^{i\Phi^R_{n2}}$ can be expressed explicitly using \eqref{PartonUbasis} and \eqref{PartonSU3basis} (also see figure~\ref{fig:Partonwirebasis}) in terms of electronic operators. It is important to acknowledge that the parton Pfaffian channels descend from electronic degrees of freedom. Electron locality forbids excitations that have non-trivial monodromy to any electronic quasiparticle. This restriction was addressed in \eqref{Pfprimaryfieldcontent} and will not be repeated here.

\section{Coupled wire models of fractional quantum Hall states}\label{sec:CWM}

In this section, we present the coupled wire models that represent the parton fractional quantum Hall (\hyperlink{FQH}{FQH}) states $\mathfrak{L}_q$ and $\mathfrak{P}_{p,q}$ (see table~\ref{tab:summary} and \eqref{LqPpqdef}). In section~\ref{sec:fillingonethird}, we construct a sequence of Abelian \FQH states $\mathfrak{L}_q$ at filling $\nu=1/3$. Each of them carries a Laughlin $U(1)_3$ charge sector and $q$ copies of neutral $SU(3)_1$ sectors that support the deconfinement of partons (except for $q=0$, which corresponds to the Laughlin state where partons are confined). In particular, $\mathcal{L}_1$ represents the \FQH state where each of the three deconfined partons fills a Landau level. Its $1+1$D boundary hosts the Dirac parton triplet conformal field theory (\hyperlink{CFT}{CFT}) $U(3)_1/\mathbb{Z}_2=U(1)_3\times SU(3)_1$, which was discussed in detail in section~\ref{sec:DiracParton}. This new parton \FQH state is topologically equivalent to and inspired by a 3D fractional topological insulator (\hyperlink{FTI}{FTI}) slab~\cite{SahooSirotaChoTeo17} with finite thickness and time-reversal (\hyperlink{TR}{TR}) symmetry breaking surfaces (see figure~\ref{fig:FTIslab}). 

In section~\ref{sec:fillingonesixth}, we construct a sequence of incompressible \FQH states $\mathfrak{P}_{p,q}$ at filling $\nu=1/6$. We speculate that they may originate from the transition between an integral quantum Hall plateau and a $\nu=n+1/3$ plateau occupied by one of the $\mathfrak{L}_q$ states. The compressible parton liquid at the transition then gains a many-body excitation energy gap by a parton pairing mechanism. It is out of the scope of this paper to address the compressible parton liquid theory that may describe an integral to fractional quantum Hall plateau transition and relate to the surface state of a \hyperlink{FTI}{FTI}. Instead, we focus on incompressible \FQH states that exhibit parton pairing. The $\mathfrak{P}_{p,q}$ state carries a $U(1)_{24}$ charge sector, which has three times the periodicity of the $U(1)_8$ charge sector of the Moore-Read Pfaffian state~\cite{MooreRead} due to the charge $e/3$ partons. Its neutral sector consists of $q$ copies of $U(1)_{12}$, each contains a neutral spin $3/2$ Dirac fermion, and $p$ Ising copies, each generated by a neutral spin $1/2$ Majorana fermion. In particular, the boundary of $\mathfrak{P}_{-1,1}$ supports the parton Pfaffian \CFT $\mathrm{Pf}^\ast=U(1)_{24}\otimes_{\mathrm{el}}U(1)_{12}\otimes_{\mathrm{el}}\overline{\mathrm{Ising}}$, which was presented in section~\ref{sec:PartonPfaffian}. Similar to $\mathfrak{L}_1$, $\mathfrak{P}_{-1,1}$ is also equivalent to a 3D \FTI slab with a \TR symmetric and a \TR breaking surface (see figure~\ref{fig:FTIslab}). Moreover, as shown in section~\ref{sec:gluingsplitting}, the parton Pfaffian \CFT is exactly half of the Dirac parton triplet. This infers that the parton Landau levels $\mathcal{L}_1$ can be split into a pair of parton Pfaffian \FQH states $\mathfrak{P}_{-1,1}$. In other words, $\mathfrak{P}_{-1,1}$ is symmetric under a parton ``particle-hole" conjugation, because the ``particle" state identical to the ``hole" state, which is obtained from subtracting the parton Pfaffian $\mathfrak{P}_{-1,1}$ from the parton Landau levels $\mathcal{L}_1$. This new notion of parton ``particle-hole" conjugation will be discussed in the next section.

The construction of these exactly solvable \FQH models relies on the coupled wire method. It was pioneered by Kane, Mukhopadhyay and Lubensky~\cite{KaneMukhopadhyayLubensky02} in modelling the Laughlin~\cite{Laughlin83} and Haldane-Halperin hierarchy~\cite{Haldane83,Halperin84} \FQH states. The construction begins with a 2D array of parallel metallic electron wires under a magnetic field. Under the Lorentz gauge $A_{\mathsf{x}}=-B\mathsf{y}$, the Fermi momenta of the wires are displaced $k_f\to k_f+(e/\hbar c)B\mathsf{y}$. Many-body interactions are restricted only to momentum preserving combinations of inter-wire and intra-wire electron backscatterings. A combination with unbalanced momentum contains an oscillating factor $e^{ik_{\mathrm{total}}\mathsf{x}}$ and vanishes upon the integration of $\mathsf{x}$. These backscattering combinations may involve multiple electrons, and take the form of ${c^{\sigma_1}_{a_1}}^\dagger\ldots{c^{\sigma_n}_{a_n}}^\dagger c^{\sigma_{n+1}}_{b_1}\ldots c^{\sigma_{2n}}_{b_n}$, where $\sigma=R,L$ labels the propagating direction of the plane wave electron modes $c^\sigma_a$ at Fermi level, and $a$ is some wire index. They can be generated by higher order corrections to the interacting action in the partition function $Z=\int \mathcal{D}c\mathcal{D}c^\dagger\exp\left[i\left(S_{\mathrm{kinetic}}+S_{\mathrm{interaction}}\right)\right]$, for some bare local interaction such as $S_{\mathrm{interaction}}=\sum_{\mathsf{y}}\int\prod_{r=1,2,3,4}d\mathsf{x}_ru^{abcd}_{\mathsf{x}_1\mathsf{x}_2\mathsf{x}_3\mathsf{x}_4}c^\dagger_a(\mathsf{x}_1)c^\dagger_b(\mathsf{x}_2)c_c(\mathsf{x}_3)c_d(\mathsf{x}_4)$, where $c_a(\mathsf{x})$ is the electron (annihilation) operator at position $\mathsf{x}$. The relevance of these backscattering combinations in the renormalization group sense depends on forward scattering interactions. In this paper, we do not address the origins and energetics of these backscattering terms. Instead, we take the strong coupling limit and study the gapped topological state associated to a given backscattering Hamiltonian. These models are exactly solvable in the sense that they consist of mutually commuting and non-competing terms. They freeze all low-energy degrees of freedom in the 2D bulk but leave behind chiral 1D boundary modes.

\subsection{Filling one-third}\label{sec:fillingonethird}

\begin{figure}[htbp]
\centering\includegraphics[width=0.5\textwidth]{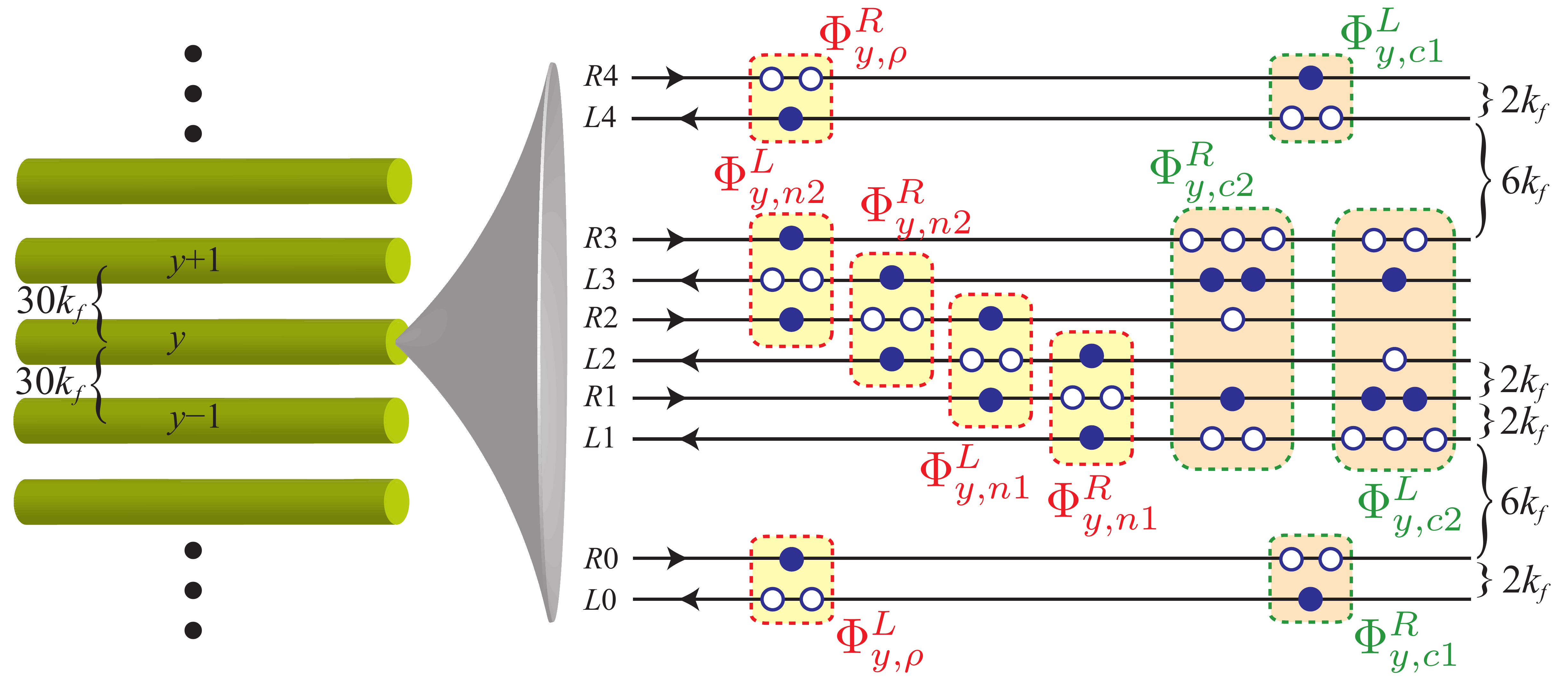}
\caption{Bundle arrangement and basis transformation for the $\nu=1/3$ parton FQH state. The momentum displacement between channels (directed lines) and bundles (green cylinders) are shown in $nk_f$, where $2k_f$ is the displacement between $R$ and $L$-movers on same wire at Fermi level. Dashed boxes represent products of electronic operators that correspond to \eqref{PartonUbasis} and \eqref{PartonSU3basis}. {\color{blue}$\bullet$} and {\color{blue}$\circ$} represents electron creation and annihilation operators respectively.}\label{fig:Partonwirebasis}
\end{figure}

We begin with an array of electronic bundles. Each bundle consists of five wires, and each wire carries a right ($R$) and a left ($L$) moving electronic Dirac fermion channel that are separated in momentum space by $2k_f$ at Fermi level. We label each bundle by an integer $y$ which represents its vertical position. The wires within a bundle are labeled by $a=0,\ldots,4$. We bosonize the Dirac fermions so that the electron (annihilation) operator at $(y,a,\sigma)$, for $\sigma=R,L=+,-$, is represented by the vertex operator \begin{align}c_{ya}^\sigma(\mathsf{x})\sim\exp\left[i\left(\tilde\Phi^\sigma_{ya}(\mathsf{x})+k^\sigma_{ya}\mathsf{x}\right)\right],\label{Abelianbosonization}\end{align} where $k^\sigma_{ya}$ is the $\mathsf{x}$-momentum of the Dirac fermion. The bosonized variables obey the equal-time commutation relation (\hyperlink{ETCR}{ETCR}) \begin{align}\left[\partial_{\mathsf{x}}\tilde\Phi_{ya}^\sigma(\mathsf{x}),\tilde\Phi_{y'a'}^{\sigma'}(\mathsf{x}')\right]=2\pi i\sigma\delta^{\sigma\sigma'}\delta_{yy'}\delta_{aa'}\delta(\mathsf{x}-\mathsf{x}').\label{ETCRParton}\end{align} The complete commutation relation that includes the zero modes is presented in \eqref{appETCRParton} in appendix~\ref{app:ETCR}.

The filling fraction of the system under a perpendicular magnetic field ${\bf B}=B\hat{\bf z}$ is \begin{align}\nu=\frac{N_e}{N_{\Phi_B}}=\frac{5(2k_f)/(2\pi)}{Bd/(hc/e)}=\frac{5(2k_f)}{\left(\frac{eB}{\hbar c}\right)d},\end{align} where $N_e=5(2k_f)/(2\pi)$ is the number electrons per unit length in a bundle of 5 wires, $N_{\Phi_B}=Bd/\phi_0$ is the number of fluxes (in units of the flux quantum $\phi_0=hc/e$) between adjacent bundles per length, and $d$ is the bundle displacement in $\mathsf{y}$. Under the Lorentz gauge ${\bf A}=-B\mathsf{y}\hat{\bf x}$, the field shifts the $\mathsf{x}$-momentum \begin{align}k^\sigma_{ya}=\frac{e}{\hbar c}B\mathsf{y}+\sigma k_f=\frac{5(2k_f)}{\nu}\frac{\mathsf{y}}{d}+\sigma k_f\label{Partonfillingkx}\end{align} of each electronic channel according to its vertical position $\mathsf{y}=\mathsf{y}(y,a)$. In the coupled wire model, we arrange the wires so that the channel momentums are given by \begin{align}\begin{array}{*{20}l}k^\sigma_{y0}=(30y+\sigma)k_f\\k^\sigma_{ya}=(30y+4a+4+\sigma)k_f,\quad\mbox{for $a=1,2,3$}\\k^\sigma_{y4}=(30y+24+\sigma)k_f\end{array}.\end{align} This pattern of momentum shifting is shown in figure~\ref{fig:Partonwirebasis}. In particular, the shift of $\mathsf{x}$-momentum between adjacent bundles in this configuration is \begin{align}\delta k=k^\sigma_{y+1,a}-k^\sigma_{ya}=30k_f.\end{align} This fixes the filling fraction to be $\nu=1/3$ by comparing with \eqref{Partonfillingkx} and $d=\mathsf{y}(y+1,a)-\mathsf{y}(y,a)$.

Next, within each bundle, we perform a basis transformation and group the electronic channels into three counter-propagating pairs of electrically charged $U(1)_3$ sectors \begin{align}&\begin{array}{*{20}l}\Phi^R_{y,\rho}=2\tilde\Phi^R_{y4}-\tilde\Phi^L_{y4}\\\Phi^R_{y,c1}=2\tilde\Phi^R_{y0}-\tilde\Phi^L_{y0}\\\Phi^R_{y,c2}=-\tilde\Phi^R_{y1}+\tilde\Phi^R_{y2}+3\tilde\Phi^R_{y3}+2\tilde\Phi^L_{y1}-2\tilde\Phi^L_{y3}\end{array},\nonumber\\&\begin{array}{*{20}l}\Phi^L_{y,\rho}=2\tilde\Phi^L_{y0}-\tilde\Phi^R_{y0}\\\Phi^L_{y,c1}=2\tilde\Phi^L_{y4}-\tilde\Phi^R_{y4}\\\Phi^L_{y,c2}=3\tilde\Phi^L_{y1}+\tilde\Phi^L_{y2}-\tilde\Phi^L_{y3}-2\tilde\Phi^R_{y1}+2\tilde\Phi^R_{y3}\end{array},\label{PartonUbasis}\end{align} and a pair of electrically neutral $SU(3)_1$ ones \begin{align}&\begin{array}{*{20}l}\Phi^R_{y,n1}=2\tilde\Phi^R_{y1}-\tilde\Phi^L_{y1}-\tilde\Phi^L_{y2}\\\Phi^R_{y,n2}=2\tilde\Phi^R_{y2}-\tilde\Phi^L_{y2}-\tilde\Phi^L_{y3}\end{array},\nonumber\\&\begin{array}{*{20}l}\Phi^L_{y,n1}=2\tilde\Phi^L_{y2}-\tilde\Phi^R_{y2}-\tilde\Phi^R_{y1}\\\Phi^L_{y,n2}=2\tilde\Phi^L_{y3}-\tilde\Phi^R_{y3}-\tilde\Phi^R_{y2}\end{array}.\label{PartonSU3basis}\end{align} These new local combinations are diagrammatically represented in figure~\ref{fig:Partonwirebasis}. They obey the equal-time commutation relation \begin{align}\left[\partial_{\mathsf{x}}\Phi_{y,\mu}^\sigma(\mathsf{x}),\Phi_{y',\mu'}^{\sigma'}(\mathsf{x}')\right]=2\pi i\sigma\delta^{\sigma\sigma'}\delta_{yy'}K_{\mu\mu'}\delta(\mathsf{x}-\mathsf{x}'),\end{align} where the $K$-matrix (suppressing all 0's) is \begin{align}K=\begin{pmatrix}3&&&&\\&3&&&\\&&3&&\\&&&2&-1\\&&&-1&2\end{pmatrix}\end{align} and the bozonized variables $\Phi^\sigma_{y,\mu}$ are ordered in $(\Phi^\sigma_{y,\rho},\Phi^\sigma_{y,c1},\Phi^\sigma_{y,c2},\Phi^\sigma_{y,n1},\Phi^\sigma_{y,n2})$. The external electromagnetic gauge transformation, that rotates the phases of electronic operators by $e^{i\tilde\Phi_{ya}}\to e^{i\Lambda}e^{i\tilde\Phi_{ya}}$, shifts the new local bosonized variables by \begin{align}\Phi^\sigma_{y,\mu}\to\Phi^\sigma_{y,\mu}+t_\mu\Lambda,\label{Partonbasisgaugetransformation}\end{align} where the charge vector is ${\bf t}=(t_\rho,t_{c1},t_{c2},t_{n1},t_{n2})=(1,1,3,0,0)$. Each combination $\Phi^\sigma_{y,\mu}$ in \eqref{PartonUbasis} and \eqref{PartonSU3basis} corresponds to a product of electronic operators $e^{i\Phi^\sigma_{y,\mu}(\mathsf{x})+ik^\sigma_{y,\mu}\mathsf{x}}$ that carries the net $\mathsf{x}$-momentum \begin{align}\begin{array}{*{20}l}k^\sigma_{y,\rho}=3(4+10y+5\sigma)k_f\\k^\sigma_{y,c1}=3(4+10y-3\sigma)k_f\\k^\sigma_{y,c2}=3(12+30y+\sigma)k_f\\k^\sigma_{y,n1}=k^\sigma_{y,n2}=0\end{array}.\label{Partonbasismomentum}\end{align}

The basis transformations \eqref{PartonUbasis} and \eqref{PartonSU3basis} were designed to facilitate the Dirac partons \eqref{DiracPartondef}. To see this, we first introduce the intra-bundle interactions for each $y$, \begin{gather}\mathcal{H}^{\mathrm{intra}}_y=u_I\cos\Theta^{\mathrm{intra}}_{y,I}+u_{II}\cos\Theta^{\mathrm{intra}}_{y,II},\label{intrabundleH}\\\begin{array}{*{20}l}\Theta^{\mathrm{intra}}_{y,I}=2\Phi^R_{y,c1}+\Phi^L_{y,c1}+\Phi^R_{y,c2}-2\Phi^L_{y,c2}\\\Theta^{\mathrm{intra}}_{y,II}=\Phi^R_{y,c1}+2\Phi^L_{y,c1}-2\Phi^R_{y,c2}+\Phi^L_{y,c2}\end{array}.\nonumber\end{gather} The two linearly independent angle variables $\Theta^{\mathrm{intra}}_{y,I}$, $\Theta^{\mathrm{intra}}_{y,II}$ satisfy the ``Haldane's nullity" gapping condition~\cite{Haldane95} $\left[\Theta^{\mathrm{intra}}_{y,l}(\mathsf{x}),\Theta^{\mathrm{intra}}_{y',l'}(\mathsf{x}')\right]=0$. (See \eqref{appintracomm} in appendix~\ref{app:ETCR}.) Thus, along each bundle $y$, $\mathcal{H}^{\mathrm{intra}}_y$ turns two (out of five) pairs of modes, $\Phi^\sigma_{y,c1},\Phi^\sigma_{y,c2}$, massive and removes them from low-energy. We also see that the interaction preserves charge $U(1)$ and $\mathsf{x}$-momentum conservation. This is because the angle variables are invariant under the $U(1)_{EM}$ gauge transformation \eqref{Partonbasisgaugetransformation} and the two sine-Gordon terms are products of electronic operators that have trivial net $\mathsf{x}$-momenta, $2k^R_{y,c1}+k^L_{y,c1}+k^R_{y,c2}-2k^L_{y,c2}=k^R_{y,c1}+2k^L_{y,c1}-2k^R_{y,c2}+k^L_{y,c2}=0$.

The intra-bundle interaction \eqref{intrabundleH} leaves behind, along each bundle $y$, three pairs of low-energy modes \begin{align}\boldsymbol\Phi^\sigma_y=(\Phi^\sigma_{y,0},\Phi^\sigma_{y,1},\Phi^\sigma_{y,2})=(\Phi^\sigma_{y,\rho},\Phi^\sigma_{y,n1},\Phi^\sigma_{y,n2}).\end{align} These modes are not affected by $\mathcal{H}^{\mathrm{intra}}_y$ because the three bosonized variables commute with the sine-Gordon angle variables $\Theta^{\mathrm{intra}}_{y,M}$. (See \eqref{appintracomm} in appendix~\ref{app:ETCR}.) It is important to acknowledge -- from \eqref{PartonUbasis} and \eqref{PartonSU3basis} -- that the three $\Phi^\sigma_{y,J}$ associate to {\em integral} combinations of electronic operators which are local. Locality only allows excitations that have non-fractional mutual monodromy with these local electronic combinations. Let $e^{-i\phi}$ be the vertex operator that creates such an excitation, where $\phi$ is some non-integral combination $b^J_\sigma\Phi^\sigma_{y,J}$. Locality requires the equal-time commutator \begin{align}\left[N^{\sigma'}_{y',J'},\phi(\mathsf{x})\right]=i\sigma\delta^{\sigma\sigma'}\delta_{yy'}K_{JJ'}b^J_\sigma\label{partonexcitationlocality}\end{align} to have integral value, where the $K$-matrix was defined in \eqref{partonKmatrix} and \begin{align}N^\sigma_{y,J}=\frac{1}{2\pi}\int\partial_{\mathsf{x}}\Phi^\sigma_{y,J}(\mathsf{x})d\mathsf{x}\end{align} is the number operator of the local electronic combination $e^{i\Phi^\sigma_{y,J}}$. The integrality of \eqref{partonexcitationlocality} comes from the fact that $N^\sigma_{y,J}$ is a physical observable and can only take integral eigenvalues in the electronic many-body Hilbert space. Locality therefore only allows vertex excitations $e^{ib^J_\sigma\Phi^\sigma_{y,J}}$, where $K_{JJ'}b^J$ are integers.

Integral combinations $\Phi_\alpha=\alpha^J_\sigma\Phi^\sigma_{y,J}$ of the local bosonized variables span the root lattice \begin{align}\Lambda_{U(1)_3\times SU(3)_1}^\sigma&=\mathrm{span}_{\mathbb{Z}}\left\{\Phi^\sigma_{y,\rho},\Phi^\sigma_{y,n1},\Phi^\sigma_{y,n2}\right\}\end{align} of the affine Kac-Moody algebra $U(1)_3\times SU(3)_1$. We define the {\em dual} bosonized variables (in the Chevalley basis), which are fractional combinations of the local ones \begin{align}\phi_y^{\sigma,J}=(K^{-1})^{JJ'}\Phi^\sigma_{y,J},\end{align} where $J=(0,1,2)=(\rho,n1,n2)$. Integral combinations $a_J^\sigma\phi_y^{\sigma,J}$ of these dual variables span a dual lattice, known as the weight lattice \begin{align}{\Lambda_{U(1)_3\times SU(3)_1}^\sigma}^\ast&=\mathrm{span}_{\mathbb{Z}}\left\{\phi_y^{\sigma,0},\phi_y^{\sigma,1},\phi_y^{\sigma,2}\right\}\nonumber\\&=K^{-1}\Lambda_{U(1)_3\times SU(3)_1}^\sigma\end{align} whose elements correspond to vertex excitations $e^{ia_J^\sigma\phi_y^{\sigma,J}}$ allowed by electron locality. The dual variables obey the equal-time commutation relation \eqref{PartonChevalleyETCR}. They can be transformed into the Cartan-Weyl basis $\tilde\phi^\sigma_{y,a}=A_a^J\phi^{\sigma,J}_y$ by \eqref{CWCtrans} that gives the Dirac fermionic partons $\pi_y^{\sigma,a}\sim e^{i\tilde\phi^\sigma_{y,a}}$.

\begin{figure}[htbp]
\centering\includegraphics[width=0.45\textwidth]{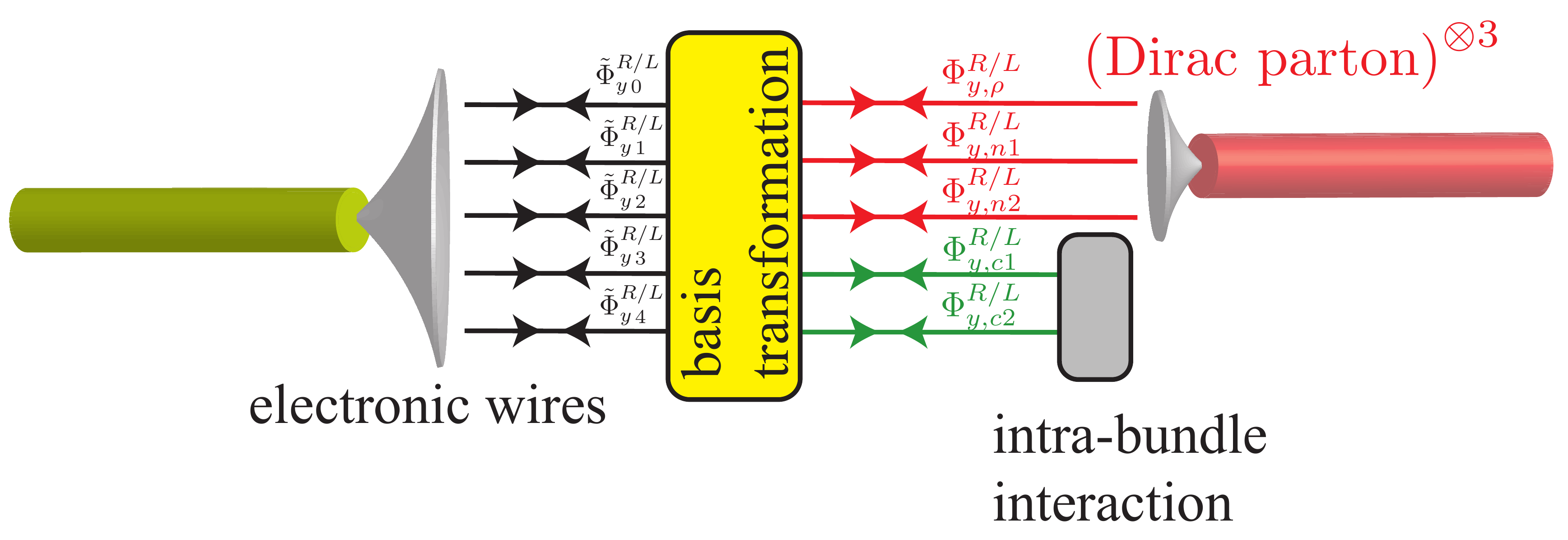}
\caption{Emergence of Dirac partons from five electronic wires. The basis transformation is defined in \eqref{PartonUbasis} and \eqref{PartonSU3basis}. The intra-bundle interaction is defined in \eqref{intrabundleH}.}\label{fig:PartonConstructIntra}
\end{figure}

The basis transformation \eqref{PartonUbasis}, \eqref{PartonSU3basis} and the intra-bundle gapping interaction \eqref{intrabundleH} now turn each bundle of five electronic wires into a counter-propagating pair of Dirac parton triplets $U(3)_1/\mathbb{Z}_3=U(1)_3\times SU(3)_1$. This process is summarized in figure~\ref{fig:PartonConstructIntra}. As promised in section~\ref{sec:DiracParton}, it is now evident that the charge $e$ fermion $\Psi_{\mathrm{el}}=e^{i3\phi^\sigma_\rho}=e^{i\Phi^\sigma_\rho}$ as well as the six neutral bosonic $SU(3)$ roots $E^{ab}=e^{i(\tilde\phi^\sigma_a-\tilde\phi^\sigma_b)}=e^{\pm i\Phi^\sigma_{n1}}$, $e^{\pm i\Phi^\sigma_{n2}}$ or $e^{\pm i(\Phi^\sigma_{n1}+\Phi^\sigma_{n2})}$ are all integral combinations of local electrons. We notice in passing that this is not the simplest way in realizing the Dirac partons. For instance, similar procedure can be applied to a bundle of four electronic wires instead of five. The current method is presented so that the neutral $SU(3)_1$ sector carries zero $\mathsf{x}$-momentum (see eq.\eqref{Partonbasismomentum}). This will become useful in the next subsection in the discussion of the parton Pfaffian FQH state.

\begin{figure}[htbp]
\centering\includegraphics[width=0.45\textwidth]{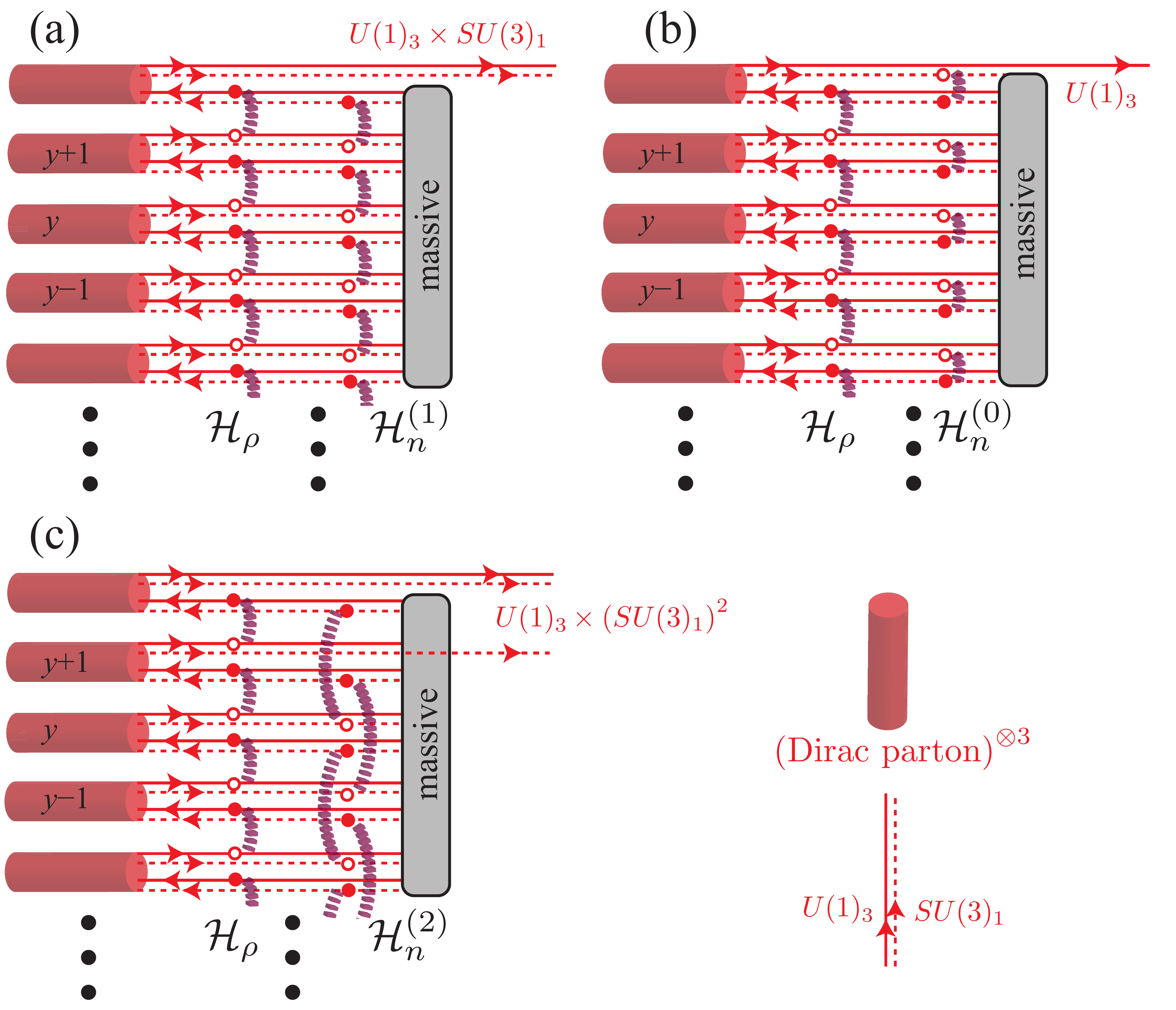}
\caption{Coupled wire models and gapless edge modes of (a) the parton state $\mathfrak{L}_1=U(1)_3\times SU(3)_1$ with $c=3/2$, (b) the Laughlin state $\mathfrak{L}_0=U(1)_3$ with $c=1$, and (c) the $\mathfrak{L}_2=U(1)_3\times(SU(3)_1)^2$ state with $c=5$. $\mathcal{H}_\rho$ and $\mathcal{H}_n^{(q)}$ are defined \eqref{LaughlinH} and \eqref{PartonSU3H} respectively.}\label{fig:PartonCoupledWire}
\end{figure}

We now describe the coupled wire construction of the $\mathfrak{L}_q$ series of \FQH state with filling $\nu=1/3$. We introduce the interwire gapping interactions \begin{align}\mathcal{H}_\rho&=u_\rho\sum_y\cos\left(\Phi^R_{y,\rho}-\Phi^L_{y+1,\rho}\right),\label{LaughlinH}\\\mathcal{H}_n^{(q)}&=u_n\sum_y\left[\sum_{j=1,2}\cos\left(\Phi^R_{y,nj}-\Phi^L_{y+q,nj}\right)\right.\label{PartonSU3H}\\&\;\;\;\;+\cos\left(\Phi^R_{y,n1}+\Phi^R_{y,n2}-\Phi^L_{y+q,n1}-\Phi^L_{y+q,n2}\right)\Big],\nonumber\end{align} where $q$ is a fixed integer. The coupled wire model Hamiltonian density contains all three interactions from \eqref{intrabundleH}, \eqref{LaughlinH} and \eqref{PartonSU3H} \begin{align}\mathcal{H}^{(q)}=\mathcal{H}_\rho+\mathcal{H}_n^{(q)}+\sum_y\mathcal{H}^{\mathrm{intra}}_y.\label{PartonCWfullH}\end{align} All the sine-Gordon potential preserves charge $U(1)_{\mathrm{EM}}$ and $\mathsf{x}$-momentum conservation. This can be verified using the charge assignment \eqref{Partonbasisgaugetransformation} and momenta \eqref{Partonbasismomentum}. The interactions introduce a finite excitation energy gap in the bulk but leaves behind gapless edge modes. See figure~\ref{fig:PartonCoupledWire} for illustration when $q=0,1,2$. In low energy, the gapless edge modes are described by the Kac-Moody \CFT \begin{align}\mathfrak{L}_q=U(1)_3\times\left[SU(3)_1\right]^q\end{align} and carry the chiral central charge $c=1+2q$. We take the notation so that the neutral sector $\left(SU(3)_1\right)^q$ is right-moving if $q>0$ and $\left(SU(3)_1\right)^q=\left(\overline{SU(3)_1}\right)^{-q}$ is left-moving if $q<0$. For example, $q=0$ corresponds to the Laughlin $\nu=1/3$ \FQH state where the topological state has no non-trivial neutral sectors and all partons are confined. We focus on the case when $q=1$ that corresponds to the parton \FQH state $U(3)_1/\mathbb{Z}_3=U(1)_3\times SU(3)_1$ at filling $\nu=1/3$ and central charge $c=3$. This has the identical topological order to a slab of fractional topological insulator with time reversal breaking and conjugate top and bottom surfaces~\cite{SahooSirotaChoTeo17} (see figure~\ref{fig:FTIslab}).

We notice that the first line in \eqref{PartonSU3H} of $\mathcal{H}^{(q)}_n$ alone can already introduce an energy gap for the neutral $SU(3)_1$ sectors in the bulk. The last term is included so that the interaction takes the form of $SU(3)_1$ Kac-Moody current backscattering \begin{align}\mathcal{H}^{(q)}_n=\frac{u_n}{2}\sum_y\sum_{\boldsymbol\alpha}E^{\boldsymbol\alpha}_{y,R}E^{-\boldsymbol\alpha}_{y+q,L},\label{PartonSU3H2}\end{align} where $\boldsymbol\alpha=\pm(1,0),\pm(0,1),\pm(1,1)$ label the six roots of $SU(3)_1$ \begin{align}E^{\boldsymbol\alpha}_{y,\sigma}=e^{i\alpha^j\Phi^\sigma_{y,nj}},\end{align} which are identical to the six current operators $E^{ab}$ in the parton basis in \eqref{SU3currentOPE} of section~\ref{sec:DiracParton} (or \eqref{appSU3generators} in appendix~\ref{app:partonalgebra}). In this case, the $SU(3)$ symmetric interaction \eqref{PartonSU3H2} introduces a finite excitation energy gap if $u_n$ is negative, so that the additional sine-Gordon term on the second line of \eqref{PartonSU3H} does not compete with the two terms on the first line. 

Lastly, we observe that the $SU(3)_1$ current operators have zero $\mathsf{x}$-momentum. This allows the backscattering interaction \eqref{PartonSU3H2} of an arbitrary hopping range $q$ to preserve momentum conservation. The coupled wire model does not provide a preference towards a particular $\mathfrak{L}_q$ phase. While the charge sector is frozen by $\mathcal{H}_\rho$, $SU(3)$ current backscattering Hamiltonians $\mathcal{H}^{(q)}_n$ with different ranges compete. A generalized coupled wire model that simultaneously includes multiple $\mathcal{H}^{(q)}_n$'s could potentially describe a $SU(3)$ spin liquid with a complex phase diagram connected by a web of topological phase transitions.

\subsection{Filling one-sixth}\label{sec:fillingonesixth}

\begin{figure}[htbp]
\centering\includegraphics[width=0.5\textwidth]{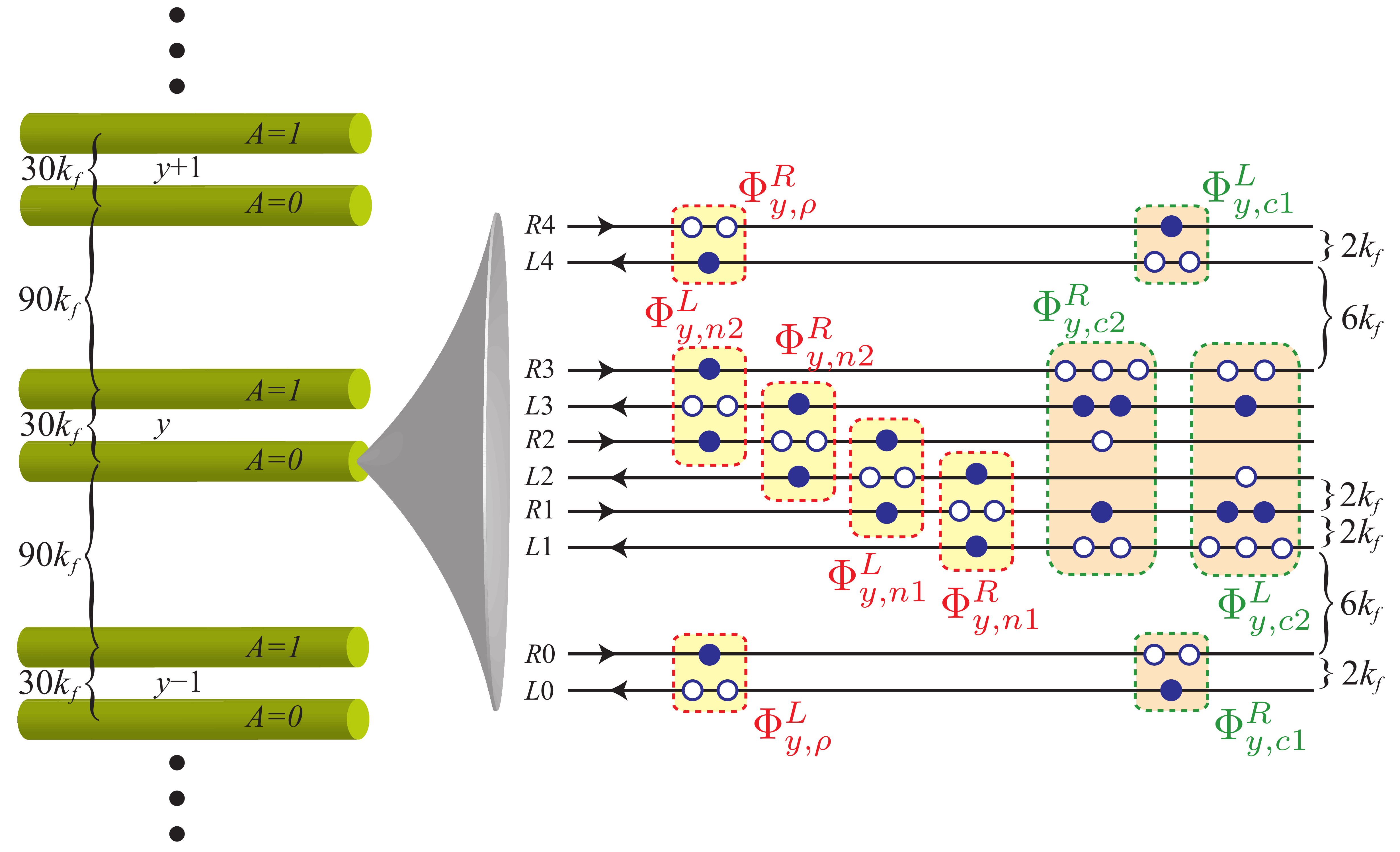}
\caption{Bundle arrangement and basis transformation for the $\nu=1/6$ $\mathrm{Pf}^\ast$ FQH state. The momentum displacement between channels (directed lines) and bundles (green cylinders) are shown in $nk_f$, where $2k_f$ is the displacement between $R$ and $L$-movers on same wire at Fermi level. Dashed boxes represent products of electronic operators. {\color{blue}$\bullet$} and {\color{blue}$\circ$} represents electron creation and annihilation operators respectively.}\label{fig:Pfwiresbasis}
\end{figure}

The model for filling one-sixth consists of half as many wires as in the one-third case. This is shown in figure~\ref{fig:Pfwiresbasis} where half the bundles from figure~\ref{fig:Partonwirebasis} are taken out. The bundles now has a staggered and dimerized configuration where translation symmetry is broken and two bundles, labeled by $A=0,1$, now form a super-unit cell. The electron (annihilation) operators are \begin{align}c^\sigma_{yAa}(\mathsf{x})\sim\exp\left[i\left(\tilde\Phi^\sigma_{yAa}(\mathsf{x})+k^\sigma_{yAa}\mathsf{x}\right)\right],\label{electronPf}\end{align} where $y$ is an integer that labels the vertical position of the 2-bundle super-unit cell, $a=0,\ldots,4$ labels the five electronic wires in each bundle, and $\sigma=R,L=+,-$ corresponds to the two counter-propagating electronic channels in each wire. The bosonized variables obey the equal-time commutation relation (\hyperlink{ETCR}{ETCR}) \begin{align}\left[\partial_{\mathsf{x}}\tilde\Phi_{yAa}^\sigma(\mathsf{x}),\tilde\Phi_{y'A'a'}^{\sigma'}(\mathsf{x}')\right]=2\pi i\sigma\delta^{\sigma\sigma'}\delta_{yy'}\delta_{AA'}\delta_{aa'}\delta(\mathsf{x}-\mathsf{x}').\label{ETCRPf}\end{align} The model is under a perpendicular magnetic field so that the $\mathsf{x}$-momenta of the electronic channels $c^\sigma_{yAa}$ are \begin{align}\begin{array}{*{20}l}k^\sigma_{yA0}=(120y+30A+\sigma)k_f\\k^\sigma_{yAa}=(120y+30A+4a+4+\sigma)k_f\\k^\sigma_{yA4}=(120y+30A+24+\sigma)k_f\end{array},\quad\mbox{for $a=1,2,3$}.\end{align}

Similar to the $\nu=1/3$ case, we first perform the basis transformation \eqref{PartonUbasis} and \eqref{PartonSU3basis} for each bundle $(y,A)$. Then, we introduce the charge $U(1)_{\mathrm{EM}}$ and momentum conserving intra-bundle gapping potential $\mathcal{H}^{\mathrm{intra}}_{y,A}$ defined in \eqref{intrabundleH} to turn each bundle into the Dirac parton triplet $U(3)_1/\mathbb{Z}_3=U(1)_3\times SU(3)_1$. (Also see figure~\ref{fig:PartonConstructIntra}.) We here repeat the transformed bosonized variables that are unaffected by $\mathcal{H}^{\mathrm{intra}}_{y,A}$. \begin{align}&\begin{array}{*{20}l}\Phi^R_{yA\rho}=2\tilde\Phi^R_{yA4}-\tilde\Phi^L_{yA4}\\\Phi^L_{yA\rho}=2\tilde\Phi^L_{yA0}-\tilde\Phi^R_{yA0}\end{array},\label{PfUbasis}\\&\begin{array}{*{20}l}\Phi^R_{yA,n1}=2\tilde\Phi^R_{yA1}-\tilde\Phi^L_{yA1}-\tilde\Phi^L_{yA2}\\\Phi^R_{yA,n2}=2\tilde\Phi^R_{yA2}-\tilde\Phi^L_{yA2}-\tilde\Phi^L_{y3}\\\Phi^L_{yA,n1}=2\tilde\Phi^L_{yA2}-\tilde\Phi^R_{yA2}-\tilde\Phi^R_{yA1}\\\Phi^L_{yA,n2}=2\tilde\Phi^L_{yA3}-\tilde\Phi^R_{yA3}-\tilde\Phi^R_{yA2}\end{array}.\label{Pfbasis}\end{align} The bosonzied variabes obey the \ETCR \begin{align}&\left[\partial_{\mathsf{x}}\Phi_{yAJ}^\sigma(\mathsf{x}),\Phi_{y'A'J'}^{\sigma'}(\mathsf{x}')\right]\nonumber\\&=2\pi i\sigma\delta^{\sigma\sigma'}\delta_{yy'}\delta_{AA'}K_{JJ'}\delta(\mathsf{x}-\mathsf{x}'),\end{align} where the $K$-matrix is \begin{align}K=\begin{pmatrix}3&&\\&2&-1\\&-1&2\end{pmatrix}\label{PartonKPfdef}\end{align} and is ordered according to $J=\rho,n1,n2=0,1,2$. The first bosonzied variable shifts by $\Phi^\sigma_{yA\rho}\to\Phi^\sigma_{yA\rho}+\Lambda$ under the external $U(1)_{\mathrm{EM}}$ transformation of an electron operator $c\to ce^{i\Lambda}$. The other two bosonized variables $\Phi^\sigma_{yA,n1},\Phi^\sigma_{yA,n2}$ corresponds to neutral sectors and are invariant under $U(1)_{EM}$. The vertices $e^{i\Phi^\sigma_{yAJ}(\mathsf{x})+ik^\sigma_{yAJ}\mathsf{x}}$ are integral combinations of electronic operators, which are diagramatically represented as the yellow boxes in figure~\ref{fig:Pfwiresbasis}. They carry the $\mathsf{x}$-momenta \begin{align}\begin{array}{*{20}l}k^\sigma_{yA\rho}=3(4+40y+10A+5\sigma)k_f,\\k^\sigma_{yA,n1}=k^\sigma_{yA,n2}=0.\end{array}\end{align}

\begin{figure}[htbp]
\centering\includegraphics[width=0.48\textwidth]{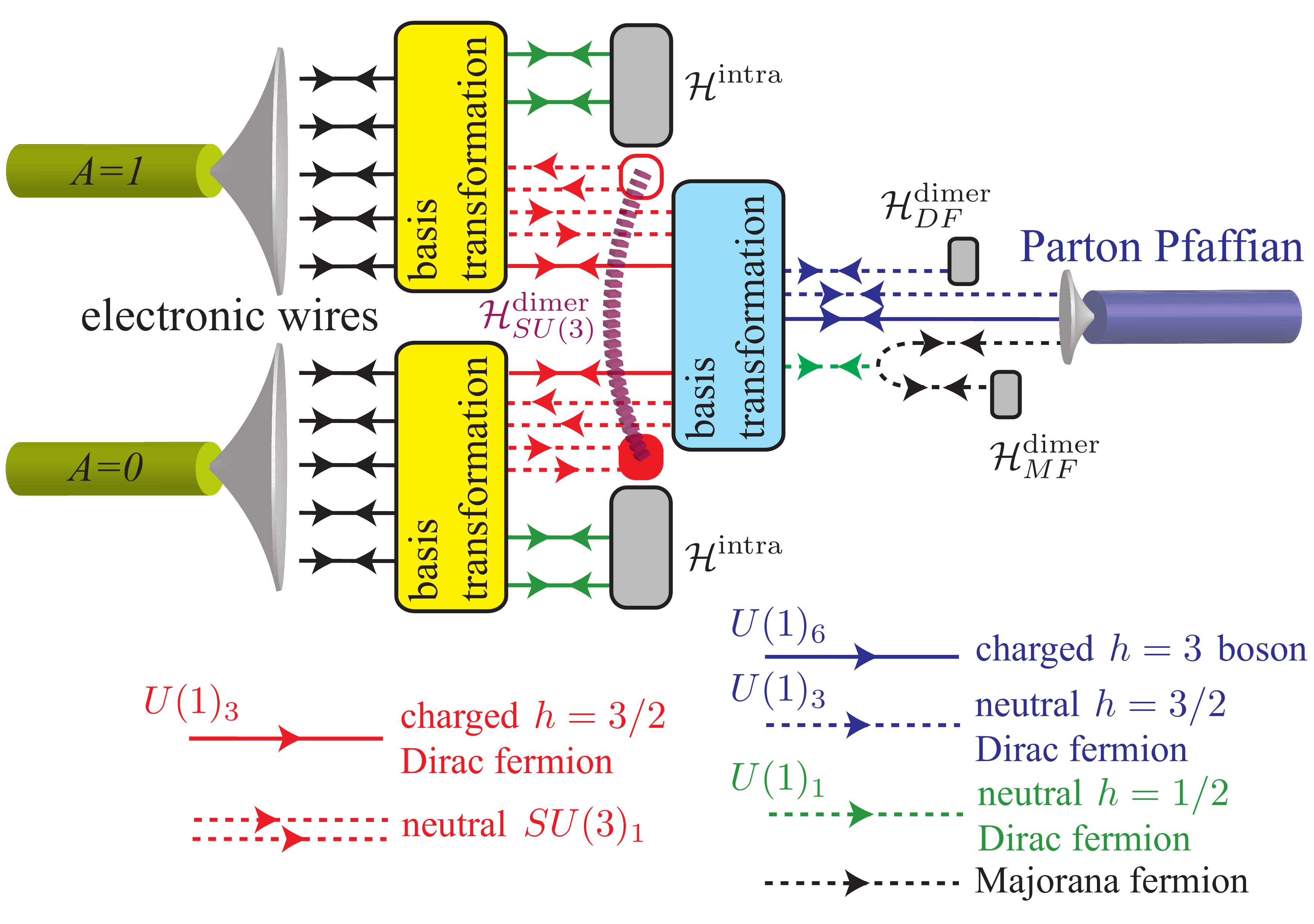}
\caption{Emergence of parton Pfaffian from ten electronic wires. Each of the pair of basis transformations in the yellow boxes is defined in \eqref{PartonUbasis} and \eqref{PartonSU3basis}. The second basis transformation in the blue box is the product of \eqref{Pfbasistrans1} and \eqref{Pfbasistrans2}. $\mathcal{H}^{\mathrm{intra}}$, $\mathcal{H}^{\mathrm{dimer}}_{SU(3)}$, $\mathcal{H}^{\mathrm{dim}}_{DF}$ and $\mathcal{H}^{\mathrm{dim}}_{MF}$ are defined in \eqref{intrabundleH}, \eqref{PfdimerSU3H}, \eqref{PfincellHDF} and \eqref{PfincellHMF} respectively.}\label{fig:PfConstructIntra}
\end{figure}

At this point, there are two counter-propagating pairs of Dirac parton triplets in each unit cell $y$ in low-energy. We introduce the intra-cell $SU(3)$ current backscattering \begin{align}&\mathcal{H}^{\mathrm{dimer}}_{y,SU(3)}=\frac{u_{III}}{2}\sum_{\boldsymbol\alpha}E^{\boldsymbol\alpha}_{y,A=0,R}E^{-\boldsymbol\alpha}_{y,A=1,L}\label{PfdimerSU3H}\\&=u_{III}\left[\sum_{j=1,2}\cos\left(\Phi^R_{y,A=0,nj}-\Phi^L_{y,A=1,nj}\right)\right.\nonumber\\&\;\;\;\;+\cos\left(\Phi^R_{y,A=0,n1}+\Phi^R_{y,A=0,n2}-\Phi^L_{y,A=1,n1}-\Phi^L_{y,A=1,n2}\right)\Big],\nonumber\end{align} where $u_{III}<0$. It mirrors the neutral inter-bundle gapping potential \eqref{PartonSU3H} and \eqref{PartonSU3H2} at $q=1$ for the $\nu=1/3$ case, except now it acts only within the dimerized super-unit cell. This leaves two counter-propagating pairs of charged modes $\Phi^\sigma_{yA\rho}$ and the two counter-propagating pairs of neutral modes $\Phi^R_{y,nj}=\Phi^R_{y,A=1,nj}$ and $\Phi^L_{y,nj}=\Phi^L_{y,A=0,nj}$, for $j=1,2$, unaffected.

Next, we perform a fractional basis transformation to the charged modes \begin{align}\begin{pmatrix}\Phi^R_{y\rho}\\\Phi^R_{y,n0}\\\Phi^L_{y\rho}\\\Phi^L_{y,n0}\end{pmatrix}=\begin{pmatrix}3/2 & 1 & -1/2 & -1 \\ -1/2 & 1 & -1/2 & 0 \\ -1/2 & -1 & 3/2 & 1 \\ -1/2 & 0 & -1/2 & 1 \end{pmatrix}
\begin{pmatrix}\Phi^R_{y,A=0,\rho}\\\Phi^R_{y,A=1,\rho}\\\Phi^L_{y,A=0,\rho}\\\Phi^L_{y,A=1,\rho}\end{pmatrix}.\label{Pfbasistrans1}\end{align} The $\rho$ ($n$) sectors are electrically charged (resp.~neutral). They obey the \ETCR \begin{align}&\left[\partial_{\mathsf{x}}\Phi^\sigma_{y\rho}(\mathsf{x}),\Phi^{\sigma'}_{y'\rho}(\mathsf{x}')\right]=2\pi i\sigma\delta^{\sigma\sigma'}\delta_{yy'}K_\rho\delta(\mathsf{x}-\mathsf{x}')\nonumber\\&\left[\partial_{\mathsf{x}}\Phi^\sigma_{y,nJ}(\mathsf{x}),\Phi^{\sigma'}_{y',nJ'}(\mathsf{x}')\right]=2\pi i\sigma\delta^{\sigma\sigma'}\delta_{yy'}K_{JJ'}\delta(\mathsf{x}-\mathsf{x}')\nonumber\\&\left[\partial_{\mathsf{x}}\Phi^\sigma_{y\rho}(\mathsf{x}),\Phi^{\sigma'}_{y',nJ}(\mathsf{x}')\right]=0,\end{align} where $K$ is the same as that in \eqref{PartonKPfdef} and \begin{align}K_\rho&=6.\end{align} They carry the $\mathsf{x}$-momenta \begin{align}k^\sigma_{y\rho}=3\left(9+40y+20\sigma\right)k_f,\quad k^\sigma_{y,nJ}=0.\end{align} It should be noticed that $\Phi^\sigma_{y\rho}$ and $\Phi^\sigma_{y,n0}$ are half-integral combinations of electronic bosonized variables -- previously referred to as ``almost local" variables in section~\ref{sec:PartonPfaffian} -- and therefore they do not associate to local electronic vertex operators. On the other hand, their sums and differences $\Phi^R_{y\rho}\pm\Phi^L_{y\rho}$, $\Phi^R_{y,n0}\pm\Phi^L_{y,n0}$ and $\Phi^\sigma_{y\rho}\pm\Phi^{\sigma'}_{y,n0}$ correspond to integral electronic combinations. 

It will be convenient later to perform a second unimodular basis transformation within the neutral sectors \begin{align}\begin{pmatrix}\Phi^R_{y,d0}\\\Phi^R_{y,d1}\\\Phi^R_{y,d2}\\\Phi^L_{y,d0}\\\Phi^L_{y,d1}\\\Phi^L_{y,d2}\end{pmatrix}=\begin{pmatrix}1&0&0&0&1&0\\0&1&-1&1&0&0\\0&2&1&1&0&0\\0&1&0&1&0&0\\1&0&0&0&1&-1\\1&0&0&0&2&1\end{pmatrix}\begin{pmatrix}\Phi^R_{y,n0}\\\Phi^R_{y,n1}\\\Phi^R_{y,n2}\\\Phi^L_{y,n0}\\\Phi^L_{y,n1}\\\Phi^L_{y,n2}\end{pmatrix}.\label{Pfbasistrans2}\end{align} This changes the \ETCR to \begin{align}&\left[\partial_{\mathsf{x}}\Phi^\sigma_{y,dJ}(\mathsf{x}),\Phi^{\sigma'}_{y',dJ'}(\mathsf{x}')\right]\nonumber\\&=2\pi i\sigma\delta^{\sigma\sigma'}\delta_{yy'}(K_d)_{JJ'}\delta(\mathsf{x}-\mathsf{x}')\end{align} where the $K_d$-matrix is now diagonal \begin{align}K_d=\begin{pmatrix}1&&\\&3&\\&&3\end{pmatrix}.\label{Kdmatrix}\end{align} The complete \ETCR between the bosonized variables $\Phi^\sigma_{y\rho},\Phi^\sigma_{y,dJ}$ can be found in \eqref{appETCRPf2} in appendix~\ref{app:ETCRpaired}.

We introduce the neutral Dirac fermions \begin{align}d_{y,J}^\sigma=e^{i\Phi^\sigma_{y,dJ}}\prod_{y'<y}(-1)^{N^d_{y'}}.\label{PfneutralDirac}\end{align} The operator $N^d_{y'}$ is a combination of local electronic number operators and is defined explicitly in \eqref{Ndoperator} in appendix~\ref{app:ETCRpaired}. It is chosen in a way to ensure mutual commutativity between the sine-Gordon angle variables in the upcoming interactions \eqref{PfincellHDF}, \eqref{PfrhogammaH}, \eqref{PfdrhoH},\eqref{PfgammadH} as well as \eqref{PfrhogammaHpq} and \eqref{PfdrhoHpq}. The fermion parity operator $(-1)^{N^d_y}$ anticommutes with the vertex operators $e^{i\Phi^\sigma_{y,dJ}}$ in the same unit-cell $y$. Similar to the Jordan-Wigner~\cite{Jordan27,JordanWigner28} fermionization of the Ising model, the string of fermion parity operators in \eqref{PfneutralDirac} ensures the mutual anticommutation relations between the neutral Dirac fermions in \eqref{PfneutralDirac} \begin{align}\left\{d_{y,J}^\sigma(\mathsf{x}),d_{y',J'}^{\sigma'}(\mathsf{x}')\right\}=\left\{d_{y,J}^\sigma(\mathsf{x}),d_{y',J'}^{\sigma'}(\mathsf{x}')^\dagger\right\}=0\end{align} for $(y,J,\mathsf{x})\neq(y',J',\mathsf{x}')$.

From \eqref{Kdmatrix}, we see that $d_{y,0}^\sigma$ is a spin $h=\pm1/2$ fermion and $d_{y,1}^\sigma,d_{y,2}^\sigma$ have spin $h=\pm3/2$. The former can be decomposed into a pair of Majorana fermions \begin{gather}d^\sigma_{y,0}=\frac{1}{\sqrt{2}}\left(\gamma^\sigma_y+i\delta^\sigma_y\right),\nonumber\\\begin{split}\gamma^\sigma_y=\sqrt{2}\cos\Phi^\sigma_{y,d0}\prod_{y'<y}(-1)^{N^d_{y'}},\\\delta^\sigma_y=\sqrt{2}\sin\Phi^\sigma_{y,d0}\prod_{y'<y}(-1)^{N^d_{y'}}.\end{split}\label{MajoranaFermion}\end{gather} We introduce the Dirac fermion backscattering dimerization \begin{align}\mathcal{H}^{\mathrm{dimer}}_{y,DF}&=-i\frac{u_{IV}}{2}{d^R_{y,2}}^\dagger d^L_{y,2}+h.c.\label{PfincellHDF}\\&=u_{IV}\cos\left(\Phi^R_{y,d2}-\Phi^L_{y,d2}\right),\nonumber\end{align} and the Majorana fermion backscattering dimerization \begin{align}\mathcal{H}^{\mathrm{dimer}}_{y,MF}&=-iu_V\delta^R_y\delta^L_y\label{PfincellHMF}\\&=u_V\left[\cos\left(\Phi^R_{y,d0}-\Phi^L_{y,d0}\right)+\cos\left(\Phi^R_{y,d0}+\Phi^L_{y,d0}\right)\right].\nonumber\end{align} (The factors of $i$ appearing in \eqref{PfincellHDF} and \eqref{PfincellHMF} are consequences of the Baker-Campbell-Hausdorff formula $e^Ae^B=e^{A+B+[A,B]/2}$ and the constant terms in the complete \ETCR \eqref{appETCRPf2} presented in appendix~\ref{app:ETCRpaired}.) These fermion bilinear potentials can be expressed as integral products of the original electron operators $c^\sigma_{yAa}$. These intra-unit cell interactions leave behind the following degrees of freedom unaffected: (a) the electrically charged spin $h=\pm3$ bosons $e^{i\Phi^\sigma_{y,\rho}}$ that generate a $U(1)_6$ sector for each propagating direction $\sigma=R,L$, (b) the electrically neutral spin $h=\pm3/2$ Dirac fermions $d^\sigma_{y,d1}$ that generate the counter-propagating pair of $U(1)_3$ sectors, and (c) the electrically neutral spin $h=\pm1/2$ counter-propagating pair of Majorana fermions $\gamma^\sigma_y$. These combine into the non-chiral parton Pfaffian \hyperlink{CFT}{CFT}. (See figure~\ref{fig:PfConstructIntra} for a summary.)

\begin{figure}[htbp]
\centering\includegraphics[width=0.45\textwidth]{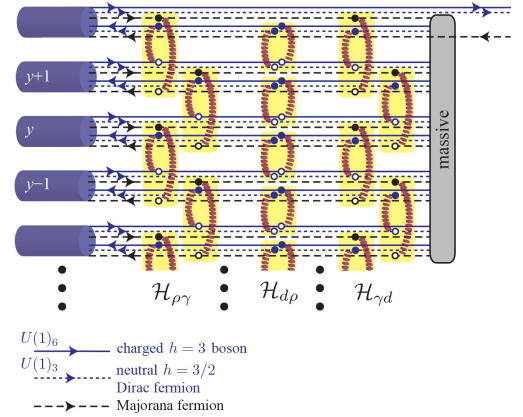}
\caption{Coupled wire model and its gapless edge modes of the $\nu=1/6$ parton Pfaffian state. $\mathcal{H}_{\rho\gamma}$, $\mathcal{H}_{d\rho}$ and $\mathcal{H}_{\gamma d}$ are defined in \eqref{PfrhogammaH}, \eqref{PfdrhoH} and \eqref{PfgammadH}.}\label{fig:PfCoupledWire}
\end{figure}

The remaining interactions in the model couple degrees of freedom between unit cells. They are \begin{widetext}\begin{align}\mathcal{H}_{\rho\gamma}&=-u_{\rho\gamma}\sum_y\cos\left(\Phi^R_{y\rho}-\Phi^L_{y+1,\rho}-\pi N^d_y\right)i\gamma^L_y\gamma^R_{y+1}\label{PfrhogammaH}\\&=\frac{u_{\rho\gamma}}{2}\sum_y\sum_{s_1,s_2=\pm}s_1s_2\cos\left(\Phi^R_{y\rho}-\Phi^L_{y+1,\rho}+s_1\Phi^L_{y,d0}+s_2\Phi^R_{y+1,d0}\right),\nonumber\\\mathcal{H}_{d\rho}&=-u_{d\rho}\sum_y\left(i{d^R_{y,1}}^\dagger d^L_{y+1,1}+h.c.\right)\cos\left(\Phi^R_{y\rho}-\Phi^L_{y+1,\rho}-\pi N^d_y\right)\label{PfdrhoH}\\&=-u_{d\rho}\sum_y\sum_{s=\pm}\cos\left(\Phi^R_{y\rho}-\Phi^L_{y+1,\rho}+s\Phi^R_{y,d1}-s\Phi^L_{y+1,d1}\right),\nonumber\\\mathcal{H}_{\gamma d}&=-u_{\gamma d}\sum_y\left(i{d^R_{y,1}}^\dagger d^L_{y+1,1}+h.c.\right)i\gamma^L_y\gamma^R_{y+1}\label{PfgammadH}\\&=u_{\gamma d}\sum_y\sum_{s_1,s_2=\pm}s_1s_2\cos\left(\Phi^R_{y,d1}-\Phi^L_{y+1,d1}+s_1\Phi^L_{y,d0}+s_2\Phi^R_{y+1,d0}\right).\nonumber\end{align}\end{widetext} The interwire potentials preserves the external $U(1)_{\mathrm{EM}}$ as well as $\mathsf{x}$-momentum. The factors of $i$ and signs $s_1$, $s_2$ in the interactions \eqref{PfrhogammaH}, \eqref{PfdrhoH} and \eqref{PfgammadH} are consequences of the Baker-Campbell-Hausdorff formula $e^Ae^B=e^{A+B+[A,B]/2}$ and the constant terms in the complete \hyperlink{ETCR}{ETCR}s \eqref{appETCRPf2} and \eqref{Ndoperatorcomm} in appendix~\ref{app:ETCRpaired}. Any two out of the three sets of interactions are enough to introduce a finite excitation energy gap in the bulk. When all three terms are present, they are not competing when the product $u_{\rho\gamma}u_{d\rho}u_{\gamma d}$ is positive. They collectively pin the ground state expectation values of the order parameters \begin{align}\begin{split}&e^{i\left\langle\Theta^\rho_{y+1/2}(\mathsf{x})\right\rangle}=e^{i\left\langle\Phi^R_{y,\rho}(\mathsf{x})-\Phi^L_{y+1,\rho}(\mathsf{x})-\pi N^d_y\right\rangle},\\&i\left\langle d^R_{y,1}(\mathsf{x})^\dagger d^L_{y+1,1}(\mathsf{x})\right\rangle,\quad
i\left\langle\gamma^L_y(\mathsf{x})\gamma^R_{y+1}(\mathsf{x})\right\rangle\end{split}\label{Pforderparameters}\end{align} to be either all positive or all negative. Figure~\ref{fig:PfCoupledWire} summarizes the three sets of interactions. 

To summarize, the coupled wire model involves the following gapping potentials \begin{align}\mathcal{H}&=\mathcal{H}_{\rho\gamma}+\mathcal{H}_{d\rho}+\mathcal{H}_{\gamma d}\nonumber\\&\;\;\;+\sum_y\mathcal{H}^{\mathrm{dimer}}_{y,MF}+\mathcal{H}^{\mathrm{dimer}}_{y,DF}+\mathcal{H}^{\mathrm{dimer}}_{y,SU(3)}\nonumber\\&\;\;\;+\sum_y\sum_{A=0,1}\mathcal{H}^{\mathrm{intra}}_{y,A}.\label{PHPfHsummary}\end{align} The first line consists of the inter-bundle terms \eqref{PfrhogammaH}, \eqref{PfdrhoH} and \eqref{PfgammadH}, which are also shown in figure~\ref{fig:PfCoupledWire}. The second line contains the terms \eqref{PfdimerSU3H}, \eqref{PfincellHDF} and \eqref{PfincellHMF} that act within a super-unit cell. The last line includes intra-bundle terms that were defined previously in \eqref{intrabundleH}. The intra-cell and intra-bundle terms are summarized in figure~\ref{fig:PfConstructIntra}. 

It is important to notice that these interactions are consistent with electron locality. Paradoxically, the potentials involve backscatterings of fractional fields. For example, $\mathcal{H}^{\mathrm{dimer}}_{y,MF}$, $\mathcal{H}^{\mathrm{dimer}}_{y,DF}$ and $\mathcal{H}_{\gamma d}$ backscatter Majorana and neutral Dirac fermions. $\mathcal{H}_{\rho\gamma}$ and $\mathcal{H}_{d\rho}$ couples the fractional charge $e$ boson $e^{i\Phi_\rho}$ between wires. However, these potentials are specifically designed so that under a basis transformation of the bosonized variables, each of them can be expressed as integral combinations of local electrons. (See \eqref{appPfincellHDFMF}, \eqref{appPfinter1} and \eqref{appPfinter1} in appendix~\ref{app:ETCRpaired}.) In particular, each of the inter-bundle terms $\mathcal{H}_{\gamma d}$, $\mathcal{H}_{\rho\gamma}$ and $\mathcal{H}_{d\rho}$ simultaneously backscatters two of the three fractional fields $e^{i\Phi_\rho}$, $d_1\sim e^{i\Phi_{d1}}$ and $\gamma\sim\cos\Phi_{d0}$. The bosonized variables $\Phi_\rho$, $\Phi_{dJ}$ are half-integral in the sense that any sum or difference between any pair is an integral combination of local electron variables. As a result, the inter-bundle potentials are local. On the other hand, potentials that backscatter a single fractional species cannot be constructed from integral combinations of electrons, and are not allowed by electron locality.

The gapping potential \eqref{PHPfHsummary} leave behind the gapless edge mode \begin{align}\mathrm{Pf}^\ast=U(1)_6\times U(1)_3\times\overline{\mbox{Majorana fermion}}.\end{align} Along the top edge in figure~\ref{fig:PfCoupledWire}, the three sectors are generated by \begin{align}2\phi^R_\rho=\Phi^R_\rho/6,\quad 2\phi^R_d=\Phi^R_{d1}/3,\quad\psi^L=\gamma^L\end{align} and are described by the Lagrangian densities \eqref{CSPfAbCFT} and \eqref{CSPfMF}. The \CFT supports the charge $(a/12)e$ Abelian primary fields \begin{align}\begin{split}\openone_{a,b}&=e^{ia\phi^R_\rho}e^{ib\phi^R_d}\\\Psi_{a,b}&=e^{ia\phi^R_\rho}e^{ib\phi^R_d}\gamma^L\end{split}\label{CWPfAb}\end{align} for even integers $a,b$ (c.f.~the notation from \eqref{Pfprimaryfieldcontent}). In the bulk, these operators $\openone_{a,b}(\mathsf{x}_0,y),\Psi_{a,b}(\mathsf{x}_0,y)$ create (or annihilate) gapped excitations in the form of vortices/kinks of the order parameters in \eqref{Pforderparameters} at $\mathsf{x}_0$. 
As a consequence of the pair backscattering structures of the inter-bundle interactions $\mathcal{H}_{\gamma d}$, $\mathcal{H}_{\rho\gamma}$ and $\mathcal{H}_{d\rho}$, they allow deconfined half-excitations that correspond to half-vortices/$\pi$-kinks of all three sectors. They associate to the charge $(a/12)e$ non-Abelian bosonic products \begin{align}\Sigma_{a,b}=e^{ia\phi^R_\rho}e^{ib\phi^R_{d1}}\sigma^L_0\label{CWPfIsing}\end{align} for odd integers $a,b$, where $\sigma^L_0$ is the Ising twist field of the Majorana fermion $\gamma^L$. Each $\Sigma_{a,b}$ has $\pi$ monodromy with all three generators $e^{i\Phi_\rho^R}$, $d_1^R\sim e^{i\Phi^R_{d1}}$ and $\gamma^L$ of $U(1)_6$, $U(1)_3$ and $\overline{\mathrm{Ising}}$. The parton Pfaffian state therefore has the extended product structure \begin{align}\mathrm{Pf}^\ast=U(1)_{24}\otimes_{\mathrm{el}}U(1)_{12}\otimes_{\mathrm{el}}\overline{\mbox{Ising}}\label{partonPforder}\end{align} described in section~\ref{sec:PartonPfaffian}. Eq.\eqref{CWPfAb} and \eqref{CWPfIsing} accounts for all primary fields. All other vortices -- such as $\openone_{a,b},\Psi_{a,b}$, for $a\not\equiv b$ modulo 2, and $\Sigma_{a,b}$, for $a\equiv b$ modulo 2 -- are non-local with respect to the electron and are confined. For example, the pair $\openone_{1,0}(\mathsf{x}_0,y),\openone_{-1,0}(\mathsf{x}_1,y)$ creates a $\pi$-kink dipole for $\Theta^\rho_{y+1/2}$ alone, and associates to a linearly diverging excitation energy $2(u_{\rho\gamma}+u_{d\rho})|\mathsf{x}_0-\mathsf{x}_1|$. The confinement of these vortices is remembered by the ``electronic tensor product" $\otimes_{\mathrm{el}}$.

\subsubsection{Variations}
The coupled wire model \eqref{PHPfHsummary} described the ``particle-hole" symmetric parton Pfaffian state. By rearranging the inter-bundle terms $\mathcal{H}_{\rho\gamma}$, $\mathcal{H}_{d\rho}$ and $\mathcal{H}_{\gamma d}$, one can construct a sequence of paired parton states at the same filling fraction $\nu=1/6$ but with different edge modes and topological order. In section~\ref{sec:fillingonethird}, the coupled wire models \eqref{PartonCWfullH} (see also figure~\ref{fig:PartonCoupledWire}) at filling $\nu=1/3$ generate a series of Abelian parton states $\mathfrak{L}_q=U(1)_3\times[SU(3)_1]^q$ by allowing $q^{\mathrm{th}}$ nearest neighbor bundle backscatterings. A similar construction can be applied for the paired parton sequence \begin{align}\mathfrak{P}_{p,q}=U(1)_{24}\otimes_{\mathrm{el}}U(1)_{12}^q\otimes_{\mathrm{el}}\mathrm{Ising}^p,\label{Pfaffiansequence}\end{align} where $p$, $q$ are integers. Negative powers associate to counter propagating sectors. For example, $U(1)_6^{-q}=\overline{U(1)_6^q}$ and $\mathrm{Ising}^{-p}=\overline{\mathrm{Ising}^p}$. For instance, the particular parton Pfaffian state described previously in \eqref{partonPforder} is $\mathrm{Pf}^\ast=\mathfrak{P}_{p=1,q=-1}$. The chiral central charge of \eqref{Pfaffiansequence} depends on its neutral sectors, and is given by \begin{align}c_{p,q}=1+q+p/2.\label{Ppqcentralcharge}\end{align} The case when both powers $p,q$ are trivial is special. It corresponds to an Abelian state with strong electronic quasiparticle pairing. It has a strongly-paired topological order and edge \CFT \begin{align}\mathrm{SP}^\ast=\mathfrak{P}_{p=0,q=0}=U(1)_{24}.\label{strongpair}\end{align} There is no gapless charge $e$ electronic quasiparticles on the boundary. Instead the smallest local electronic primary field is a charge $2e$ Cooper pair. 

\begin{figure}[htbp]
\centering\includegraphics[width=0.45\textwidth]{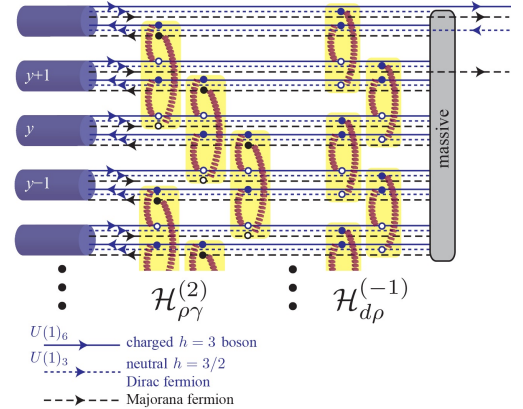}
\caption{Coupled wire model and its gapless edge modes of the $\nu=1/6$ parton Pfaffian state $\mathfrak{P}_{2,-1}$ in \eqref{Pfaffiansequence}. $\mathcal{H}_{\rho\gamma}^{(2)}$ and $\mathcal{H}_{d\rho}^{-1}$ are defined in \eqref{PfrhogammaHpq} and \eqref{PfdrhoHpq}.}\label{fig:PfCoupledWire2}
\end{figure}

We first consider the variations when the powers $p$ and $q$ are not both trivial. The general coupled wire model is \begin{align}\mathcal{H}^{(p,q)}&=\mathcal{H}^{(p)}_{\rho\gamma}+\mathcal{H}^{(q)}_{d\rho}\nonumber\\&\;\;\;+\sum_y\mathcal{H}^{\mathrm{dimer}}_{y,MF}+\mathcal{H}^{\mathrm{dimer}}_{y,DF}+\mathcal{H}^{\mathrm{dimer}}_{y,SU(3)}\nonumber\\&\;\;\;+\sum_y\sum_{A=0,1}\mathcal{H}^{\mathrm{intra}}_{y,A}.\label{PfHpqsummary}\end{align} The intra-cell and intra-bundle terms in the second and third line are identical to that of \eqref{PHPfHsummary}, and were defined in \eqref{intrabundleH}, \eqref{PfdimerSU3H}, \eqref{PfincellHDF} and \eqref{PfincellHMF} (see also figure~\ref{fig:PfConstructIntra}). The first two terms $\mathcal{H}^{(p)}_{\rho\gamma}$ and $\mathcal{H}^{(q)}_{d\rho}$ involve $p^{\mathrm{th}}$ and $q^{\mathrm{th}}$ nearest neighbor backscattering. We ignore terms like $\mathcal{H}_{\gamma d}$ in \eqref{PfgammadH} that couple only neutral modes because they are redundant for the bulk energy gap. 

The generalized inter-bundle terms are defined by \begin{widetext}\begin{align}\mathcal{H}^{(p)}_{\rho\gamma}&=\left\{\begin{diagram}-u_{\rho\gamma}\sum_yi\gamma^R_y\gamma^L_{y+p}\prod_{j=0}^{p-1}\cos\left(\Phi^R_{y+j,\rho}-\Phi^L_{y+j+1,\rho}-\pi N^d_{y+j}\right),&\quad&\mbox{for $p>0$}\\-u_{\rho\gamma}\sum_yi\gamma^R_y\gamma^L_{y+p}\prod_{j=0}^{-p-1}\cos\left(\Phi^R_{y-j-1,\rho}-\Phi^L_{y-j,\rho}-\pi N^d_{y-j-1}\right),&\quad&\mbox{for $p<0$}\\-u_{\rho\gamma}\sum_yi\gamma^R_y\gamma^L_y,&\quad&\mbox{for $p=0$}\end{diagram}\right.,\label{PfrhogammaHpq}\\\mathcal{H}^{(q)}_{d\rho}&=\left\{\begin{diagram}-u_{d\rho}\sum_y\left(i{d^R_{y,1}}^\dagger d^L_{y+q,1}+h.c.\right)\prod_{j=0}^{q-1}\cos\left(\Phi^R_{y+j,\rho}-\Phi^L_{y+j+1,\rho}-\pi N^d_{y+j}\right),&\quad&\mbox{for $q>0$}\\-u_{d\rho}\sum_y\left(i{d^R_{y,1}}^\dagger d^L_{y+q,1}+h.c.\right)\prod_{j=0}^{-q-1}\cos\left(\Phi^R_{y-j-1,\rho}-\Phi^L_{y-j,\rho}-\pi N^d_{y-j-1}\right),&\quad&\mbox{for $q<0$}\\-u_{d\rho}\sum_y\left(i{d^R_{y,1}}^\dagger d^L_{y,1}+h.c.\right),&\quad&\mbox{for $q=0$}\end{diagram}\right..\label{PfdrhoHpq}\end{align}\end{widetext} These interactions conserve charge $U(1)_{\mathrm{EM}}$ and $\mathsf{x}$-momentum. They can be re-expressed as integral combinations of electron operators (see \eqref{appPfincellHDFMF}, \eqref{appPfinter1} and \eqref{appPfinter1} in appendix~\ref{app:ETCRpaired}), and are therefore consistent with electron locality. They introduce a finite excitation energy in the bulk and leave behind the edge \CFT \eqref{Pfaffiansequence} in low-energy. An example for $p=2$ and $q=-1$ is illustrated in figure~\ref{fig:PfCoupledWire2}.

Special care is needed for the trivial case when $p=q=0$. This is because $\mathcal{H}^{(0)}_{\rho\gamma}$ and $\mathcal{H}^{(0)}_{d\rho}$ in \eqref{PfrhogammaHpq} and \eqref{PfdrhoHpq} only backscatter the Majorana and Dirac fermions $\gamma$ and $d_1$ within a super-cell. They do not involve the charge boson $e^{i\Phi_\rho}$, which still remains massless in the bulk. The charge sector can be turned massive by the boson backscattering \begin{align}\mathcal{H}_\rho^{(0,0)}=-u_\rho\sum_y\cos\left(2\Phi^R_{y\rho}-2\Phi^L_{y+1,\rho}\right),\label{PfchargeH00}\end{align} where the factor of 2 ensures that $\mathcal{H}_\rho^{(0,0)}$ is an integral combination of electron operators. By including \eqref{PfchargeH00} in the coupled wire model \eqref{PfHpqsummary} for $p=q=0$, $\mathcal{H}^{(0,0)}_\rho+\mathcal{H}^{(0,0)}$ removes all low-energy degrees of freedom in the bulk and leave behind the strongly-paired edge \CFT \eqref{strongpair}.

\subsection{The parton T-Pfaffian surface state of a fractional topological insulator}\label{sec:partonTPfaffian}

In previous works~\cite{SahooSirotaChoTeo17,ChoTeoFradkin17}, we proposed the symmetry-preserving $\mathcal{T}-\mathrm{Pf}^\ast$ surface topological order of a fractional topological insulator (\hyperlink{FTI}{FTI})~\cite{MaciejkoQiKarchZhang10,SwingleBarkeshliMcGreevySenthil11,LevinBurnellKochStern11,maciejko2012models,ye2016composite,maciejko2015fractionalized,stern2016fractional,YeChengFradkin17}. The \FTI consists of deconfined partons coupled to a discrete $\mathbb{Z}_3$ gauge theory. Each of the three parton species occupies a time-reversal (\hyperlink{TR}{TR}) symmetric fermionic topological band and hosts a parton Dirac surface state. The surface can be turned massive without breaking \TR symmetry or charge $U(1)$ conservation, and similar to the T-Pfaffian surface state~\cite{ChenFidkowskiVishwanath14}, the parton version $\mathcal{T}-\mathrm{Pf}^\ast$ admits a fractional anyonic excitation structure and hosts an Ising-like topological order. The surface quasiparticle structure was introduced in ref.~\onlinecite{SahooSirotaChoTeo17,ChoTeoFradkin17} and was reviewed in \eqref{TPfcontent} in section~\ref{sec:PartonPfaffian}. In this section, we propose a coupled wire description to this surface topological order. The construction parallels the coupled wire description by Mross, Essin and Alicea~\cite{MrossEssinAlicea15} of the T-Pfaffian surface state of a conventional topological insulator.

\begin{figure}[htbp]
\centering\includegraphics[width=0.35\textwidth]{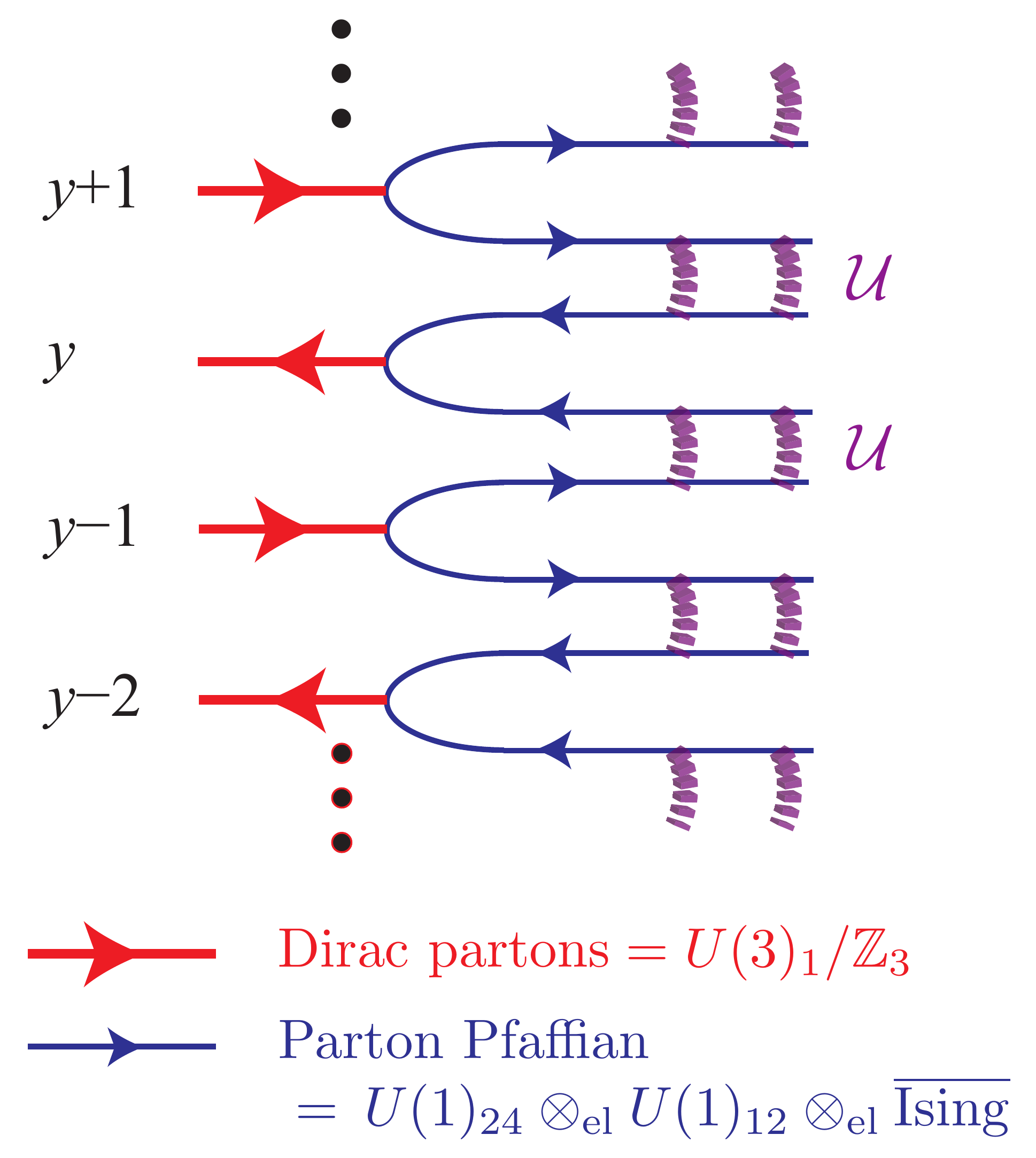}
\caption{Coupled wire model of the $\mathcal{T}-\mathrm{Pf}^\ast$ surface state of a fractional topological insulator. The energy gap is introduced by the many-body electron backscattering $\mathcal{U}$ defined in \eqref{Pfbackscattering}.}\label{fig:TPfsurface}
\end{figure}

The massless parton Dirac fermions on the surface of the \FTI can be mimicked by a 2D array of chiral parton Dirac channels, labeled by their vertical position $y$, with alternating propagating directions $\sigma=R,L=+,-=(-1)^y$. These channels are represented by the red wires in figure~\ref{fig:TPfsurface}. Each channel carries a parton Dirac \CFT $U(3)_1/\mathbb{Z}_3=U(1)_3\times SU(3)_1$ described in section~\ref{sec:DiracParton}. It consists of three parton Dirac fermions that propagate in a single-direction. The emergence of these parton channels can be facilitated by an antiferromagnetic stripe order on the \FTI surface. It introduces a time-reversal symmetry-breaking parton energy gap on each stripe, where the Dirac mass $m(y)=m_0\mathrm{sgn}[\sin(\pi y)]$ flips sign across adjacent stripes. This removes the 2D parton Dirac surface fermions and leaves behind a chiral Dirac channel $U(3)_1/\mathbb{Z}_3$ along each 1D interface. The unbalanced chirality is allowed by \TR symmetry-breaking. On the other hand, the 2D array collectively recovers an antiferromagnetic time-reversal ({\color{blue}\hypertarget{AFTR}{AFTR}}) symmetry, which combines the local \TR conjugation with the half-translation $y\to y+1$. 

The chiral parton Dirac fermion at wire $y$ can be bosonized into $\pi^a_y(\mathsf{x})\sim e^{i\tilde\phi_{y,a}(\mathsf{x})}$, for $a=1,2,3$. The bosonized variables $\tilde{\phi}^a_y$ fields satisfy the equal-time commutation relation (\hyperlink{ETCR}{ETCR}) \begin{align}\left[\partial_{\mathsf{x}}\tilde\phi_{y,a}(\mathsf{x}),\tilde\phi_{y',a'}(\mathsf{x}')\right]=2\pi i(-1)^y\delta_{y,y'}\delta_{a,a'}\delta(\mathsf{x}-\mathsf{x}').\end{align} The symmetry-preserving many-body gapping interaction relies on the bipartition of each parton Dirac channel into a pair of parton Pfaffians, $U(3)_1/\mathbb{Z}_3=\mathrm{Pf}^\ast\boxtimes\mathrm{Pf}^\ast$. The splitting allows the parton Pfaffian channels to be backscattered independently in opposite directions (see figure~\ref{fig:TPfsurface}). This splitting basis transformation was introduced in section~\ref{sec:gluingsplitting} (see figure~\ref{fig:splitting} for a summary). It requires a channel reconstruction that extends the parton Dirac \CFT by two counter-propagating pairs of electron modes, $c^\sigma_{y,b}(\mathsf{x})\sim e^{i\tilde\Phi^\sigma_{y,b}(\mathsf{x})}$, for $\sigma=R,L=+,-$ and $b=0,1$. The electronic bosonized variables obey the \ETCR \begin{align}&\left[\partial_{\mathsf{x}}\tilde\Phi^{\sigma}_{y,b}(\mathsf{x}),\tilde\Phi^{\sigma '}_{y',b'}(\mathsf{x}')\right]=2\pi i\sigma\delta_{y,y'}\delta_{b,b'}\delta^{\sigma,\sigma'}\delta(\mathsf{x}'-\mathsf{x}).\end{align} 

The \AFTR symmetry, represented by the anti-unitary operator $\mathcal{T}$, sends the partons and the auxiliary electrons from the $y^{\mathrm{th}}$ wire to the $(y+1)^{\mathrm{th}}$ one \begin{align}\begin{split}\mathcal{T}\tilde\phi_{y,a}\mathcal{T}^{-1}&=-\tilde\phi_{y+1,a}+\frac{(-1)^{y}}{2}\pi,\\\mathcal{T}\tilde\Phi^\sigma_{y,b}\mathcal{T}^{-1}&=-\tilde\Phi^\sigma_{y+1,b}+\frac{(-1)^{y}}{2}\pi.\end{split}\label{TRparton}\end{align} The \AFTR operator squares to \begin{align}\mathcal{T}^2=(-1)^F\mathrm{translation}_{y\to y+2},\end{align} where $(-1)^F$ is the total fermion parity operator. 
The additional phases to the \AFTR transformation are included so that $\mathcal{T}^{2}\tilde\phi_{y,a}\mathcal{T}^{-2}=(-1)^F\tilde{\phi}_{y+2,a}(-1)^F=\tilde{\phi}_{y+2,a}+(-1)^y\pi$ and similarly for the electronic bosonized variables $\tilde\Phi^\sigma_{y,b}$. 

The basis transformations \eqref{splittingbasistrans1}, \eqref{splittingbasistrans2}, \eqref{splittingsd} and \eqref{splittingB} together with the backscattering \eqref{splittingHrho} split the chiral parton Dirac \CFT $U(3)_1/\mathbb{Z}_3$ along each interface into a pair of parton Pfaffians $\mathrm{Pf}^\ast_{A/B}=U(1)_{24}^{A/B}\otimes_{\mathrm{el}}U(1)_{12}^{A/B}\otimes_{\mathrm{el}}\overline{\mathrm{Ising}}^{A/B}$. The bosonized variables $\phi^{A/B}_\rho$, $\phi^{A/B}_d$ and $\phi_0$ in $U(1)_{24}^{A/B}$, $U(1)_{12}^{A/B}$ and $\overline{SO(2)_1}=\overline{\mathrm{Ising}}^A\times\overline{\mathrm{Ising}}^B$ are described by the Lagrangian densities (up to non-universal velocity terms) \begin{gather}\mathcal{L}_{\mathrm{tot}}=\sum_y\left[\left(\sum_{C=A,B}\mathcal{L}^C_{y,\rho}+\mathcal{L}^C_{y,d}\right)+\mathcal{L}_{y,0}\right],\\\begin{split}&\mathcal{L}^C_{y,\rho}=\frac{1}{2\pi}(-1)^y24\partial_{\mathrm{t}}\phi^C_{y,\rho}\partial_{\mathrm{x}}\phi^C_{y,\rho},\\&\mathcal{L}^C_{y,d}=\frac{1}{2\pi}(-1)^y12\partial_{\mathrm{t}}\phi^C_{y,d}\partial_{\mathrm{x}}\phi^C_{y,d},\\&\mathcal{L}_{y,0}=\frac{1}{2\pi}(-1)^{y+1}4\partial_{\mathrm{t}}\phi_{y,0}\partial_{\mathrm{x}}\phi_{y,0}.\end{split}\nonumber\end{gather} The neutral Dirac fermion $e^{i\phi_{y,0}}\sim\cos(2\phi_{y,0})+i\sin(2\phi_{y,0})$ can be split into Majorana components \begin{align}\begin{split}&\psi^A_y=\sqrt{2}\cos(2\phi_{y,0})\prod_{y'<y}(-1)^{N_{y'}},\\&\psi^B_y=\sqrt{2}\sin(2\phi_{y,0})\prod_{y'<y}(-1)^{N_{y'}},\end{split}\end{align} where the Jordon-Wigner string is a product of some electronic number operators $N_y$ similar to that in \eqref{MajoranaFermion} and it ensures the mutual anti-commutation relations between Majorana fermions in different wires. The bosonized variables and Majorana fermions transform according to the antiferromagnetic time-reversal symmetry \eqref{TRparton} \begin{gather}\mathcal{T}\phi^{A/B}_{y,\rho}\mathcal{T}^{-1}=-\phi^{A/B}_{y+1,\rho}+\frac{(-1)^y}{8}\pi,\nonumber\\\mathcal{T}\phi^{A/B}_{y,d}\mathcal{T}^{-1}=-\phi^{A/B}_{y+1,d},\quad\mathcal{T}\phi_{y,0}\mathcal{T}^{-1}=-\phi_{y+1,0},\nonumber\\\mathcal{T}\psi^A_y\mathcal{T}^{-1}=\psi^A_{y+1},\quad\mathcal{T}\psi^B_y\mathcal{T}^{-1}=-\psi^B_{y+1}.\label{TRPfaffian}\end{gather}

There are three primitive electronic quasiparticles $\openone_{12,\pm6}=e^{i12\phi_\rho}e^{\pm i6\phi_d}$, $\Psi_{12,0}=e^{i12\phi_\rho}\psi$ in each parton Pfaffian channel. They all carries electric charge $e$ and were defined in \eqref{Pflocalel} in section~\ref{sec:PartonPfaffian} and \eqref{elecQPsplitting} in section~\ref{sec:gluingsplitting}. The inter-wire gapping interactions in figure~\ref{fig:TPfsurface} consists of the backscatterings of these electronic quasiparticles. \begin{gather}\mathcal{U}=\sum_y\mathcal{U}_{y+1/2}=\sum_y\left(\mathcal{U}^1_{y+1/2}+\mathcal{U}^2_{y+1/2}+\mathcal{U}^3_{y+1/2}\right)\label{Pfbackscattering}\end{gather} where the interactions between each pair of wires are \begin{align}\mathcal{U}^1_{y+1/2}&=-u_1(-1)^y{\openone_{12,6,y+1}^A}^\dagger\openone_{12,6,y}^B+h.c.\nonumber\\&=-\tilde{u}_1(-1)^y\cos\left(\Theta^\rho_{y+1/2}+\Theta^d_{y+1/2}\right),\nonumber\\\mathcal{U}^2_{y+1/2}&=-u_2(-1)^y{\openone_{12,-6,y+1}^A}^\dagger\openone_{12,-6,y}^B+h.c.\nonumber\\&=-\tilde{u}_2(-1)^y\cos\left(\Theta^\rho_{y+1/2}-\Theta^d_{y+1/2}\right),\nonumber\\\mathcal{U}^3_{y+1/2}&=-u_3(-1)^yi{\Psi_{12,0,y+1}^A}^\dagger\Psi_{12,0,y}^B+h.c.\nonumber\\&=-\tilde{u}_3(-1)^y\cos(\Theta^\rho_{y+1/2})i\psi^A_{y+1}\psi^B_y.\end{align} The sine-Gordon angle variables are $\Theta^\rho_{y+1/2}=12\phi_{y,\rho}-12\phi_{y+1,\rho}+\pi N_y$ and $\Theta^d_{y+1/2}=6\phi_{y,d}-6\phi_{y+1,d}+\pi N_y$. They transform according to the \AFTR symmetry $\mathcal{T}\Theta^\rho_{y+1/2}\mathcal{T}^{-1}=-\Theta^\rho_{y+3/2}+3\pi (-1)^y$ and $\mathcal{T}\Theta^d_{y+1/2}\mathcal{T}^{-1}=-\Theta^d_{y+3/2}$. Together with the \AFTR transformation \eqref{TRPfaffian} that sends $i\psi^A_{y+1}\psi^B_y\to i\psi^A_{y+2}\psi^B_{y+1}$, this shows the collection of inter-wire backscatterings \eqref{Pfbackscattering} preserves the antiferromagnetic time-reversal symmetry. Moreover, they all preserve charge $U(1)$ conservation because each term simply brings a charge $e$ electronic quasiparticle from one wire to the next.

The coupled wire model \eqref{Pfbackscattering} resembles the one that defines the paired parton Pfaffian state in \eqref{PfrhogammaH}, \eqref{PfdrhoH} and \eqref{PfgammadH}. The model is exactly solvable and the three gapping interactions are non-competing when $\tilde{u}_1\tilde{u}_2>0$ and $\tilde{u}_3\neq0$ so that they pin the ground state expectation values $\langle i\psi^A_{y+1}\psi^B_y\rangle=(-1)^{m_\psi}$ and $\langle\Theta^\rho_{y+1/2}\rangle=m_\rho\pi$, $\langle\Theta^\rho_{y+1/2}\rangle=m_d\pi$, where $(-1)^{m_\psi+m_\rho}=\mathrm{sgn}(\tilde{u}_3)$ and $(-1)^{m_\rho+m_d}=\mathrm{sgn}(\tilde{u}_1)$. Therefore they freeze all low-energy degrees of freedom on the surface and introduce a finite excitation energy gap in strong coupling.

\begin{figure}[htbp]
\centering\includegraphics[width=0.35\textwidth]{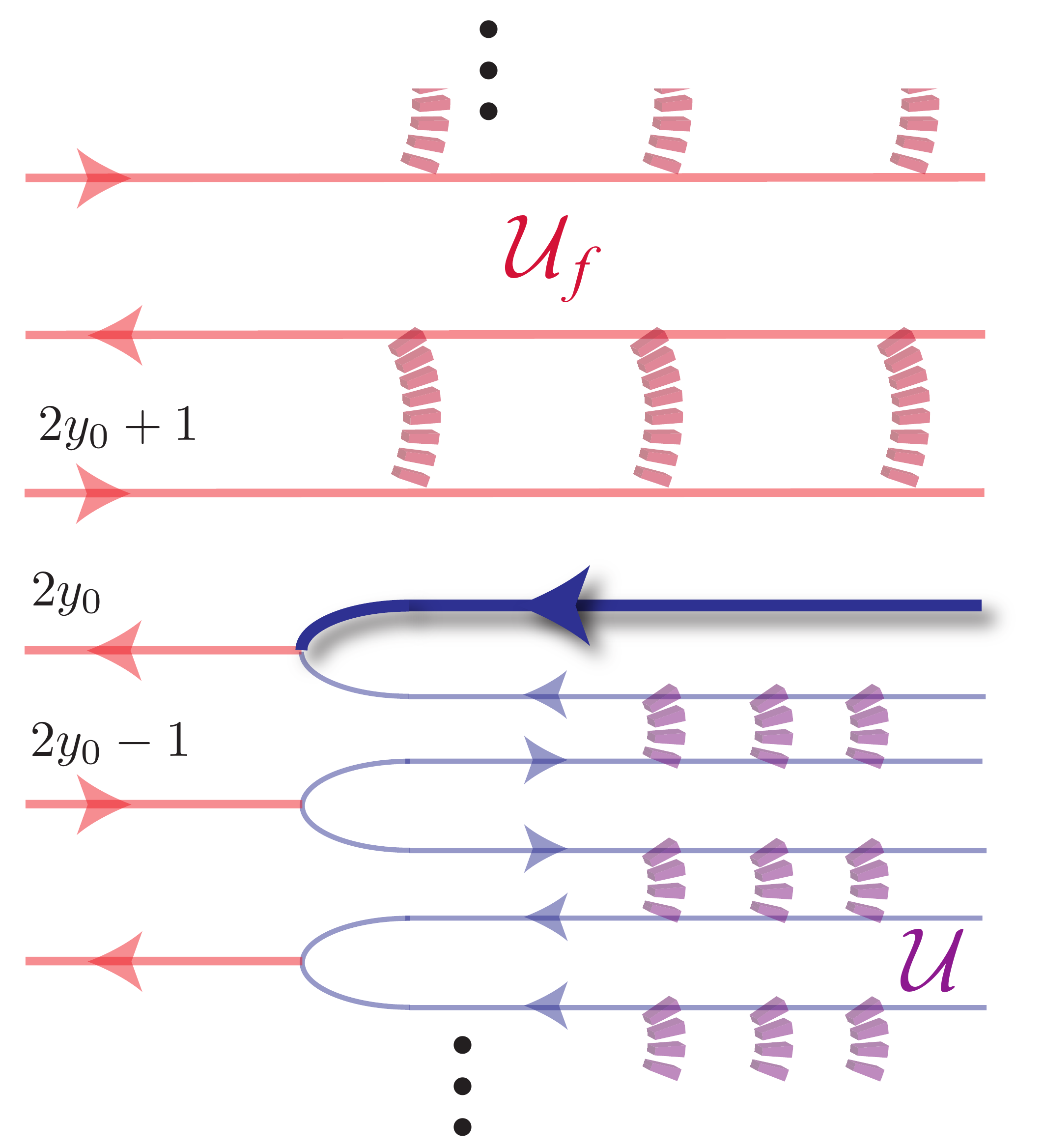}
\caption{A parton Pfaffian CFT (highligted blue line) at the domain wall separating two gapped regions with symmetry-preserving interaction $\mathcal{U}$ and AFTR symmetry breaking interaction $\mathcal{U}_f$.}\label{fig:TPfsurface2}
\end{figure}

The array of parton Dirac channels $U(1)_3\times SU(3)_1$ can also model a gapped \FTI surface that violates the \AFTR symmetry. The ``ferrimagnetic" interaction is given by \begin{align}\mathcal{U}_f=&-u_f\sum_y\Big[\cos\left(3\phi^0_{2y-1,\pi}-3\phi^0_{2y,\pi}\right)\nonumber\\&\;\;\;+\sum_{a\neq b}E^{ab}_{2y-1}E^{ab}_{2y}\Big]\nonumber\\&-m_e\sum_y\sum_{b=0,1}{c^R_{y,b}}^\dagger c^L_{y,b}+h.c.\label{Uf}\end{align} where $3\phi^0_{y,\pi}=\tilde\phi_{y,1}+\tilde\phi_{y,2}+\tilde\phi_{y,3}$ is the electronic bosonized variable of the charge sector $U(1)_3$ and $E^{ab}_y=e^{i(\tilde\phi_{y,a}-\tilde\phi_{y,b})}$ are the roots of the neutral $SU(3)_1$ sector (see section~\ref{sec:DiracParton}). The last line consists of intra-wire electronic backscatterings that removes the two pairs of counter-propagating auxiliary electrons from low-energy. \eqref{Uf} dimerizes the array of parton Dirac channels and introduces a finite excitation energy gap on the surface.

When the \AFTR symmetry-breaking surface is juxtaposed with the parton $\mathcal{T}-\mathrm{Pf}^\ast$ surface enabled with interaction $\mathcal{U}$ in \eqref{Pfbackscattering}, they leave behind a parton Pfaffian \CFT at the domain wall interface (see figure~\ref{fig:TPfsurface2}). This verifies one of the propositions in our previous work~\cite{SahooSirotaChoTeo17} that a \FTI slab with a symmetry-preserving top surface and a time-reversal breaking bottom surface hosts a parton Pfaffian \CFT along the boundary (see figure~\ref{fig:FTIslab}). The bulk-boundary correspondence implies the topological equivalence between the \FTI thin film (treated as a quasi-2D topological phase) and the particle-hole symmetric paired parton \FQH state $\mathfrak{P}_{1,-1}=\mathrm{Pf}^\ast$.

\section{Particle-hole symmetry for partons}\label{sec:particlehole}
Particle-hole (\hyperlink{PH}{PH}) conjugation~\cite{Girvin84,Son15,BarkeshliMulliganFisher15,WangSenthil16,BalramJain17,NguyenGolkarRobertsSon18} inverts a \FQH state of electrons to a \FQH state of holes in the lowest Landau level ({\color{blue}\hypertarget{LLL}{LLL}}). The \PH conjugate of a \FQH state $\mathfrak{F}$ is a new \FQH state obtained by subtracting $\mathfrak{F}$ from the \hyperlink{LLL}{LLL}. The ``subtraction" is topologically captured by a product \begin{align}\mathrm{PH}_{\mathrm{LLL}}(\mathfrak{F})=U(1)_1\otimes\overline{\mathfrak{F}}\label{PHLLL}\end{align} between the lowest Landau level $U(1)_1$ and the time-reversal conjugate $\overline{\mathfrak{F}}$ of $\mathfrak{F}$. This product describes the bulk topological order as well as the gapless edge conformal field theory (\hyperlink{CFT}{CFT}). For instance, the time-reversal conjugate $\overline{\mathfrak{F}}$ refers to the same edge \CFT as $\mathfrak{F}$ except with the opposite propagating direction. The \hyperlink{LLL}{LLL}'s electric and thermal Hall transport, which are specified by the filling fraction $\nu_{\mathrm{LLL}}=1$ and edge chiral central charge $c_{\mathrm{LLL}}=c_R-c_L=1$ in \eqref{sigmakappaxy}, are reduced by the subtraction \begin{align}\begin{split}\nu\left(\mathrm{PH}_{\mathrm{LLL}}(\mathfrak{F})\right)=1-\nu(\mathfrak{F}),\\c\left(\mathrm{PH}_{\mathrm{LLL}}(\mathfrak{F})\right)=1-c(\mathfrak{F}).\label{nucPHLLLflip}\end{split}\end{align} For example, this relates \FQH states in the Jain's sequence~\cite{Jain90} $\mathfrak{J}^+_p$ with filling $\nu=p/(2p+1)$ and central charge $c=p$ to another sequence $\mathfrak{J}^-_{p+1}$ with filling $\nu=(p+1)/(2(p+1)-1)$ and central charge $c=1-p$. The former (latter) has an effective Chern-Simons description $\mathcal{L}=\frac{1}{2\pi}K_{IJ}a_I\wedge da_J$ with the $p\times p$ $K$-matrix $(K_{\mathfrak{J}^+_p})_{IJ}=\delta_{IJ}+2$ [resp.~$(p+1)\times(p+1)$ $K$-matrix $(K_{\mathfrak{J}^-_{p+1}})_{IJ}=-\delta_{IJ}+2$]. The two are related by \eqref{PHLLL} which identifies \begin{align}K_{\mathfrak{J}^-_{p+1}}=G^T\left[1\oplus(-K_{\mathfrak{J}^+_p})\right]G=G^T\begin{pmatrix}1&\\&-K_{\mathfrak{J}^+_p}\end{pmatrix}G,\end{align} where 1 is the $K$-matrix of the lowest Landau level $U(1)_1$ and $G$ can be chosen to be the $GL(p+1,\mathbb{Z})$ transformation \begin{align}G=\left(\begin{smallmatrix}1&2&2&\ldots&2\\&-1&&&\\&&-1&&\\&&&\ddots&\\&&&&-1\end{smallmatrix}\right)_{(p+1)\times(p+1)}.\end{align} 

Particle-hole symmetry acts within the half-filled Landau level. It flips between \FQH states that shares the same filling at $\nu=1/2$. For example, \eqref{PHLLL} conjugates the Read-Moore Pfaffian state~\cite{MooreRead} $\mathrm{Pf}_1=U(1)_8\otimes_{\mathrm{el}}\mathrm{Ising}$ to the anti-Pfaffian state~\cite{LevinHalperinRosenow07,LeeRyuNayakFisher07} $\mathrm{PH}_{\mathrm{LLL}}(\mathrm{Pf})=U(1)_1\otimes\overline{U(1)_8}\otimes_{\mathrm{el}}\overline{\mathrm{Ising}}$, which is equivalent to $\mathrm{Pf}_{-3}=U(1)_8\otimes_{\mathrm{el}}\overline{\mathrm{Ising}}^3$ up to anyon condensation~\cite{BaisSlingerlandCondensation}. The equivalence can be explicitly demonstrated by the basis transformation \begin{align}\begin{pmatrix}2\phi^R\\\Phi^L_n\end{pmatrix}=\begin{pmatrix}1&-1\\-1&2\end{pmatrix}\begin{pmatrix}\tilde\Phi^R_{\mathrm{LLL}}\\2\phi^L\end{pmatrix},\label{PfAPftrans}\end{align} where $e^{i\tilde\Phi^R_{\mathrm{LLL}}}$ is the edge chiral $U(1)_1$ Dirac electron of the \hyperlink{LLL}{LLL}, $e^{i4\phi^R}$ and $e^{i4\phi^L}$ are the charge $e$ spin 1 bosons in $U(1)_8$ and $\overline{U(1)_8}$, and $d^L\sim e^{i\Phi^L_n}\sim\gamma^L+i\delta^L$ is a spin $1/2$ neutral Dirac fermion, which can be split into a pair of Majorana fermions and correspond to a pair of $\overline{\mathrm{Ising}}$ \hyperlink{CFT}{CFT}s. More recently, a particle-hole symmetric Pfaffian state $\mathrm{Pf}_{-1}=U(1)_8\otimes_{\mathrm{el}}\overline{\mathrm{Ising}}$ was proposed by Son~\cite{Son15} that is invariant under the particle-hole conjugation \eqref{PHLLL}, and has the identical topological structure of anyon excitations as the time-reversal symmetric T-Pfaffian surface state~\cite{ChenFidkowskiVishwanath14} of a 3D topological insulator. The equivalence between $\mathrm{Pf}_{-1}$ and its conjugate $\mathrm{PH}_{\mathrm{LLL}}(\mathrm{Pf}_{-1})=U(1)_1\otimes\overline{U(1)_8}\otimes_{\mathrm{el}}\mathrm{Ising}$ can be shown by a basis transformation similar to \eqref{PfAPftrans}, which leaves behind a $L$-moving neutral Dirac fermion $d^L\sim\gamma^L+i\delta^L$ that reduces to a single Majorana fermion $\gamma^L$ upon removing $\delta^L$ from low-energy by backscattering to the $R$-moving Ising sector. Later, Kane, Stern and Halperin~\cite{KaneSternHalperin17} generalized a sequence of Pfaffian \FQH states $\mathrm{Pf}_p=U(1)_8\otimes_{\mathrm{el}}\mathrm{Ising}^p$ at filling $\nu=1/2$ with chiral central charge $c=1+p/2$ using the coupled wire construction. Particle-hole symmetry acts closely as an involution \begin{align}\mathrm{PH}_{\mathrm{LLL}}(\mathrm{Pf}_p)=\mathrm{Pf}_{-2-p}.\end{align}

In this section, we discuss the topological aspects of a speculative emergent particle-hole symmetry based on partons. In this case, the analogue of a filled Landau level, which defines the base of the conventional \PH symmetry \eqref{PHLLL}, is one of the parton \FQH states $\mathfrak{L}_{q_0}=U(1)_3\times[SU(3)_1]^q$ at filling $\nu=1/3$ described in section~\ref{sec:fillingonethird}. In particular, the Dirac parton triplet $\mathfrak{L}_1=U(3)_1/\mathbb{Z}_3$ consists of filled Landau levels of the three deconfined partons $\pi^a$ that constitute the electron, $\Psi_{\mathrm{el}}=\pi^1\pi^2\pi^3$. The parton particle-hole conjugate is a \FQH state of parton quasi-hole excitations in $\mathfrak{L}_{q_0}$. The \PH conjugation of a \FQH state $\mathfrak{F}$ can be topologically captured by its subtraction from $\mathfrak{L}_{q_0}$ \begin{align}\mathrm{PH}_{\mathfrak{L}_{q_0}}(\mathfrak{F})=\mathfrak{L}_{q_0}\boxtimes\overline{\mathfrak{F}},\label{PHLq0}\end{align} where $\overline{\mathfrak{F}}$ is the time-reversal of $\mathfrak{F}$, and $\boxtimes$ is some reduced tensor product that involves the identification of partons in $\mathfrak{L}_{q_0}$ and $\overline{\mathfrak{F}}$ through an anyon condensation~\cite{BaisSlingerlandCondensation} process described below. Eq.\eqref{PHLq0} dictates the relationships of electric and energy transport between a conjugate pair (c.f.~\eqref{nucPHLLLflip} for the conventional case) \begin{align}\begin{split}&\nu\left(\mathrm{PH}_{\mathfrak{L}_{q_0}}(\mathfrak{F})\right)=\nu(\mathfrak{L}_{q_0})-\nu(\mathfrak{F})=\frac{1}{3}-\nu(\mathfrak{F}),\\&c\left(\mathrm{PH}_{\mathfrak{L}_{q_0}}(\mathfrak{F})\right)=c(\mathfrak{L}_{q_0})-c(\mathfrak{F})=1+2q_0-c(\mathfrak{F}).\end{split}\label{nucPHLq0flip}\end{align} We speculate that there may be an underlying microscopic description of the parton \PH conjugation that support \eqref{PHLq0}. It would involve an antiunitary \PH conjugation operator that flips between parton particles and holes. In the coupled wire mdoel, the operator would transform the bosonized variables from wires to wires $\mathcal{C}\phi_{y,\alpha}^\sigma\mathcal{C}^{-1}=c^{\sigma\sigma'}_{y-y',\alpha\alpha'}\phi^\sigma_{y',\alpha'}$. A similar construction has been proposed recently by Fuji and Furusaki~\cite{FujiFurusaki18} that applies to the conventional \PH conjugation.

In this section, we focus on the topological aspects of parton \PH conjugation instead of its microscopic origin. In particular, we focus on the conjugation among the paired parton \FQH states $\mathfrak{P}_{p,q}=U(1)_{24}\otimes_{\mathrm{el}}U(1)_{12}^q\otimes_{\mathrm{el}}\mathrm{Ising}^p$ at the \hyperlink{PH}{PH}-symmetric filling $\nu=1/6$ described in section~\ref{sec:fillingonesixth}. We will show that the parton \PH conjugation $\mathrm{PH}_{\mathfrak{L}_{q_0}}$ defined in \eqref{PHLq0} acts as an involution within the paired parton sequence \begin{align}&\mathrm{PH}_{\mathfrak{L}_{q_0}}\left(\mathfrak{P}_{p,q}\right)=\mathfrak{P}_{p',q'},\label{PHLq0Ppq}\\&\mbox{where }\left\{\begin{array}{*{20}l}p'=-p+2(q_0-1+(-1)^{q_0})\\q'=-q+q_0-(-1)^{q_0}\end{array}\right..\nonumber\end{align} We will demonstrate the \PH action on the edge \CFT and infer its effect on topological order through the bulk-boundary correspondence. 

The general structure of the $\mathrm{PH}_{\mathfrak{L}_{q_0}}$ conjugation can be illustrated by two special cases, $q_0=1$ or $0$. The first is based on Dirac parton triplet \begin{align}\mathfrak{L}_1=U(3)_1/\mathbb{Z}_3=U(1)_3\times SU(3)_1,\label{partonLL}\end{align} which describes the completely filled Landau levels for the three deconfined partons (see section~\ref{sec:DiracParton} for the edge \CFT content). The second is based on the Laughlin state \begin{align}\mathfrak{L}_0=\mathrm{Laughlin}=U(1)_3,\label{Laughlin}\end{align} which is the phase where partons are confined. Both were constructed as coupled wire models with filling fraction $\nu=1/3$ in section~\ref{sec:fillingonethird}. The former has central charge $c=3$ and the latter has $c=1$ on the boundary. The \PH conjugate of a paired parton state $\mathfrak{P}_{p,q}$ is the subtraction of $\mathfrak{P}_{p,q}$ from the parton Landau levels \eqref{partonLL} or from the Laughlin state \eqref{Laughlin}, \begin{align}\begin{split}&\mathrm{PH}_\pi\left(\mathfrak{P}_{p,q}\right)\equiv\mathfrak{L}_1\boxtimes\overline{\mathfrak{P}_{p,q}}=\mathfrak{P}_{-2-p,2-q},\\&\mathrm{PH}_\lambda\left(\mathfrak{P}_{p,q}\right)\equiv\mathfrak{L}_0\boxtimes\overline{\mathfrak{P}_{p,q}}=\mathfrak{P}_{-p,-1-q},\end{split}\label{PHpartonLaughlin}\end{align} where $\pi$ ($\lambda$) stands for parton (resp.~Laughlin). The gapless edge of the \FQH state consists of a forward propagating $U(3)_1/\mathbb{Z}_3$ (or $U(1)_3$) \CFT and a backward propagating $\mathfrak{P}_{p,q}$. They can be turned into $\mathfrak{P}_{-2-p,2-q}$ (resp.~$\mathfrak{P}_{-p,-1-q}$) via basis transformations and electron backscatterings. A paired parton state is \PH symmetric if it is topologically equivalent to its conjugate. 

We first illustrate the particle-hole action of the parton Pfaffian state $\mathrm{Pf}^\ast=U(1)_{24}\otimes_{\mathrm{el}}U(1)_{12}\otimes\overline{\mathrm{Ising}}$ with respect to the parton Landau levels \eqref{partonLL}. We saw in section~\ref{sec:gluingsplitting} that a pair of parton Pfaffian states can be glued into the parton Landau levels, $U(3)_1/\mathbb{Z}_3=\mathrm{Pf}^\ast\boxtimes\mathrm{Pf}^\ast$. By formally subtracting a parton Pfaffian state from both sides of the equation, we expect the parton Pfaffian state to be \PH symmetric, i.e.~$\mathrm{PH}_\pi(\mathrm{Pf}^\ast)=\mathrm{Pf}^\ast$. Here we demonstrate the equality explicity.

The edge \CFT of the particle-hole conjugate $\mathrm{PH}_\pi(\mathrm{Pf}^\ast)=[U(1)_3\times SU(3)_1]\times\overline{\mathrm{Pf}^\ast}$ is described by the Lagrangian density (suppressing non-universal velocity terms) \begin{align}\mathcal{L}&=\mathcal{L}_{U(1)_3}^R+\mathcal{L}_{SU(3)}^R+\mathcal{L}_\rho^L+\mathcal{L}_d^L+\mathcal{L}^R_{\mathrm{Ising}},\\\mathcal{L}^R_{U(1)_3}&=\frac{1}{2\pi}3\partial_{\mathsf{t}}\phi^{R,0}_\pi\partial_{\mathsf{x}}\phi^{R,0}_\pi,\nonumber\\\mathcal{L}^R_{SU(3)}&=\frac{1}{2\pi}K_{ij}^{SU(3)}\partial_{\mathsf{t}}\phi^{R,i}_\pi\partial_{\mathsf{x}}\phi^{R,j}_\pi,\nonumber\\\mathcal{L}^L_\rho&=-\frac{1}{2\pi}24\partial_{\mathsf{t}}\phi^L_\rho\partial_{\mathsf{x}}\phi^L_\rho,\nonumber\\\mathcal{L}^L_d&=-\frac{1}{2\pi}12\partial_{\mathsf{t}}\phi^L_d\partial_{\mathsf{x}}\phi^L_d,\nonumber\\\mathcal{L}^R_{\mathrm{Ising}}&=i\psi^R(\partial_{\mathsf{t}}-v\partial_{\mathsf{x}})\psi^R.\nonumber\end{align} The first three bosonized variables $\phi^0_\pi,\phi^1_\pi,\phi^2_\pi$ generate the parton Dirac triplet in the Chevalley basis (c.f.~Lagrangian density \eqref{CSLagrangian}), where $K^{SU(3)}_{ij}$ is the Cartan matrix of $SU(3)$ given by the lower $2\times2$ block of \eqref{partonKmatrix}. The remaining degrees of freedom associates to the counter-propagating parton Pfaffian \CFT (c.f.~Lagrangian densities \eqref{CSPfAbCFT} and \eqref{CSPfMF}). The external $U(1)_{\mathrm{EM}}$ shifts $\phi^{R,0}_\pi\to\phi^{R,0}_\pi+\Lambda/3$ and $\phi^L_\rho\to\phi^L_\rho+\Lambda/12$ when a charge $e$ electronic quasiparticle is transformed under $c\to e^{i\Lambda}c$, and leaves the other neutral fields unchanged.

We first perform a basis transformation among the charged Laughlin $U(1)_3$ sector $\phi^{R,0}_\pi$ and the paired $\overline{U(1)_{24}}$ sector $\phi^L_\rho$, \begin{align}\begin{pmatrix}\phi^R_\rho\\\phi^L_{d'}\end{pmatrix}=\frac{1}{2}\begin{pmatrix}1&-1\\-1&2\end{pmatrix}\begin{pmatrix}\phi^{R,0}_\pi\\2\phi^L_\rho\end{pmatrix}.\label{partonpairLR}\end{align} This turns the charged sectors into $U(1)_{24}\times\overline{U(1)_{12}}$, \begin{gather}\mathcal{L}^R_{U(1)_3}+\mathcal{L}^L_\rho=\mathcal{L}^R_\rho+\mathcal{L}^L_{d'},\\\begin{split}\mathcal{L}^R_\rho&=\frac{1}{2\pi}24\partial_{\mathsf{t}}\phi^R_\rho\partial_{\mathsf{x}}\phi^R_\rho,\\\mathcal{L}^L_{d'}&=-\frac{1}{2\pi}12\partial_{\mathsf{t}}\phi^L_{d'}\partial_{\mathsf{x}}\phi^L_{d'}.\end{split}\nonumber\end{gather} Under the external $U(1)_{\mathrm{EM}}$ transformation that changes $c\to e^{i\Lambda}c$ for a charge $e$ electronic quasiparticle, $\phi^R_\rho$ is shifted by $\phi^R_\rho\to\phi^R_\rho+\Lambda/12$ while $\phi^L_{d'}$ is unaltered and thus is electrically neutral.

Next, we combine the neutral sectors $\overline{U(1)_{12}}$ and $SU(3)_1$ -- generated by $\phi^L_{d'}$ and $\phi^{R,1}_\pi,\phi^{R,2}_\pi$ respectively -- together, and perform the basis transformation \begin{align}2\begin{pmatrix}\phi^L_n\\\phi^{R,1}_d\\\phi^{R,2}_d\end{pmatrix}=B\begin{pmatrix}2\phi^L_{d'}\\\phi^{R,1}_\pi\\\phi^{R,2}_\pi\end{pmatrix}\label{splittingB2}\end{align} using the unimodular $B$ transformation defined in \eqref{Bmatirx}. This turns the neutral sectors into $\overline{SO(2)_1}\times U(1)_{12}\times U(1)_{12}$, \begin{gather}\mathcal{L}^L_{d'}+\mathcal{L}^R_{SU(3)}=\mathcal{L}_{SO(2)}^L+\mathcal{L}_{d1}^R+\mathcal{L}^R_{d2},\label{PHneutralsplitting}\\\begin{split}\mathcal{L}_{SO(2)}^L&=-\frac{1}{2\pi}4\partial_{\mathsf{t}}\phi^L_n\partial_{\mathsf{x}}\phi^L_n+\mbox{velocity terms}\\&=i\psi^L(\partial_{\mathsf{t}}+v\partial_{\mathsf{x}})\psi^L+i\gamma^L(\partial_{\mathsf{t}}+v\partial_{\mathsf{x}})\gamma^L,\nonumber\\\mathcal{L}^R_{dj}&=\frac{1}{2\pi}12\partial_{\mathsf{t}}\phi^{R,j}_d\partial_{\mathsf{x}}\phi^{R,j}_d,\end{split}\nonumber\end{gather} for $j=1,2$, where we have decomposed the spin-$1/2$ neutral Dirac fermion $e^{i2\phi^L_n}$ in $\overline{SO(2)_1}$ into Majorana components $(\psi^L+i\gamma^L)/\sqrt{2}$.

The basis transformations \eqref{partonpairLR} and \eqref{splittingB2} can be applied not only to the parton Pfaffian state $\mathrm{Pf}^\ast$ but also to any paired parton state $\mathfrak{P}_{p,q}$. For the $\mathrm{PH}_\pi$ conjugate of $\mathrm{Pf}^\ast$, the total Lagrangian density is now $\mathcal{L}=\mathcal{L}^R_\rho+\left(\mathcal{L}^R_{d1}+\mathcal{L}^R_{d2}+\mathcal{L}^L_d\right)+\mathcal{L}^R_{\mathrm{Ising}}+\mathcal{L}_{SO(2)}$. Here each of the $d$ sectors carries a spin $h=\pm3/2$ Dirac fermion $e^{i6\phi_d}$, and $\mathrm{Ising}$ and $\overline{SO(2)_1}$ carry Majorana fermions. We consider the gapping potentials \begin{align}\mathcal{H}_d&=-u_d\cos\left(6\phi^{R,2}_d-6\phi^L_d\right)\label{PHH1}\\&=-u_d\cos\left(6\phi^L_d+6\phi^L_{d'}+3\phi^{R,2}_\pi\right)\nonumber\\&\sim\frac{u_d}{2}d_Ld'_L(\pi^1_R)^\dagger(\pi^2_R)^\dagger(\pi^3_R)^2+h.c.,\nonumber\\\mathcal{H}_n&=iu_n\gamma^L\psi^R\label{PHH2}\end{align} to cancel unprotected counter-propagating modes. 

$\mathcal{H}_d$ backscatters the pair of Dirac fermions $d_L\sim e^{i6\phi^L_d}$ and $d'_L\sim e^{i6\phi^L_{d'}}$ to the two $SU(3)$ current operators $E^{13}_R\sim\pi^1_R(\pi^3_R)^\dagger$ and $E^{23}_R\sim\pi^2_R(\pi^3_R)^\dagger$, where $\pi^a_R$ are the deconfined partons in the filled parton Landau levels. (The identification $e^{-i3\phi^{R,2}_\pi}=E^{13}E^{23}$ can be shown from the basis transformation \eqref{CWCtrans}.) The $SU(3)$ current operators can be expressed as integral combinations of electron operators. On the other hand, the neutral Dirac fermions $d_L$ and $d'_L$ are non-local as they both involve half-electronic operators. However, when combined together, $d_Ld'_L\sim e^{i(12\phi^L_\rho+6\phi^L_d)}e^{-i3\phi^{R,0}_\pi}$ is an integral electronic combination. This is because $e^{i(12\phi^L_\rho+6\phi^L_d)}$ is just the charge $e$ electronic quasiparticle $\openone_{12,6}$ in \eqref{elecQPsplitting}, and is also the electronic combination $e^{i\Phi_\rho}d$ that is backscattered by $\mathcal{H}_{d\rho}$ in \eqref{PfdrhoH}. Moreover, $e^{i3\phi^{R,0}_\pi}\sim\pi^1\pi^2\pi^3=\lambda^3$ is simply the charge $e$ electronic quasiparticle in the Laughlin sector. The Dirac fermion backscattering \eqref{PHH1} can be rewritten as \begin{align}\mathcal{H}_d&=-u_d\cos\left\{3\left[(4\phi^L_\rho+2\phi^L_d)+(-\phi^{R,0}_\pi+\phi^{R,2}_\pi)\right]\right\}\nonumber\\&=-u_d\cos\left\{3\left[(4\phi^L_\rho+2\phi^L_d)-\tilde\phi_3^R\right]\right\},\label{PHH1cond}\end{align} where from \eqref{CWCtrans}, $\tilde\phi^R_3=\phi^{R,0}_\pi-\phi^{R,2}_\pi$ is the bosonized variable of the parton $\pi^3=e^{i\tilde\phi_3^R}$ in the parton Landau levels $\mathfrak{L}_1$. Eq.\eqref{PHH1cond} therefore condenses the electrically neutral bosonic parton pair \begin{align}\openone_{4,2}\otimes(\pi^3)^\dagger\sim e^{i(4\phi^L_\rho+2\phi^L_d)}e^{-i\tilde\phi^R_3}\label{PHH1condensate}\end{align} where $\openone_{4,2}$ is the charge $e/3$ fermionic parton in $\overline{\mathrm{Pf}^\ast}$. 

$\mathcal{H}_n$ in \eqref{PHH2} backscatters between the Majorana fermions $\gamma^L$ and $\psi^R$. Neither of the Majorana's is local electronic, but the combination $\gamma^L\psi^R\sim\psi^R\sin\left(12\phi^L_\rho-3\phi^{R,0}_\pi+\phi^{R,1}_\pi+\phi^{R,2}_\pi\right)$ is. This is because $\psi^Re^{i12\phi^L_\rho}$ is the charge $e$ electronic quasiparticle $\Psi_{12,0}$ in $\overline{\mathrm{Pf}^\ast}$ (see 
\eqref{elecQPsplitting}), and is also the electronic combination backscattered by $\mathcal{H}_{\rho\gamma}$ in \eqref{PfrhogammaH}. The remaining $e^{i(-3\phi^{R,0}_\pi+\phi^{R,1}_\pi+\phi^{R,2}_\pi)}$ can be decomposed into the electronic quasiparticles $e^{-i3\phi^{R,0}_\pi}\sim\lambda^3$ in the Laughlin sector and the current operator $E^{13}\sim\pi^1(\pi^3)^\dagger$ in the neutral $SU(3)$ sector. Hence, \eqref{PHH2} can be rewritten in the form of the backscattering \begin{align}\mathcal{H}_n&\sim u_n\Psi_{12,0}(\lambda^3)^\dagger E^{13}\end{align} between the electronic quasiparticles $\Psi_{12,0}$ and $\lambda^3(E^{13})^\dagger$ in the parton Pfaffian $\overline{\mathrm{Pf}^\ast}$ and the parton Landau level $\mathfrak{L}_1$. 

Eq.\eqref{PHH1} and \eqref{PHH2} leaves behind the low-energy degrees of freedom $\mathcal{L}^R_\rho+\mathcal{L}^R_{d1}+\mathcal{L}^L_\psi$, where $\mathcal{L}^L_\psi$ is the Lagrangian density for the remaining Majorana fermion $\psi^L$ in $\mathcal{L}^L_{SO(2)}$ in \eqref{PHneutralsplitting}. This matches exactly with the Lagrangian densities \eqref{CSPfAbCFT} and \eqref{CSPfMF} for the parton Pfaffian state. This completes the proof of the particle-hole symmetry \begin{align}\mathrm{PH}_\pi\left(\mathrm{Pf}^\ast\right)\equiv\mathfrak{L}_1\boxtimes\overline{\mathrm{Pf}^\ast}=\mathrm{Pf}^\ast.\end{align} In the process, the gapping potential \eqref{PHH1} and \eqref{PHH2} remove unprotected low-energy degrees of freedom, namely $d_Ld'_L$, $e^{i3\phi^{R,2}_\pi}$, $\gamma^L$ and $\psi^R$, along the boundary. 
The reduced tensor product notation $\boxtimes$ reminds the cancellation of unprotected boundary modes. Equivalently, it signifies the condensation of \eqref{PHH1condensate}, which identifies the partons in $\mathfrak{L}_1$ and $\overline{\mathrm{Pf}^\ast}$.

Similar procedure can be carried out for a general paired parton state $\mathfrak{P}_{p,q}$ in \eqref{Pfaffiansequence} and the particle-hole conjugation with respect to the parton Landau levels \eqref{partonLL} is \begin{align}\mathrm{PH}_\pi\left(\mathfrak{P}_{p,q}\right)\equiv U(3)_1/\mathbb{Z}_3\otimes\overline{\mathfrak{P}_{p,q}}=\mathfrak{P}_{-p-2,2-q}.\label{PHparton}\end{align} The edge \CFT reconstruction and bulk condensation can be summarized by \begin{align}\begin{diagram}U(1)_3\otimes SU(3)_1\otimes\overline{U(1)_{24}\otimes_{\mathrm{el}}U(1)_{12}^q\otimes_{\mathrm{el}}\mathrm{Ising}^p}\\\dTo^{\eqref{partonpairLR}}\\U(1)_{24}\otimes_{\mathrm{el}}\overline{U(1)_{12}}\otimes SU(3)_1\otimes_{\mathrm{el}}\overline{U(1)_{12}^q}\otimes_{\mathrm{el}}\mathrm{Ising}^{-p}\\\dTo^{\eqref{splittingB2}}\\U(1)_{24}\otimes_{\mathrm{el}}\overline{SO(2)_1}\otimes_{\mathrm{el}}U(1)_{12}^2\otimes_{\mathrm{el}}\overline{U(1)_{12}^q}\otimes_{\mathrm{el}}\mathrm{Ising}^{-p}\\\dTo^{\mathrm{condensation}}\\U(1)_{24}\otimes_{\mathrm{el}}U(1)_{12}^{2-q}\otimes_{\mathrm{el}}\mathrm{Ising}^{-p-2}.\end{diagram}\label{PHflowchart1}\end{align} The \PH conjugation is consistent with the chiral central charge \eqref{Ppqcentralcharge} of $\mathfrak{P}_{p,q}$ \begin{align}\mathrm{PH}_\pi(c_{p,q})\equiv3-c_{p,q}=c_{-p-2,2-q}\end{align} where the parton Landau levels has the net central charge $c=3$.

Particle-hole conjugation can also be defined with respect to the Laughlin $\nu=1/3$ FQH state, where the partons are confined and the neutral $SU(3)_1$ sector does not appear. It relates \begin{align}\mathrm{PH}_\lambda\left(\mathfrak{P}_{p,q}\right)\equiv U(1)_3\otimes\overline{\mathfrak{P}_{p,q}}=\mathfrak{P}_{-p,-1-q}.\label{PHLaughlin}\end{align} The procedure can be carried out similar to the previous case and is summarized below. \begin{align}\begin{diagram}U(1)_3\otimes\overline{U(1)_{24}\otimes_{\mathrm{el}}U(1)_{12}^q\otimes_{\mathrm{el}}\mathrm{Ising}^p}\\\dTo^{\eqref{partonpairLR}}\\U(1)_{24}\otimes_{\mathrm{el}}\overline{U(1)_{12}}\otimes_{\mathrm{el}}\overline{U(1)_{12}^q}\otimes_{\mathrm{el}}\mathrm{Ising}^{-p}\\\dTo^{\mathrm{condensation}}\\U(1)_{24}\otimes_{\mathrm{el}}U(1)_{12}^{-1-q}\otimes_{\mathrm{el}}\mathrm{Ising}^{-p}.\end{diagram}\label{PHflowchart2}\end{align} Again, the conjugation is consistent with the chiral central charge \begin{align}\mathrm{PH}_\lambda(c_{p,q})\equiv1-c_{p,q}=c_{-p,-1-q}\end{align} where the Laughlin \FQH state has central charge $c=1$.

From \eqref{PHparton} and \eqref{PHLaughlin}, we see that the parton Pfaffian state $\mathrm{Pf}^\ast=\mathfrak{P}_{-1,1}$ is the only \PH symmetric state with respect to the parton Landau levels, and there is no paired parton state that is \PH symmetric with respect to the Laughlin state given integer $p,q$. With the hindsight from the T-Pfaffian surface state~\cite{ChenFidkowskiVishwanath14} of a topological insulator and its relationship with the \PH symmetric Pfaffian \FQH state~\cite{Son15}, it is perhaps not a surprise that the parton Pfaffian state $\mathrm{Pf}^\ast=\mathfrak{P}_{-1,1}$ is \PH symmetric. In previous works~\cite{SahooSirotaChoTeo17,ChoTeoFradkin17} (reviewed in \eqref{TPfcontent} in section~\ref{sec:PartonPfaffian} as well as in section~\ref{sec:partonTPfaffian}), we proposed the symmetry preserving gapped surface topological order, $\mathcal{T}-\mathrm{Pf}^\ast$, of a 3D fractional topological insulator, which host deconfined Dirac parton excitations in the bulk. The $\mathcal{T}-\mathrm{Pf}^\ast$ anyon content has three times the periodicity and one-third the charge assignment as the conventional T-Pfaffian state. It is a subset of the parton Pfaffian $\mathrm{Pf}^\ast$ topological order, which can be supported by a thin slab of fractional topological insulator with a $\mathcal{T}-\mathrm{Pf}^\ast$ top surface and a time-reversal breaking gapped bottom surface. It is therefore not a coincidence that when the parton Pfaffian $\mathrm{Pf}^\ast$ topological order is supported by a \FQH state in 2D, it exhibits a \PH symmetry in the context of partons. This example may be one of many dualities between 2D \FQH states with a generalized notion of \PH symmetry and 3D symmetry enriched topological phases.

We conclude this section by generalizing the particle-hole conjugations $\mathrm{PH}_\pi=\mathrm{PH}_{\mathfrak{L}_1}$ and $\mathrm{PH}_\lambda=\mathrm{PH}_{\mathfrak{L}_0}$ in \eqref{PHpartonLaughlin} to the arbitrary base \begin{align}\mathfrak{L}_{q_0}=U(1)_3\otimes[SU(3)_1]^{q_0}.\end{align} We now show that the \PH symmetry $\mathrm{PH}_{\mathfrak{L}_{q_0}}$ acts as the involution \eqref{PHLq0Ppq} within the paired parton \FQH sequence $\mathfrak{P}_{p,q}$. Like in \eqref{PHflowchart1} and \eqref{PHflowchart2}, the topological action that turns $\mathrm{PH}_{\mathfrak{L}_{q_0}}(\mathfrak{P}_{p,q})\equiv\mathfrak{L}_{q_0}\times\overline{\mathfrak{P}_{p,q}}$ into $\mathfrak{P}_{p',q'}$ involves a series of basis transformations and condensations (or edge backscatterings). We begin with $q_0\geq0$. This process can be summarized by \begin{widetext}\begin{align}\begin{diagram}U(1)_3\otimes SU(3)_1^{q_0}\otimes\overline{U(1)_{24}\otimes_{\mathrm{el}}U(1)_{12}^q\otimes_{\mathrm{el}}\mathrm{Ising}^p}\\\dTo^{\eqref{partonpairLR}}\\U(1)_{24}\otimes_{\mathrm{el}}\left(\overline{U(1)_{12}}\otimes SU(3)_1^{q_0}\right)\otimes_{\mathrm{el}}\overline{U(1)_{12}^q}\otimes_{\mathrm{el}}\mathrm{Ising}^{-p}\\\dTo^{\eqref{PHprodB}}_{\prod_{j=1}^{q_0}B_j}\\U(1)_{24}\otimes_{\mathrm{el}}\left(SO(2)_1^{q_0-1+(-1)^{q_0}}\otimes_{\mathrm{el}}U(1)_{12}^{q_0-(-1)^{q_0}}\right)\otimes_{\mathrm{el}}\overline{U(1)_{12}^q}\otimes_{\mathrm{el}}\mathrm{Ising}^{-p}\\\dTo^{\mathrm{condensation}}\\U(1)_{24}\otimes_{\mathrm{el}}U(1)_{12}^{-q+q_0-(-1)^{q_0}}\otimes_{\mathrm{el}}\mathrm{Ising}^{-p+2(q_0-1+(-1)^{q_0})},\end{diagram}\label{PHflowchart3}\end{align} where the sequence of basis transformation \begin{align}\prod_{j=1}^{q_0}B_j:\overline{U(1)_{12}}\otimes[SU(3)_1]^{q_0}\to[SO(2)_1]^{q_0-1+(-1)^{q_0}}\otimes_{\mathrm{el}}[U(1)_{12}]^{q_0-(-1)^{q_0}}\label{PHtranssequence}\end{align}\end{widetext} is defined by the following. The first transformation $B_1$ is identical to \eqref{splittingB2}, which used the unimodular $B$ matrix defined in \eqref{Bmatirx}. It turns $\overline{U(1)_{12}}$ and one of the $SU(3)_1$'s into \begin{align}B:\overline{U(1)_{12}}\times SU(3)_1\to\overline{SO(2)_1}\times[U(1)_{12}]^2.\end{align} Next, the transformation $B_2$ takes the $\overline{SO(2)_1}$ sector -- which is generated by $\phi^L_n$ in the Lagrangian density in \eqref{PHneutralsplitting} -- and another $SU(3)_1$ into \begin{gather}B':\overline{SO(2)_1}\times SU(3)_1\to[SO(2)_1]^2\times\overline{U(1)_{12}}\\2\begin{pmatrix}\phi^{R,1}_n\\\phi^{R,2}_n\\\phi^L_d\end{pmatrix}=B'\begin{pmatrix}2\phi^L_n\\\phi^{R,1}_\pi\\\phi^{R,2}_\pi\end{pmatrix}\nonumber\\B'=\begin{pmatrix}1&1&3\\-1&-1&-2\\0&-1&-1\end{pmatrix}^{-1}=\begin{pmatrix}-1&-2&1\\-1&-1&-1\\1&1&0\end{pmatrix}\nonumber\end{gather} where $\phi^{R,1}_n,\phi^{R,2}_n,\phi^L_d$ generate $[SO(2)_1]^2\times\overline{U(1)_{12}}$ with the Lagrangian densities \begin{align}\mathcal{L}_{SO(2)_1}^{R,1}+\mathcal{L}_{SO(2)_1}^{R,2}&=\frac{1}{2\pi}4\left(\partial_{\mathsf{t}}\phi^{R,1}_n\partial_{\mathsf{x}}\phi^{R,1}_n+\partial_{\mathsf{t}}\phi^{R,2}_n\partial_{\mathsf{x}}\phi^{R,2}_n\right)\nonumber\\\mathcal{L}_d^L&=-\frac{1}{2\pi}12\partial_{\mathsf{t}}\phi^L_d\partial_{\mathsf{x}}\phi^L_d.\end{align} The transformation equates \begin{gather}\begin{pmatrix}-1&&\\&2&-1\\&-1&2\end{pmatrix}=(B')^T\begin{pmatrix}1&&\\&1&\\&&-3\end{pmatrix}B',\nonumber\\\mathcal{L}_{SO(2)_1}^{R,1}+\mathcal{L}_{SO(2)_1}^{R,2}+\mathcal{L}_d^L=\mathcal{L}_{SO(2)_1}^L+\mathcal{L}_{SU(3)_1}^R.\end{gather} The sequence of basis transformation \eqref{PHtranssequence} composes of series \begin{align}&\overline{U(1)_{12}}\times[SU(3)_1]^{q_0}\nonumber\\&\xrightarrow{B}[U(1)_{12}]^2\times\overline{SO(2)_1}\times[SU(3)_1]^{q_0-1}\nonumber\\&\xrightarrow{B'}[U(1)_{12}]^2\times[SO(2)_1]^2\times\overline{U(1)_{12}}\times[SU(3)_1]^{q_0-2}\nonumber\\&\xrightarrow{B}\ldots\to\label{PHprodB}\\&\left\{\begin{array}{*{20}l}[U(1)_{12}]^{q_0}\times[SO(2)_1]^{q_0}\times\overline{U(1)_{12}}&\mbox{if $q_0$ is even}\\{}[U(1)_{12}]^{q_0+1}\times[SO(2)_1]^{q_0-1}\times\overline{SO(2)_1}&\mbox{if $q_0$ is odd}\end{array}\right.\nonumber\\&\xrightarrow{\mathrm{backscattering}}[U(1)_{12}]^{q_0-(-1)^{q_0}}\times[SO(2)_1]^{q_0-1+(-1)^{q_0}}\nonumber\end{align} where the backscattering Hamiltonian \begin{align}\mathcal{H}_{\mathrm{b.c.}}=\left\{\begin{array}{*{20}l}-u_d\cos(6\phi^{R,q_0}_d-6\phi^L_d)&\mbox{if $q_0$ is even}\\-u_n\cos(2\phi^{R,q_0-1}_n-2\phi^L_n)&\mbox{if $q_0$ is even}\end{array}\right.\label{PHH3}\end{align} introduces a mass gap for the counter-propagating pair $U(1)_{12}\times\overline{U(1)_{12}}$ or $SO(2)_1\times\overline{SO(2)_1}$ and removes them from low-energy. Similar to \eqref{PHH1} and \eqref{PHH2}, \eqref{PHH3} can be expressed as integral combination of electronic operators. 

The last arrow in \eqref{PHflowchart3} involves backscattering interactions similar to \eqref{PHH1} and \eqref{PHH2} that introduce a mass gap for the counter-propagating conjugate pairs $U(1)_{12}\times\overline{U(1)_{12}}$'s and $\mathrm{Ising}\times\overline{\mathrm{Ising}}$'s along the system edge, where the forward moving $\mathrm{Ising}$ sectors are provided by splitting the spin $h=1/2$ neutral Dirac fermion $e^{i2\phi_n^R}$ in each $SO(2)_1$ into a pair of Majorana fermions. Analogous to \eqref{PHH1condensate}, the edge backscattering is equivalent to condensing electrically neutral parton pairs between $\mathfrak{L}_{q_0}$ and $\overline{\mathfrak{P}_{p,q}}$ in the bulk. This completes the proof for the particle-hole symmetry action \eqref{PHLq0Ppq} for $q_0\geq0$.

When $q_0$ is negative, the center arrow in \eqref{PHflowchart3} needs to be modified. This is because the sequence of basis transformation \eqref{PHprodB} relies on the presence of counter-propagating sectors, and $\overline{U(1)_{12}}\times[SU(3)_1]^{q_0}$ consists entirely of backward propagating sectors when $q_0<0$. In this case, one can first introduce the additional counter-propagating pair $U(1)_{12}\times\overline{U(1)_{12}}$ to the system edge. This can be achieved by turning off the Dirac fermion backscattering $\mathcal{H}^{\mathrm{dimer}}_{y,DF}$ in \eqref{PfincellHDF} along the wire at the system boundary. Next, one can apply the sequence \eqref{PHprodB} of $B$ and $B'$ transformations to turn \begin{align}\begin{split}&U(1)_{12}\times\overline{SU(3)_1}^{|q_0|}\\&\to\overline{U(1)_{12}}^{|q_0|-(-1)^{q_0}}\times\overline{SO(2)_1}^{|q_0|-1+(-1)^{q_0}}.\end{split}\end{align} After condensing/backscattering counter-propagating pairs, the \PH conjugate $\mathrm{PH}_{\mathfrak{L}_{q_0}}(\mathfrak{P}_{p,q})$ becomes $\mathfrak{P}_{p'',q''}$ where \begin{align}\begin{array}{*{20}l}p''=-p+2(q_0+1-(-1)^{q_0})\\q''=-q+q_0+(-1)^{q_0}-2\end{array},\label{p''q''}\end{align} which is identical to \eqref{PHLq0Ppq} when $q_0$ is even. When $q_0$ is odd, \eqref{p''q''} and \eqref{PHLq0Ppq} differs by $U(1)_{12}^4\otimes_{\mathrm{el}}\overline{SO(2)_1}^4$, which can be described by a bosonized Lagrangian density with the $K$-matrix $K=12\openone_4\oplus(-4\openone_4)$. A mass gap can be introduced by the sine-Gordon potential \begin{align}\mathcal{H}=-u\sum_{j=1}^4\cos\left({\bf n}_j^TK\boldsymbol\phi\right)\end{align} where the null-vectors can be chosen to be \begin{align}\begin{pmatrix}--&{\bf n}_1^T&--\\--&{\bf n}_2^T&--\\--&{\bf n}_3^T&--\\--&{\bf n}_4^T&--\end{pmatrix}=\begin{pmatrix}1&0&0&0&0&1&-1&1\\0&1&0&0&1&0&-1&-1\\0&0&1&0&1&-1&0&1\\0&0&0&1&1&1&1&0\\\end{pmatrix}.\end{align} Equivalently, $U(1)_{12}^4$ can be  reduced to $SO(2)_1^4$ by condensing the following collection of mutually local bosons \begin{align}\begin{split}&e^{i4(\phi_d^2+\phi_d^3+\phi_d^4)},e^{i4(\phi_d^1-\phi_d^3+\phi_d^4)},\\&e^{i4(-\phi_d^1-\phi_d^2+\phi_d^4)},e^{i4(\phi_d^1-\phi_d^2+\phi_d^3)}.\end{split}\end{align}

Lastly, we noticed that the general \PH conjugation \eqref{PHLq0Ppq} admits a symmetric paired parton\FQH state when $q_0$ is odd \begin{align}\mathrm{PH}_{\mathfrak{L}_{q_0}}(\mathfrak{P}_{p,q})=\mathfrak{P}_{p,q},\quad\mbox{for }\left\{\begin{array}{*{20}l}p=q_0-1+(-1)^{q_0}\\q=[q_0-(-1)^{q_0}]/2\end{array}\right..\end{align} It would be interesting to associate each of these \PH symmetric \FQH states to the symmetry-preserving surface state of a fractional topological insulator.

\section{Conclusion and discussion}\label{sec:conclusion}
We theoretically proposed two new sequences of electronic fractional quantum Hall (\hyperlink{FQH}{FQH}) states, one at filling fractions $\nu=1/3$ and another at $\nu=1/6$. They were summarized in table~\ref{tab:summary}. The first sequence consists of the Abelian states that are characterized by the bulk topological order and edge conformal field theories (\hyperlink{CFT}{CFT}s) $\mathfrak{L}_q=U(1)_3\otimes[SU(3)_1]^q$. The charged $U(1)_3$ sector is responsible for the electric Hall transport and is identical to the Laughlin~\cite{Laughlin83} $\nu=1/3$ \FQH state. The electrically neutral $SU(3)_1$ sectors allow the deconfinement of partons, which are fermionic quasiparticle excitations that carry the fractional electric charge $e/3$. The presence of these chiral neutral sectors causes an imbalance between electric and thermal Hall transport, and leads to the violation of the Wiedemann-Franz law, $c/\nu=\left(\kappa_{xy}/\sigma_{xy}\right)\left[\pi^{2}k_{B}^{2}/\left(3e^{2}\right)\right]T\neq1$. The parton Abelian states $\mathfrak{L}_q$ can be distinguished by their chiral central charge $c=1+2q$, which specifies the thermal Hall conductance, $\kappa_{xy}=c\pi^2k_B^2T/(3h)$.

The second sequence generically consists of non-Abelian \FQH states at filling $\nu=1/6$ that support charge $e/12$ Ising anyons. They are characterized by the bulk topological order and edge \CFT $\mathfrak{P}_{p,q}=U(1)_{24}\otimes_{\mathrm{el}}U(1)_{12}^q\otimes_{\mathrm{el}}\mathrm{Ising}^p$. The $U(1)_{24}$ sector couples to the external electromagnetic $U(1)_{\mathrm{EM}}$ symmetry and is responsible for the electric Hall transport $\sigma_{xy}=(1/6)e^2/h$. The $U(1)_{12}$ sectors are electrically neutral and each carries a spin $3/2$ Dirac fermion, which is an emergent fractional excitation. The Ising sectors are generated by spin $1/2$ Majorana fermions. The neutral Dirac and Majorana fermions can be combined with the charge $e$ boson in $U(1)_{24}$ to represent the local electronic quasiparticles. Deconfined Ising anyons are products of $\pi$-fluxes (also referred to as twist fields) of all sectors. The sequence generalizes one of the $Q$-Pfaffian states, which is identical to $\mathfrak{P}_{1,0}$, proposed by Read and Moore~\cite{MooreRead}. Like the Abelian sequence at $\nu=1/3$, the generalization here also supports deconfined parton quasiparticles (see eq.\eqref{Pfaffianpartons}). These paired parton \FQH states have distinct thermal Hall transports, which are specified by the chiral central charge $c=1+q+p/2$.

Contrary to the more common phenomenological slave-fermion mean-field approach, we presented an exact description of partons. In section~\ref{sec:partonCFTs}, we introduced the \hyperlink{CFT}{CFT}s that described the gapless low-energy parton degrees of freedom, which appeared on the $1+1$D edges of the \FQH states and serve as the building blocks of the coupled wire \FQH models. Through the bulk-boundary correspondence, these \hyperlink{CFT}{CFT}s dictate the anyon excitation structures and the topological orders of the $2+1$D \FQH states that host them as boundary modes. The anyon types in the 2D bulk has a one-to-one correspondence to the primary fields of the 1D edge \hyperlink{CFT}{CFT}. The fusion rules between bulk anyons are identical to those that describe the operator product expansions between edge primary fields. Anyons' spins equate to primary fields' conformal scaling dimensions modulo 1. The topological $S$-matrix~\cite{Kitaev06}, which encodes the quantum dimensions as well as the mutual monodromy of bulk anyons, is identical to the modular $S$-matrix that represents the modular $S$ transformation~\cite{bigyellowbook} of the characters in the \CFT partition function. The fusion and spin data of the Dirac parton \CFT $\mathfrak{L}_1=U(3)_1/\mathbb{Z}_3=U(1)_3\otimes SU(3)_1$ and the parton Pfaffian \CFT $\mathfrak{P}_{1,-1}=\mathrm{Pf}^\ast=U(1)_{24}\otimes_{\mathrm{el}}U(1)_{12}\otimes_{\mathrm{el}}\overline{\mathrm{Ising}}$ were presented in section~\ref{sec:DiracParton} and \ref{sec:PartonPfaffian}. In section~\ref{sec:gluingsplitting}, we demonstrated the gluing and splitting of these parton degrees of freedom. They were summarized by the reduced tensor product $\mathfrak{L}_1=\mathrm{Pf}^\ast\boxtimes\mathrm{Pf}^\ast$. This was the parton generalization of the splitting and gluing of the electronic Dirac fermion and the \PH symmetric Pfaffian pair, $U(1)_1=\mathrm{Pf}\boxtimes\mathrm{Pf}$, where $\mathrm{Pf}=U(1)_8\otimes\overline{\mathrm{Ising}}$. The gluing of the pair of parton Pfaffian theories was carried out by a sine-Gordon Hamiltonian \eqref{gluingH} that facilitated the condensation of the bosonic collection \eqref{condensedbosonsPF2} of anyon pairs. The bipartitioning of the Dirac parton triplet $\mathfrak{L}_1$ was enabled by a sequence of fractional basis transformations, which were summarized in figure~\ref{fig:splitting}. 

Understanding the parton \hyperlink{CFT}{CFT}s allowed the coupled wire model construction~\cite{KaneMukhopadhyayLubensky02} of the $2+1$D parton \FQH states $\mathfrak{L}_q$ and $\mathfrak{P}_{p,q}$, which were presented in section~\ref{sec:fillingonethird} and \ref{sec:fillingonesixth}. These models were constructed from a 2D interacting array of metallic electron wires. The ballistic wires were arranged in a particular periodic spatial configuration (see figure~\ref{fig:Partonwirebasis} and \ref{fig:Pfwiresbasis}) so that in the presence of a perpendicular magnetic field and at the filling fraction $\nu=1/3$ and $\nu=1/6$, certain combinations of many-body inter-wire backscattering interactions became favorable as they preserved momentum conservation. In strong coupling, these interactions opened a finite excitation energy gap that froze all low-energy electronic degrees of freedom in the 2D bulk. At the same time, they left behind gapless parton degrees of freedom along system edges that were described by the aforementioned \hyperlink{CFT}{CFT}s. A summary can be found in figure~\ref{fig:PartonCoupledWire}, \ref{fig:PfCoupledWire} and \ref{fig:PfCoupledWire2}. The backscattering interactions were designed so that the model Hamiltonians were all exactly solvable. They consisted of mutually commuting interaction terms, which independently froze mutually decoupled order parameters. Following a similar coupled wire construction~\cite{MrossEssinAlicea15} that described the T-Pfaffian surface state of a conventional topological insulator, we presented a model in section~\ref{sec:partonTPfaffian} that illustrated the symmetry-preserving many-body gapping of the surface of a fractional topological insulator. We conjectured that the surface model should carry a parton T-Pfaffian ($\mathcal{T}-\mathrm{Pf}^\ast$) topological order, which was discussed in our previous works~\cite{SahooSirotaChoTeo17,ChoTeoFradkin17} and was reviewed in \eqref{TPfcontent} in section~\ref{sec:PartonPfaffian} as well as in section~\ref{sec:partonTPfaffian}. In particular, the parton surface topological order is a subset of the parton Pfaffian topological order $\mathfrak{P}_{1,-1}=\mathrm{Pf}^\ast$, which exhibits an emergent parton particle-hole symmetry.

In section~\ref{sec:particlehole}, we presented the emergent parton particle-hole (\hyperlink{PH}{PH}) conjugations among the paired parton \FQH state $\mathfrak{P}_{p,q}$ at filling $\nu=1/6$. The \PH conjugations were defined by ``subtracting" any one of the paired parton \FQH state $\mathfrak{P}_{p,q}$ from an arbitrarily given Abelian parton \FQH state $\mathfrak{L}_{q_0}$ at filling $\nu=1/3$. The ``subtraction" was topologically defined by taking a reduced tensor product between the Abelian state with the time-reversal conjugate of the paired parton state, $\mathrm{PH}_{\mathfrak{L}_{q_0}}(\mathfrak{P}_{p,q})\equiv\mathfrak{L}_{q_0}\boxtimes\overline{\mathfrak{P}_{p,q}}$. The conjugation action produced another paired parton state and was summarized in \eqref{PHLq0Ppq}. In particular, since $\mathfrak{L}_1$ represented the filled parton Landau levels, the \PH conjugation based on $\mathfrak{L}_1$ naturally generalized the notion of particle-hole symmetry from the context of electrons to partons. We also showed that the parton Pfaffian state $\mathfrak{P}_{1,-1}=\mathrm{Pf}^\ast=U(1)_{24}\otimes_{\mathrm{el}}U(1)_{12}\otimes_{\mathrm{el}}\overline{\mathrm{Ising}}$ is \PH symmetric in the basis of the parton Landau levels $\mathfrak{L}_1$.

We conclude this paper by identifying some unaddressed issues, speculations and implications. First, the many-body interacting coupled wire models were topologically oriented without paying attention to energetics. The inter-wire backscatterings were introduced to demonstrate the structure of the ground state and how the low-energy electronic degrees of freedom can be frozen out in the bulk while leaving behind gapless edge modes. The many-body interacting terms, although allowed by charge and momentum conservation, are generically irrelevant in the renormalization group sense. They can become energetically favorable in the presence of forward electron scatterings that modify the velocities of the bosonized variables, i.e.~the Luttinger liquid parameters. In a more realistic setting, these intricate inter-wire backscattering interactions may emerge as higher-order correction terms to a more conventional interacting action, such as an array of two-body interacting chains or ladders. However, such approaches generically generate competing backscattering interactions that render the model unsolvable. On the other hand, we anticipate our topological construction to inspire lattice or continuum models that can be numerically analyzed and address naturally occurring interactions in materials.

Second, the topological order of the presented coupled wire models relies heavily on the bulk-boundary correspondence. It can also be addressed in a closed toric geometry with no edges by the algebra of Wilson loops, which can be generated by strings of electron intra- and inter-wire tunneling in the vertical direction and sliding operators in the horizontal direction. The degenerate ground states form an irreducible representation of the Wilson algebra. Anyon excitations manifest as kinks in the order parameters pinns by the backscattering interactions can be created by open Wilson strings. Their spin and braiding statistics can be determined by the intersection phases from interchanging string operators. These derivations are omitted in the scope of this paper and we defer the continuation of this discussion to future works.

Third, the presentation of the particle-hole conjugation and symmetry lacks a microscopic description, which involves the short-range non-local anti-unitary charge conjugation action on the wire bosonized variables $\mathcal{C}\phi_{y,\alpha}\mathcal{C}^{-1}=C_\alpha^\beta(y-y')\phi_{y',\beta}+\kappa_{y,\alpha}$, where $\mathcal{C}$ is the \PH operator and $C_\alpha^\beta(y-y')$, $\kappa_{y,\alpha}$ are c-numbers. Related description of electronic \PH symmetry and particle-vortex duality in the coupled wire setting has been discussed by Mross, Alicea and Motrunich~\cite{MrossAliceaMotrunich16,MrossAliceaMotrunich16PRL,MrossAliceaMotrunich17} and by Fuji and Furusaki~\cite{FujiFurusaki18}. We anticipate a similar microscopic description can be applied in the context of partons. In addition, we expect a general correspondence to hold between \PH symmetric paired parton \FQH states in two dimensions and time-reversal symmetric fractional topological insulators (\hyperlink{FTI}{FTI}s) in three dimensions. For instance, the previously proposed~\cite{SahooSirotaChoTeo17,ChoTeoFradkin17} correspondence between the parton Pfaffian state $\mathfrak{P}_{1,-1}=\mathrm{Pf}^\ast$ and a particular \hyperlink{FTI}{FTI}~\cite{maciejko2012models}, which hosts bulk partons coupled with a $\mathbb{Z}_3$ gauge theory, is one existing example.

The proposed parton \FQH states could in principle be verified in materials. The chiral central charges $c=1+2q$ for the Abelian $\mathfrak{L}_q$ states and $c=1+q+p/2$ for the paired parton $\mathfrak{P}_{p,q}$ states corresponds to distinct thermal Hall signatures~\cite{KaneFisher97,Cappelli01,Kitaev06,MrossOregSternMargalitHeiblum18} (see \eqref{sigmakappaxy}). Recently, thermal Hall conductance has been measured~\cite{BanerjeeHeiblumRosenblattOregFeldmanSternUmansky17,BanerjeeHeiblumUmanskyFeldmanOregStern18} in the Laughlin particle state at filling $\nu=1/3$ and hole state at $\nu=2/3$ as well as the Pfaffian state at $\nu=5/2$, suggesting \PH symmetry~\cite{Son15} at the $5/2$ plateau. Similar thermal Hall observations at filling $\nu=N\pm1/3$ and $\nu=N\pm1/6$, if such plateau exists, may provide indications to one of these parton \FQH states.

\acknowledgments
GYC thank the support from BK 21 plus project at POSTECH in Korea. JCYT is supported by the National Science Foundation under Grant No.~DMR-1653535.

\appendix

\section{Kac-Moody algebra}\label{app:KacMoody}
In this appendix, we review the algebraic properties of bosonized variables. We demonstrate the basic arithmetic principles in carrying out operator product expansions (\hyperlink{OPE}{OPE}s) of vertex operators and derive the $SU(3)_1$ Kac-Moody algebra (also known as an affine Lie algebra or a Wess-Zumino-Witten (WZW) algebra) encountered in section~\ref{sec:DiracParton}. We present the equal-time commutation relations (\hyperlink{ETCR}{ETCR}s) between bosonized variables in the coupled wire setting in section~\ref{sec:CWM}, and pay special attention to the commutation relations between zero modes. 

\subsection{The \texorpdfstring{$U(1)_3\times SU(3)_1$}{U(1)xSU(3)} parton algebra}\label{app:partonalgebra}
The bosonization of the three parton Dirac fermions $\pi_a=e^{i\tilde\phi_a}$, for $a=1,2,3$, described in section~\ref{sec:DiracParton} is based on the time-ordered correlation \eqref{TOC} between the three bosonized variables. \begin{align}\langle\tilde\phi_a(\mathsf{z})\tilde\phi_b(\mathsf{w})\rangle=-\delta_{ab}\log(\mathsf{z}-\mathsf{w})+\frac{i\pi}{2}S_{ab}\label{appTOC}\end{align} where $\mathsf{z},\mathsf{w}\sim e^{\tau+i\mathsf{x}}$ are complex space-time parameter in a radially ordered geometry, and the constant factor \begin{align}S=(S_{ab})_{3\times3}=\begin{pmatrix}0&1&-1\\-1&0&1\\1&-1&0\end{pmatrix}\label{appKleinS}\end{align} is set to ensure anticommutation relations between mutual parton fermions. The correlation \eqref{appTOC} is equivalent to the equal-time commutation relation (\hyperlink{ETCR}{ETCR}) \begin{align}\left[\tilde\phi_a(\mathsf{x}),\tilde\phi_a(\mathsf{x}')\right]&=i\pi\delta_{ab}\mathrm{sgn}(x-x')+i\pi S_{ab},\label{apppartonETCR}\end{align} which implies \eqref{ETCRpartons} upon differentiation with respect to $\mathsf{x}$, where $\mathrm{sgn}(s)=s/|s|$ when $s\neq0$ or $0$ when $s=0$. 

The operator product expansions (\hyperlink{OPE}{OPE}) between a general pair of normal ordered vertex operators can be evaluated according to \begin{align}e^{A(\mathsf{z})}e^{B(\mathsf{w})}&=e^{A(\mathsf{z})+B(\mathsf{w})+\langle A(\mathsf{z})B(\mathsf{w})\rangle}\\&=e^{A(\mathsf{w})+\partial A(\mathsf{w})(\mathsf{z}-\mathsf{w})+\ldots+B(\mathsf{w})+\langle A(\mathsf{z})B(\mathsf{w})\rangle},\nonumber\end{align} where $A,B$ are linear combinations of the bosonized variables $\widetilde{\phi}_a$. At equal time, this is equivalent to the Baker-Campbell-Hausdorff formula \begin{align}e^{A(\mathsf{x})}e^{B(\mathsf{x}')}=e^{A(\mathsf{x})+B(\mathsf{x}')+\frac{1}{2}\left[A(\mathsf{x}),B(\mathsf{x}')\right]}\label{BCHformula}\end{align} where all higher-order commutators vanish because $\left[A(\mathsf{x}),B(\mathsf{x}')\right]$ is a $c$-number. For instance, the \OPE between a pair of parton Dirac fermions is \begin{align}\pi^a(z)\left(\pi^b(w)\right)^\dagger&=e^{i\tilde\phi_a(\mathsf{z})-i\tilde\phi_b(\mathsf{w})-\delta_{ab}\log(\mathsf{z}-\mathsf{w})+i\pi S_{ab}/2}\nonumber\\&=\delta_{ab}\left[\frac{1}{\mathsf{z}-\mathsf{w}}+i\partial\tilde\phi_a(w)\right]\\&\;\;\;+iS_{ab}e^{i(\tilde\phi_a(\mathsf{w})-\tilde\phi_b(\mathsf{w}))}+\ldots,\nonumber\end{align} where higher-order non-singular pieces are suppressed in the limit $\mathsf{z}\to\mathsf{w}$. The $S_{ab}$ factor ensures fermions with distinct flavors anticommutes \begin{align}\pi^a(\mathsf{z})\pi^b(\mathsf{w})&=e^{-i\pi S_{ab}}\pi^b(\mathsf{w})\pi^a(\mathsf{z})=-\pi^b(\mathsf{w})\pi^a(\mathsf{z}).\end{align}

The $U(3)_1/\mathbb{Z}_3=U(1)_3\times SU(3)_1$ affine Lie algebra that generate the parton triplet \eqref{DiracPartondef} is generated by the normal ordered current operators \begin{align}\begin{split}&H^a(\mathsf{z})=\pi^a(\mathsf{z})\pi^a(\mathsf{z})^\dagger=i\partial\tilde\phi_a(\mathsf{z})\\&E^{ab}(\mathsf{z})=-iS_{ab}\pi^a(\mathsf{z})\pi^b(\mathsf{z})^\dagger=e^{i\left(\tilde\phi_a(\mathsf{z})-\tilde\phi_b(\mathsf{z})\right)}\end{split}\label{appSU3generators}\end{align} where $a\neq b$. The Cartan generators $H^a$ of $U(3)$ can be rotated into the Cartan generators of the diagonal charge $U(1)_3$ sector $H_\rho=i(\partial\tilde\phi_1+\partial\tilde\phi_2+\partial\tilde\phi_3)/\sqrt{3}$ and the neutral $SU(3)_1$ sector $H_{n1}=i(\partial\tilde\phi_1-\partial\tilde\phi_2)/\sqrt{2}$ and $h_{n2}=i(\partial\tilde\phi_1+\partial\tilde\phi_2-2\partial\tilde\phi_3)/\sqrt{6}$. They obey the \hyperlink{OPE}{OPE}s \begin{align}H^a(\mathsf{z})H^b(\mathsf{w})&=\frac{\delta^{ab}}{(\mathsf{z}-\mathsf{w})^2}+:H^a(\mathsf{w})H^b(\mathsf{w}):+\ldots\nonumber\\H_\rho(\mathsf{z})H_\rho(\mathsf{w})&=\frac{1}{(\mathsf{z}-\mathsf{w})^2}+:H_\rho(\mathsf{w})H_\rho(\mathsf{w}):+\ldots\nonumber\\H_{ni}(\mathsf{z})H_{nj}(\mathsf{w})&=\frac{\delta_{ij}}{(\mathsf{z}-\mathsf{w})^2}+:H_{ni}(\mathsf{w})H_{nj}(\mathsf{w}):+\ldots\nonumber\\H_\rho(\mathsf{z})H_{nI}(\mathsf{w})&=:H_\rho(\mathsf{w})H_{nI}(\mathsf{w}):+\ldots.\label{appSU3currentOPE1}\end{align}

The six $SU(3)$ roots $E^{ab}$ are raising and lowering operators and follow the singular \hyperlink{OPE}{OPE}s with the Cartan generators \begin{align}H^a(\mathsf{z})E^{bc}(\mathsf{w})&=i\langle\tilde\phi_a(\mathsf{z})\tilde\phi_d(\mathsf{w})\rangle\frac{\delta}{\delta\tilde\phi_d({\mathsf{w}})}E^{bc}(\mathsf{w})+\ldots\nonumber\\&=\frac{\delta^{ac}-\delta^{bc}}{\mathsf{z}-\mathsf{w}}E^{bc}(\mathsf{w})+:H^a(\mathsf{w})E^{bc}(\mathsf{w}):+\ldots\label{appSU3currentOPE2}\end{align} A pair of raising and lowering operators obey the \hyperlink{OPE}{OPE}s \begin{align}&E^{ab}(\mathsf{z})E^{cd}(\mathsf{w})\nonumber\\&=i^{S_{ad}+S_{bc}-S_{ac}-S_{bd}}\frac{e^{i\left(\tilde\phi_a(\mathsf{z})-\tilde\phi_b(\mathsf{z})+\tilde\phi_c(\mathsf{w})-\tilde\phi_d(\mathsf{w})\right)}}{(\mathsf{z}-\mathsf{w})^{ad+bc-ac-bd}}\nonumber\\&=\delta^{ad}\delta^{bc}\left[\frac{1}{(\mathsf{z}-\mathsf{w})^2}+\frac{H^a(\mathsf{w})-H^b(\mathsf{w})}{\mathsf{z}-\mathsf{w}}\right.\label{appSU3currentOPE3}\\&\quad\quad\quad+\left.\left(H^a(\mathsf{w})-H^b(\mathsf{w})\right)^2+\frac{i}{2}\left(\partial^2\tilde\phi_a(\mathsf{w})-\partial^2\tilde\phi_b(\mathsf{w})\right)\right]\nonumber\\&\;\;\;+if^{(ab)(cd)(ef)}E^{ef}(\mathsf{w})\left[\frac{1}{\mathsf{z}-\mathsf{w}}+\left(H^a(\mathsf{w})-H^b(\mathsf{w})\right)\right]+\ldots\nonumber\end{align} where $f^{(ab)(cd)(ef)}=\varepsilon^{abd}\delta^{bc}\delta^{ae}\delta^{df}-\varepsilon^{abc}\delta^{ad}\delta^{bf}\delta^{ce}$ is the structure factor of $SU(3)$ where $a\neq b$, $c\neq d$ and $e\neq f$. Here the constant ``cocycle" factor $i^{S_{ad}+S_{bc}-S_{ac}-S_{bd}}$ at the first equality in \eqref{appSU3currentOPE3} originates from the constant non-singular term $S_{ab}$ in the time-ordered correlation \eqref{appTOC}. It is responsible for the antisymmetry of the structure factor, $f^{(ab)(cd)(ef)}=-f^{(cd)(ab)(ef)}$. Eq.\eqref{appSU3currentOPE1}, \eqref{appSU3currentOPE2} and \eqref{appSU3currentOPE3} recover the current algebra \OPE in \eqref{SU3currentOPE} if keeping only singular terms. The two Cartan generators $H_{n1}$, $H_{n2}$ and the six roots $E^{ab}$ form the $SU(3)$ current algebra at level 1 \begin{align}J_\alpha(\mathsf{z})J_\beta(\mathsf{w})=\frac{\delta_{\alpha\beta}}{(\mathsf{z}-\mathsf{w})^2}+\frac{if^{\alpha\beta\gamma}}{\mathsf{z}-\mathsf{w}}J_\gamma(\mathsf{w})+\ldots\end{align} where $J_\alpha$ are the eight $SU(3)$ current generators and $f^{\alpha\beta\gamma}$ is the full structure factor of $SU(3)$. 

The conformal embedding of $U(1)_3\times SU(3)_1$ into $U(3)_1$ is demonstrated by the splitting of the energy-momentum tensor \begin{align}T_{U(3)_1}=T_{U(1)_3}+T_{SU(3)_1}.\end{align} The full energy-momentum tensor of the parton triplet is the normal ordered product \begin{align}T_{U(3)_1}(\mathsf{z})=\sum_{a=1}^3\pi^a(\mathsf{z})^\dagger\partial\pi^a(\mathsf{z})=-\frac{1}{2}\sum_{a=1}^3\partial\tilde\phi^a(\mathsf{z})\partial\tilde\phi^a(\mathsf{z}).\end{align} The energy-momentum tensor of the Laughlin $U(1)_3$ sector is \begin{align}T_{U(1)_3}(\mathsf{z})&=-\frac{1}{2}H_\rho(\mathsf{z})H_\rho(\mathsf{z})\\&=-\frac{1}{6}\left[\sum_{a=1}^3(\partial\tilde\phi_a(\mathsf{z}))^2+2\sum_{1\leq a<b\leq3}\partial\tilde\phi_a(\mathsf{z})\partial\tilde\phi_b(\mathsf{z})\right]\nonumber\end{align} and the energy-momentum tensor of the neutral $SU(3)_1$ sector is given by the Sugawara form \begin{align}&T_{SU(3)_1}(\mathsf{z})\nonumber\\&=\frac{1}{2(1+h_{SU(3)})}\left[\sum_{j=1,2}H_{nj}(\mathsf{z})H_{nj}(\mathsf{z})+\sum_{a\neq b}E^{ab}(\mathsf{z})E^{ba}(\mathsf{z})\right]\nonumber\\&=-\frac{1}{3}\left[\sum_{a=1}^3(\partial\tilde\phi_a(\mathsf{z}))^2-\sum_{1\leq a<b\leq3}\partial\tilde\phi_a(\mathsf{z})\partial\tilde\phi_b(\mathsf{z})\right]\end{align} where $h_{SU(3)}=3$ is the dual Coxeter number. The mutual \OPE between $T_{U(1)_3}$ and $T_{SU(3)1}$ is non-singular, and therefore the two sectors decouple.

\subsection{Equal-time commutation relations and Klein factors}\label{app:ETCR}
In this appendix, we set the equal-time commutation relations (\hyperlink{ETCR}{ETCR}) between the bosonized variables in the coupled wire models in section~\ref{sec:CWM}. In particular, we present the commutation relations between zero modes that correspond to constant terms $C_{\alpha\beta}$ in \begin{align}[\phi^\alpha(\mathsf{x}),\phi^\beta(\mathsf{x}')]=i\pi(K^{-1})^{\alpha\beta}\mathrm{sgn}(\mathrm{x}-\mathrm{x}')+i\pi C_{\alpha\beta}.\end{align} these terms drop out upon differentiation and are absent in the canonical \ETCR $[\partial_{\mathsf{x}}\Phi_\alpha(\mathsf{x}),\Phi_\beta(\mathsf{x}')]=2\pi i(K^{-1})^{\alpha\beta}\delta(\mathrm{x}-\mathrm{x}')$ set by the ``$p\dot{q}$" term of the Lagrangian density $\mathcal{L}=(1/2\pi)K_{\alpha\beta}\partial_{\mathsf{t}}\phi^\alpha\partial_{\mathsf{x}}\phi^\alpha$. However, they are necessary for a consistent multi-component bosonization scheme. For example, the constant piece involving the antisymmetric matrix $S_{ab}$ in the \ETCR \eqref{apppartonETCR} for parton bosonized variables $\tilde\phi_a$ ensures the anticommutation relations between fermionic parton operators $\pi^a=e^{i\tilde\phi_a}$ of distinct flavors. The constant term is also essential to the antisymmetry of the structure factor in the $SU(3)_1$ current algebra \eqref{appSU3currentOPE2}. In the coupled wire construction, similar constant terms must be carefully instated to the \ETCR of bosonized variables to uphold the appropriate algebraic relations between physical operators.

\subsubsection{The Abelian parton sequence \texorpdfstring{$\mathfrak{L}_q$}{Lq}}\label{app:ETCRAbelian}
Here we define the \ETCR for the bosonized variables in the coupled wire model for the sequence of Abelian \FQH states at filling one-third described in section~\ref{sec:fillingonethird}. The model is based on an array of electronic bundles, each labeled by an integer $y$ that reflects its vertical position, and each consists of 5 counter-propagating pairs of electronic channels, labeled by the wire index $a=0,\ldots,4$ and propagation direction $\sigma=R,L=+,-$. The electronic operators are bosonized according to $c^\sigma_{ya}(\mathsf{x})=e^{i(\tilde\Phi^\sigma_{ya}(\mathsf{x})+k^\sigma_{ya}\mathsf{x})}$. The bosonized variables obey the \ETCR \begin{align}\left[\tilde\Phi_{ya}^\sigma(\mathsf{x}),\tilde\Phi_{y'a'}^{\sigma'}(\mathsf{x}')\right]&=i\pi\sigma\delta^{\sigma\sigma'}\delta_{yy'}\delta_{aa'}\mathrm{sgn}(\mathsf{x}-\mathsf{x}')\nonumber\\&\;\;\;+i\pi C^{\sigma\sigma'}_{yay'a'}.\label{appETCRParton}\end{align} The first term corresponds to the canonical \ETCR \eqref{ETCRParton} upon differentiation with respect to $\mathsf{x}$, and the constant term on the second line is antisymmetric, \begin{align}C^{\sigma\sigma'}_{yay'a'}=-C^{\sigma'\sigma}_{y'a'ya},\label{appKleinantisymm1}\end{align} because of the antisymmetry of the commutator. In order for the electron operators to obey mutual fermionic statistics, the constant $C^{\sigma\sigma'}_{yay'a'}$ must be an odd integer whenever $(y,a,\sigma)\neq(y',a',\sigma')$ so that, from the Baker-Campbell-Hausdorff formula \eqref{BCHformula}, \begin{align}e^{i\tilde\Phi^\sigma_{ya}(\mathsf{x})}e^{i\tilde\Phi^{\sigma'}_{y'a'}(\mathsf{x}')}&=e^{\left[i\tilde\Phi^\sigma_{ya}(\mathsf{x}),i\tilde\Phi^{\sigma'}_{y'a'}(\mathsf{x}')\right]}e^{i\tilde\Phi^{\sigma'}_{y'a'}(\mathsf{x}')}e^{i\tilde\Phi^\sigma_{ya}(\mathsf{x})}\nonumber\\&=e^{-i\pi C^{\sigma\sigma'}_{yay'a'}}e^{i\tilde\Phi^{\sigma'}_{y'a'}(\mathsf{x}')}e^{i\tilde\Phi^\sigma_{ya}(\mathsf{x})}\nonumber\\&=-e^{i\tilde\Phi^{\sigma'}_{y'a'}(\mathsf{x}')}e^{i\tilde\Phi^\sigma_{ya}(\mathsf{x})}\label{appfermionanticomm1}\end{align} between electronic vertex operators with distinct channel labels.

The bosonization of the electronic channels can alternatively be carried out by using the {\em bare} bosonized variables $_\flat\tilde\Phi^\sigma_{ya}(\mathsf{x})$ that obey the decoupled \ETCR \begin{align}\left[_\flat\tilde\Phi_{ya}^\sigma(\mathsf{x}),{_\flat\tilde\Phi}_{y'a'}^{\sigma'}(\mathsf{x}')\right]&=i\pi\sigma\delta^{\sigma\sigma'}\delta_{yy'}\delta_{aa'}\mathrm{sgn}(\mathsf{x}-\mathsf{x}').\end{align} They are related to the previous bosonized variables by \begin{align}\tilde\Phi^\sigma_{ya}(\mathsf{x})&={_\flat\tilde\Phi}^\sigma_{ya}(\mathsf{x})+\frac{i\pi}{2}\sum_{y'a'\sigma'}C^{\sigma\sigma'}_{yay'a'}\sigma'N^{\sigma'}_{y'a'}.\end{align} Here, $N^\sigma_{ya}=\int d\mathsf{x}\partial_\mathsf{x}\tilde\Phi^\sigma_{ya}(\mathsf{x})/(2\pi i)$ is the electron number operator for the fermion channel \begin{align}c^\sigma_{ya}(\mathsf{x})=e^{i\tilde\Phi^\sigma_{ya}(\mathsf{x})}=\kappa^\sigma_{ya}e^{i{_\flat\tilde\Phi}^\sigma_{ya}(\mathsf{x})},\end{align} and the constant operators \begin{align}\kappa^\sigma_{ya}=e^{\frac{i\pi}{2}\sum_{y'a'\sigma'}C^{\sigma\sigma'}_{yay'a'}N^{\sigma'}_{y'a'}}\end{align} are referred to as Klein factors.

The constant terms $C^{\sigma\sigma'}_{yay'a'}$ are chosen to be \begin{gather}C^{\sigma\sigma'}_{yay'a'}=\left\{\begin{array}{*{20}c}L^{\sigma\sigma'}_{aa'}&\mbox{if $y=y'$}\\M^{\sigma\sigma'}_{aa'}&\mbox{if $y<y'$}\\-M^{\sigma'\sigma}_{a'a}&\mbox{if $y>y'$}\end{array}\right.\label{appKlein1}\\\mbox{where}\quad\begin{split}&L^{\sigma\sigma'}_{aa'}=L^{++}_{aa'}\sigma\delta^{\sigma\sigma'}+L^{+-}_{aa'}\mathrm{sgn}(\sigma-\sigma'),\\&M^{\sigma\sigma'}_{aa'}=M^{++}_{aa'}\sigma\delta^{\sigma\sigma'}+M^{+-}_{aa'}\mathrm{sgn}(\sigma-\sigma')\end{split}\nonumber\end{gather} are decomposed into the $5\times5$ matrices \begin{align}L^{++}&=\begin{pmatrix} 0 & -1 & -1 & -1 & -1 \\ 1 & 0 & -1 & 1 & -1 \\ 1 & 1 & 0 & -1 & -1 \\ 1 & -1 & 1 & 0 & -1 \\ 1 & 1 & 1 & 1 & 0\end{pmatrix}\nonumber\\L^{+-}&=\begin{pmatrix}-1 & -1 & -1 & -1 & 1 \\ -1 & -1 & 1 & 1 & 1 \\ -1 & 1 & 1 & -1 & 1 \\ -1 & 1 & -1 & 1 & 1 \\ 1 & 1 & 1 & 1 & 1\end{pmatrix}\nonumber\\M^{++}&=\begin{pmatrix} 1 & 1 & 1 & 1 & 1 \\ -1 & 1 & -1 & 1 & 1 \\ -1 & 1 & 1 & -1 & 1 \\ -1 & -1 & 1 & 1 & 1 \\ -1 & -1 & -1 & -1 & 1\end{pmatrix}\nonumber\\M^{+-}&=\begin{pmatrix}1 & 1 & 1 & 1 & 1 \\ 1 & -1 & 1 & 1 & -1 \\ 1 & 1 & 1 & -1 & -1 \\ 1 & 1 & -1 & 1 & -1 \\ 1 & -1 & -1 & -1 & -1\end{pmatrix}\label{appMLmatrices}\end{align} with their rows, columns ordered according to $a,a'=0,\ldots,4$ respectively. The constant factor \eqref{appKlein1} can also be expressed as \begin{align}&C^{\sigma\sigma'}_{yay'a'}\nonumber\\&=\left[L^{++}_{aa'}\sigma\delta^{\sigma\sigma'}+L^{+-}_{aa'}\mathrm{sgn}(\sigma-\sigma')\right]\delta_{yy'}\nonumber\\&\;\;\;+\left[(M^{++}_{aa'}-\delta_{aa'})\sigma\delta^{\sigma\sigma'}+M^{+-}_{aa'}\mathrm{sgn}(\sigma-\sigma')\right](1-\delta_{yy'})\nonumber\\&\;\;\;-\delta_{aa'}\sigma\delta^{\sigma\sigma'}\mathrm{sgn}(y-y').\end{align} The antisymmtry relation \eqref{appKleinantisymm1} is satisfied because \begin{gather}(L^{++})^T=-L^{++},\quad(L^{+-})^T=L^{+-},\\(M^{++}-\openone)^T=-(M^{++}-\openone),\quad(L^{+-})^T=L^{+-}.\nonumber\end{gather} The anticommutation relation \eqref{appfermionanticomm1} between mutual electron operators holds because $C^{\sigma\sigma'}_{yay'a'}=\pm1$ so that $e^{-i\pi C^{\sigma\sigma'}_{yay'a'}}=-1$ for $(y,a,\sigma)\neq(y',a',\sigma')$. In addition to the antisymmetries, the Kac-Moody algebra \eqref{appETCRParton} also respects the time-reversal symmetry \begin{align}\left[\mathcal{T}\tilde\Phi_{ya}^\sigma(\mathsf{x})\mathcal{T}^{-1},\mathcal{T}\tilde\Phi_{y'a'}^{\sigma'}(\mathsf{x}')\mathcal{T}^{-1}\right]=-\left[\tilde\Phi_{ya}^\sigma(\mathsf{x}),\tilde\Phi_{y'a'}^{\sigma'}(\mathsf{x}')\right],\label{TRETCR}\end{align} although the symmetry is eventually broken by the backscattering interactions. Here, the time-reversal operator $\mathcal{T}$ is antiunitary and transforms the bosonized variables according to \begin{align}\mathcal{T}\tilde\Phi^\sigma_{ya}(\mathsf{x})\mathcal{T}^{-1}=-\tilde\Phi^{-\sigma}_{ya}(\mathsf{x})-i\pi\vartheta^\sigma_{ya}\end{align} where $\vartheta^\sigma_{ya}$ are constant real numbers that satisfy $\vartheta^+_{ya}+\vartheta^-_{ya}=1$ so that the transformation is in agreement with $\mathcal{T}^2=(-1)^N$, where $N=\sum_{ya\sigma}\int d{\mathsf{x}}\partial_{\mathsf{x}}\tilde\Phi_{ya}^\sigma/(2\pi i)$ is the total electron number operator. 

The signs of the matrix entries in \eqref{appMLmatrices} are chosen so that the sine-Gordon angle variables involved in the coupled wire models \eqref{PartonCWfullH} mutually commute. This ensures the Hamiltonians \eqref{PartonCWfullH} for the Abelian parton \FQH states $\mathcal{L}_q$ are exactly solvable. In \eqref{PartonUbasis} and \eqref{PartonSU3basis}, we transformed the bosonized variables from $(\tilde\Phi^\sigma_{y0},\ldots,\tilde\Phi^\sigma_{y4})$ to the new basis $(\Phi^\sigma_{y,\rho},\Phi^\sigma_{y,c1},\Phi^\sigma_{y,c2},\Phi^\sigma_{y,n1},\Phi^\sigma_{y,n2})$. The Dirac fermion channels $\Phi_{c1},\Phi_{c2}$ become massive under the intra-bundle backscattering interactions \eqref{intrabundleH}, which pins the two sine-Gordon angle variables $\Theta^{\mathrm{intra}}_{y,I},\Theta^{\mathrm{intra}}_{y,II}$. The \ETCR \eqref{appETCRParton} with the choice of the constant terms $C^{\sigma\sigma'}_{yay'a'}$ presented in \eqref{appKlein1} and \eqref{appMLmatrices} warrants the commutation relations \begin{align}\begin{split}&\left[\Theta^{\mathrm{intra}}_{y,l}(\mathsf{x}),\Theta^{\mathrm{intra}}_{y',l'}(\mathsf{x}')\right]=0,\quad\left[\Theta^{\mathrm{intra}}_{y,l}(\mathsf{x}),\Phi^\sigma_{y',\rho}(\mathsf{x}')\right]=0,\\&\left[\Theta^{\mathrm{intra}}_{y,l}(\mathsf{x}),\Phi^\sigma_{y',n1}(\mathsf{x}')\right]=0,\quad\left[\Theta^{\mathrm{intra}}_{y,l}(\mathsf{x}),\Phi^\sigma_{y',n2}(\mathsf{x}')\right]=0\end{split}\label{appintracomm}\end{align} for $l,l'=I,II$. 

The remaining three bosonized variables $(\Phi^\sigma_{y,\rho},\Phi^\sigma_{y,n1},\Phi^\sigma_{y,n2})$ corresponding to the $U(3)_1/\mathbb{Z}_3=U(1)_3\times SU(3)_1$ parton Dirac triplet obey the commutation relations \begin{align}\left[\Phi^\sigma_{y,\rho}(\mathsf{x}),\Phi^{\sigma'}_{y'\rho}(\mathsf{x}')\right]&=3\pi i\sigma\delta^{\sigma\sigma'}\delta_{yy'}\mathrm{sgn}(\mathsf{x}-\mathsf{x}')\nonumber\\&\;\;\;+3\pi i\mathrm{sgn}(\sigma-\sigma')\nonumber\\&\;\;\;-3\pi i\sigma\delta^{\sigma\sigma'}\mathrm{sgn}(y-y')\nonumber\\\left[\Phi^\sigma_{y,nj}(\mathsf{x}),\Phi^{\sigma'}_{y',nj'}(\mathsf{x}')\right]&=i\pi\sigma\delta^{\sigma\sigma'}\delta_{yy'}(K_{SU(3)})_{jj'}\mathrm{sgn}(\mathsf{x}-\mathsf{x}')\nonumber\\&\;\;\;+i\pi S^{\sigma\sigma'}_{yjy'j'}\nonumber\\\left[\Phi^\sigma_{y,\rho}(\mathsf{x}),\Phi^{\sigma'}_{y',nj}(\mathsf{x}')\right]&=0\label{appETCRU3}\end{align} where $K_{SU(3)}=2\openone_{2\times2}-\sigma_x$ is the Cartan matrix of $SU(3)$. The constant terms in the neutral sector are 
\begin{gather}S^{\sigma\sigma'}_{yjy'j'}=\left\{\begin{array}{*{20}c}(S_0)^{\sigma\sigma'}_{jj'}&\mbox{if $y=y'$}\\(S_1)^{\sigma\sigma'}_{jj'}&\mbox{if $y<y'$}\\-(S_1)^{\sigma'\sigma}_{j'j}&\mbox{if $y>y'$}\end{array}\right.\label{appKlein2}\\\mbox{where}\quad\begin{split}&(S_0)^{\sigma\sigma'}_{jj'}=(S_0)^{++}_{jj'}\sigma\delta^{\sigma\sigma'}+(S_0)^{+-}_{jj'}\mathrm{sgn}(\sigma-\sigma')\\&(S_1)^{\sigma\sigma'}_{jj'}=(S_1)^{++}_{jj'}\sigma\delta^{\sigma\sigma'}+(S_1)^{+-}_{jj'}\mathrm{sgn}(\sigma-\sigma')\end{split}\nonumber\end{gather} and the constants are grouped into the $2\times2$ matrices \begin{gather}\begin{split}(S_0)^{++}=\begin{pmatrix}0&-3\\3&0\end{pmatrix},\quad(S_0)^{+-}=\begin{pmatrix}2&2\\2&-4\end{pmatrix},\\(S_1)^{++}=\begin{pmatrix}2&-4\\2&2\end{pmatrix},\quad(S_1)^{+-}=\begin{pmatrix}2&2\\2&-4\end{pmatrix}.\end{split}\end{gather} Equivalently, the constant terms in the neutral sector can also be expressed as \begin{align}S^{\sigma\sigma'}_{yjy'j'}&=3\sigma\delta^{\sigma\sigma'}\mathrm{sgn}(j-j')+(S_0)^{+-}_{jj'}\mathrm{sgn}(\sigma-\sigma')\nonumber\\&\;\;\;-(K_{SU(3)})_{jj'}\sigma\delta^{\sigma\sigma'}\mathrm{sgn}(y-y').\label{appKleinS0}\end{align} The \ETCR \eqref{appETCRU3} guarantees the commutativity \begin{align}&\left[\Theta^\rho_{y,y+1}(\mathsf{x}),\Theta^\rho_{y',y'+1}(\mathsf{x}')\right]=\left[\Theta^\rho_{y,y+1}(\mathsf{x}),\Theta^{nj}_{y',y'+q}(\mathsf{x}')\right]\nonumber\\&=\left[\Theta^{nj}_{y,y+q}(\mathsf{x}),\Theta^{nj'}_{y',y'+q}(\mathsf{x}')\right]=0\end{align} of the sine-Gordon angle variables \begin{align}&\Theta^\rho_{y,y+1}=\Phi^R_{y,\rho}-\Phi^L_{y+1,\rho},\\&\Theta^{n1}_{y,y+q}=\Phi^R_{y,n1}-\Phi^L_{y+q,n1},\quad\Theta^{n2}_{y,y+q}=\Phi^R_{y,n2}-\Phi^L_{y+q,n2}\nonumber\end{align} in the inter-bundle backscattering interactions \eqref{LaughlinH} and \eqref{PartonSU3H}. 

\subsubsection{The paired parton sequence \texorpdfstring{$\mathfrak{P}_{p,q}$}{Ppq}}\label{app:ETCRpaired}
Here we define the equal-time commutation relation (\hyperlink{ETCR}{ETCR}) for the bosonized variables in the coupled wire model for the sequence of paired parton \FQH states at filling one-sixth presented in section~\ref{sec:fillingonesixth}. The model is based on an array of bundles, each consists of five electronic wires. The bundles are arranged in a 2-bundle unit cell (see figure~\ref{fig:Pfwiresbasis}). The electron (annihilation) operators are bosonized according to $c_{yAa}^\sigma(\mathsf{x})=e^{i(\tilde\Phi_{yAa}^\sigma+k^\sigma_{yAa}\mathsf{x})}$, where the integer $y$ labels the unit cell, $A=0,1$ labels the two bundles, $a=0,\ldots,4$ designates the five electronic wires, and $\sigma=R,L=+,-$ denotes the forward and backward channels along a wire. The \ETCR for the electronic bosonized variables can be deduced directly from the previous appendix~\ref{app:ETCRAbelian}, which applies to the model at filling one-third that contains twice as many wires (see figure~\ref{fig:Partonwirebasis}). Restricting to the current system, \begin{align}\left[\tilde\Phi_{yAa}^\sigma(\mathsf{x}),\tilde\Phi_{y'A'a'}^{\sigma'}(\mathsf{x}')\right]&=i\pi\sigma\delta^{\sigma\sigma'}\delta_{yy'}\delta_{AA'}\delta_{aa'}\mathrm{sgn}(\mathsf{x}-\mathsf{x}')\nonumber\\&\;\;\;+i\pi C^{\sigma\sigma'}_{yAay'A'a'}.\label{appETCRPf}\end{align}\begin{gather}C^{\sigma\sigma'}_{yAay'A'a'}=\left\{\begin{array}{*{20}l}L^{\sigma\sigma'}_{aa'}&\mbox{if $y=y'$, $A=A'$}\\M^{\sigma\sigma'}_{aa'}&\mbox{if $y+A/2<y'+A'/2$}\\-M^{\sigma'\sigma}_{a'a}&\mbox{if $y+A/2>y'+A'/2$}\end{array}\right.\nonumber\end{gather} where the $L$ and $M$ constants were defined in \eqref{appKlein1} and \eqref{appMLmatrices}.

We first observe that the intra-cell sine-Gordon terms $\mathcal{H}^{\mathrm{intra}}$ and $\mathcal{H}^{\mathrm{dimer}}_{SU(3)}$ defined in \eqref{intrabundleH} and \eqref{PfdimerSU3H} (see also figure~\ref{fig:PfConstructIntra}) are contained in the previous coupled wire model \eqref{PartonCWfullH} at filling one-third. The commutativity between the sine-Gordon angle variables were confirmed in appendix~\ref{app:ETCRAbelian}. These gapping terms leave behind two counter-propagating pairs of charge (neutral) channels $\Phi^\sigma_{yA\rho}$ (resp.~$\Phi^\sigma_{y,nj}$, for $j=1,2$). From \eqref{appETCRU3}, they obey the \ETCR \begin{align}&\left[\Phi^\sigma_{yA\rho}(\mathsf{x}),\Phi^{\sigma'}_{y'A'\rho}(\mathsf{x}')\right]\nonumber\\&=3\pi i\sigma\delta^{\sigma\sigma'}\delta_{AA'}\delta_{yy'}\mathrm{sgn}(\mathsf{x}-\mathsf{x}')\nonumber\\&\;\;\;+3\pi i\mathrm{sgn}(\sigma-\sigma')\nonumber\\&\;\;\;-3\pi i\sigma\delta^{\sigma\sigma'}\left[\delta_{yy'}\mathrm{sgn}(A-A')+\mathrm{sgn}(y-y')\right]\nonumber\\&\left[\Phi^\sigma_{y,nj}(\mathsf{x}),\Phi^{\sigma'}_{y',nj'}(\mathsf{x}')\right]\nonumber\\&=i\pi\sigma\delta^{\sigma\sigma'}\delta_{yy'}(K_{SU(3)})_{jj'}\mathrm{sgn}(\mathsf{x}-\mathsf{x}')+i\pi S^{\sigma\sigma'}_{yjy'j'}\nonumber\\&\left[\Phi^\sigma_{yA\rho}(\mathsf{x}),\Phi^{\sigma'}_{y',nj'}(\mathsf{x}')\right]=0\end{align} where $S^{\sigma\sigma'}_{yjy'j'}$ was given in \eqref{appKleinS0}.

Next, we performed the basis transformations \eqref{Pfbasistrans1} and \eqref{Pfbasistrans2} (also see figure~\ref{fig:PfConstructIntra}) to
change $(\Phi^\sigma_{yA\rho},\Phi^\sigma_{y,nj})$ to $(\Phi^\sigma_{y\rho},\Phi^\sigma_{y,dJ})$, where $e^{i\Phi^\sigma_{y\rho}}$ is a charge $e$ boson, and $e^{i\Phi^\sigma_{y,dJ}}$ are neutral Dirac fermions. The new bosonized variables obey the \ETCR \begin{align}\left[\Phi^\sigma_{y\rho}(\mathsf{x}),\Phi^{\sigma'}_{y'\rho}(\mathsf{x}')\right]&=6\pi i\sigma\delta^{\sigma\sigma'}\delta_{yy'}\mathrm{sgn}(\mathsf{x}-\mathsf{x}')\nonumber\\&\;\;\;+6\pi i(2-\delta_{yy'})\mathrm{sgn}(\sigma-\sigma')\nonumber\\&\;\;\;-12\pi i\sigma\delta^{\sigma\sigma'}\mathrm{sgn}(y-y')\nonumber\\\left[\Phi^\sigma_{y,dJ}(\mathsf{x}),\Phi^{\sigma'}_{y',dJ'}(\mathsf{x}')\right]&=i\pi\sigma\delta^{\sigma\sigma'}\delta_{yy'}(K_d)_{JJ'}\mathrm{sgn}(\mathsf{x}-\mathsf{x}')\nonumber\\&\;\;\;+i\pi\Sigma^{\sigma\sigma'}_{yJy'J'}\nonumber\\\left[\Phi^\sigma_{y\rho}(\mathsf{x}),\Phi^{\sigma'}_{y',dJ}(\mathsf{x}')\right]&=i\pi c^{\sigma\sigma'}_{y\rho y'J}\label{appETCRPf2}\end{align} where $K_d=\mathrm{diag}(1,3,3)$ is the diagonal matrix \eqref{Kdmatrix}, and $J,J'=0,1,2$. The constant terms $\Sigma^{\sigma\sigma'}_{yJy'J'}$ and $c^{\sigma\sigma'}_{y\rho y'J}$ are \begin{gather}\Sigma^{\sigma\sigma'}_{yJy'J'}=\left\{\begin{array}{*{20}c}(\Sigma_0)^{\sigma\sigma'}_{JJ'}&\mbox{if $y=y'$}\\(\Sigma_1)^{\sigma\sigma'}_{JJ'}&\mbox{if $y<y'$}\\-(\Sigma_1)^{\sigma'\sigma}_{J'J}&\mbox{if $y>y'$}\end{array}\right.\nonumber\\c^{\sigma\sigma'}_{y\rho y'J}=\left\{\begin{array}{*{20}c}(c_0)^{\sigma\sigma'}_J&\mbox{if $y=y'$}\\0&\mbox{if $y<y'$}\\(c_1)^{\sigma\sigma'}_J&\mbox{if $y>y'$}\end{array}\right.\\
\mbox{where}\quad\begin{split}&(\Sigma_0)^{\sigma\sigma'}_{JJ'}=(\Sigma_0)^{++}_{JJ'}\sigma\delta^{\sigma\sigma'}+(\Sigma_0)^{+-}_{JJ'}\mathrm{sgn}(\sigma-\sigma')\\&(\Sigma_1)^{\sigma\sigma'}_{JJ'}=(\Sigma_1)^{++}_{JJ'}\sigma\delta^{\sigma\sigma'}+(\Sigma_1)^{+-}_{JJ'}\mathrm{sgn}(\sigma-\sigma')\end{split}\nonumber\\
\mbox{and}\quad\begin{split}&(c_0)^{\sigma\sigma'}_J=(c_0)^{++}_J\sigma\delta^{\sigma\sigma'}+(c_0)^{+-}_J\mathrm{sgn}(\sigma-\sigma')\\&(c_1)^{\sigma\sigma'}_J=(c_1)^{++}_J\sigma\delta^{\sigma\sigma'}+(c_1)^{+-}_J\mathrm{sgn}(\sigma-\sigma')\end{split}\nonumber\end{gather} and the constants are grouped into the rank 3 matrices and vectors \begin{gather}\begin{split}(\Sigma_0)^{++}=\begin{pmatrix}0&3&-3\\-3&0&-9\\3&9&0\end{pmatrix},\quad(\Sigma_0)^{+-}=\begin{pmatrix}1&-3&3\\-3&-9&3\\3&3&9\end{pmatrix},\\(\Sigma_1)^{++}=\begin{pmatrix}-2&0&-6\\0&6&-6\\6&12&6\end{pmatrix},\quad(\Sigma_1)^{+-}=\begin{pmatrix}-2&-6&0\\0&-6&6\\6&6&12\end{pmatrix},\end{split}\nonumber\\
\begin{split}&(c_0)^{++}=(c_0)^{+-}=(6,6,6)\\&(c_1)^{++}=-(c_1)^{+-}=(-3,3,3)\end{split}\label{appSigmaMcV}\end{gather}

Having establishing the \hyperlink{ETCR}{ETCR}s, we now move on to the definition of the number operator $N^d_y$ that was used in \eqref{PfneutralDirac} for the definition of the neutral Dirac fermions \begin{align}d_{y,J}^\sigma&=e^{i\Phi^\sigma_{y,dJ}}\prod_{y'<y}e^{i\pi N^d_{y'}}=e^{i\Phi^\sigma_{y,dJ}+i\pi\sum_{y'<y}N^d_{y'}}.\label{appPfneutralDirac}\end{align} Within the same unit-cell $y$, the odd off-diagonal entries of $\Sigma_0^{++}$ and the odd entries of $\Sigma_0^{+-}$ in \eqref{appSigmaMcV} guarantee the mutual anticommutation relations between vertex operators \begin{align}e^{i\Phi^\sigma_{y,dJ}}e^{i\Phi^{\sigma'}_{y,dJ'}}&=e^{-[\Phi^\sigma_{y,dJ},\Phi^{\sigma'}_{y,dJ'}]}e^{i\Phi^{\sigma'}_{y,dJ'}}e^{i\Phi^\sigma_{y,dJ}}\nonumber\\&=-e^{i\Phi^{\sigma'}_{y,dJ'}}e^{i\Phi^\sigma_{y,dJ}}.\end{align} However, the vertex operators commute when they occupy different unit-cells because the entries of $\Sigma_1^{++}$ and $\Sigma_1^{+-}$ in \eqref{appSigmaMcV} are even. To facilitate the mutual fermionic anticommutation relations between the operators in \eqref{appPfneutralDirac}, we choose the number operator to be \begin{align}N^d_y&=N^R_{y,A=0,\rho}+5N^R_{y,A=1,\rho}\nonumber\\&\;\;\;-3N^L_{y,A=0,\rho}-3N^L_{y,A=1,\rho},\label{Ndoperator}\\\mbox{where }&N^\sigma_{yA\rho}=\frac{1}{2\pi}\int d\mathsf{x}\partial_{\mathsf{x}}\Phi^R_{yA\rho}.\nonumber
\end{align} It is a linear combination of the local electronic number operators $N^R_{yA\rho}=2N^R_{y,A,a=4}-N^L_{y,A,a=4}$ and $N^L_{yA\rho}=2N^L_{y,A,a=0}-N^R_{y,A,a=0}$, where $N^\sigma_{yAa}$ are the number operators for the electrons $c^\sigma_{yAa}$ in \eqref{electronPf} that constitute the coupled wire model. The number operator obeys the following \ETCR with the bosonized variables \begin{align}&\left[N^d_y,\Phi^\sigma_{y'\rho}(\mathsf{x})\right]=6i\delta_{yy'},\nonumber\\&\left[N^d_y,\Phi^\sigma_{y',dJ}(\mathsf{x})\right]=i(6-\sigma(c_1)^{++}_J)\delta_{yy'},\label{Ndoperatorcomm}\end{align} where $(c_1)^{++}_J=-3,3,3$, for $J=0,1,2$, is the same vector given in \eqref{appSigmaMcV}. In particular, the odd commutator in the second line ensures \begin{align}d^\sigma_{y,dJ}d^{\sigma'}_{y',dJ'}&=e^{-[\Phi^\sigma_{y,dJ},\pi N^d_y]}d^{\sigma'}_{y',dJ'}d^\sigma_{y,dJ}\nonumber\\&=-d^{\sigma'}_{y',dJ'}d^\sigma_{y,dJ}\end{align} for $y<y'$. The particular combination of \eqref{Ndoperator} is chosen so that the sine-Gordon angle variables $\Theta^\rho_{y+1/2}=\Phi^R_{y\rho}-\Phi^\rho_{y+1,\rho}-\pi N^d_y$ appearing in the interactions \eqref{PfrhogammaH}, \eqref{PfdrhoH}, \eqref{PfrhogammaHpq} and \eqref{PfdrhoHpq} commutes with the neutral Dirac fermions \begin{align}\left[\Theta^\rho_{y+1/2}(\mathsf{x}),d^\sigma_{y',J}(\mathsf{x}')\right]=0.\end{align}

The intra-bundle Dirac/Majorana fermion backscatterings \eqref{PfincellHDF}, \eqref{PfincellHMF} and the inter-bundle interactions \eqref{PfrhogammaH}, \eqref{PfdrhoH} and \eqref{PfgammadH} as well as \eqref{PfrhogammaHpq} and \eqref{PfdrhoHpq} can be expressed using local electronic operators. This can be verified by using the basis transformations \eqref{Pfbasistrans1} and \eqref{Pfbasistrans2} that relates the local electronic bosonized variables $\Phi^\sigma_{y\rho A}$, $\Phi^\sigma_{y,nj}$ for $A=0,1$, $j=1,2$ to the fractional bosonized variables $\Phi^\sigma_{y\rho}$, $\Phi^\sigma_{y,dJ}$, for $J=0,1,2$. The fractional variables are half-integral in the sense that any sum or difference between any pair of $\Phi^\sigma_{y\rho}$ and $\Phi^\sigma_{y,dJ}$ is an integral combination of local electronic ones. Here, for concreteness, we express the interactions in terms of the local variables $\Phi^\sigma_{y\rho A}$, $\Phi^\sigma_{y,nj}$, which in turn are integral combinations of the fundamental electrons $\Phi^\sigma_{yAa}$, for $a=0,\ldots,5$, that constitute the coupled wire model as shown in \eqref{PfUbasis} and \eqref{Pfbasis}. Intra-bundle Dirac/Majorana fermion backscatterings, such as \eqref{PfincellHMF} and \eqref{PfincellHMF}, depend on the bosonized variables \begin{gather}\begin{array}{*{20}l}\Phi^R_{y\rho}+\Phi^L_{y\rho}=&\sum_{\sigma=\pm}\Phi^\sigma_{y,A=0,\rho}\\\Phi^R_{y\rho}-\Phi^L_{y\rho}=&\sum_{\sigma=\pm}2\sigma\left(\Phi^\sigma_{y,A=0,\rho}+\Phi^\sigma_{y,A=1,\rho}\right)\\\Phi^R_{y,d0}+\Phi^L_{y,d0}=&\sum_{\sigma=\pm}\left(-\Phi^\sigma_{y,A=0,\rho}+\Phi^\sigma_{y,A=1,\rho}+\Phi^\sigma_{y,n1}\right)\\\Phi^R_{y,d0}-\Phi^L_{y,d0}=&\sum_{\sigma=\pm}\sigma\left(\Phi^\sigma_{y,A=1,\rho}-\Phi^\sigma_{y,n1}\right)\\\Phi^R_{y,d1}-\Phi^L_{y,d1}=&\sum_{\sigma=\pm}\sigma\left(-\Phi^\sigma_{y,A=1,\rho}+\Phi^\sigma_{y,n1}-\Phi^\sigma_{y,n2}\right)\\\Phi^R_{y,d2}-\Phi^L_{y,d2}=&\sum_{\sigma=\pm}\sigma\left(-\Phi^\sigma_{y,A=1,\rho}+2\Phi^\sigma_{y,n1}+\Phi^\sigma_{y,n2}\right)\end{array}\label{appPfincellHDFMF}\end{gather} 
Inter-bundle interactions, such as \eqref{PfrhogammaH}, \eqref{PfdrhoH} and \eqref{PfgammadH} as well as \eqref{PfrhogammaHpq} and \eqref{PfdrhoHpq}, can be broken down using the following integral combinations of electronic variables \begin{gather}\begin{array}{*{20}l}\Phi^\sigma_{y\rho}+\Phi^\sigma_{y,d0}=&\Phi^\sigma_{y,A=0,\rho}+2\Phi^\sigma_{y,A=1,\rho}-\Phi^{-\sigma}_{y,A=0,\rho}\\&-\Phi^{-\sigma}_{y,A=1,\rho}+\Phi^{-\sigma}_{y,n1}\\\Phi^\sigma_{y\rho}-\Phi^\sigma_{y,d0}=&2\Phi^\sigma_{y,A=0,\rho}-\Phi^{-\sigma}_{y,A=1,\rho}-\Phi^{-\sigma}_{y,n1}\\\Phi^\sigma_{y\rho}+\Phi^{-\sigma}_{y,d0}=&\Phi^\sigma_{y,A=0,\rho}+\Phi^\sigma_{y,A=1,\rho}+\Phi^\sigma_{y,n1}-\Phi^{-\sigma}_{y,A=0,\rho}\\\Phi^\sigma_{y\rho}-\Phi^{-\sigma}_{y,d0}=&2\Phi^\sigma_{y,A=0,\rho}+\Phi^\sigma_{y,A=1,\rho}\\&-\Phi^\sigma_{y,n1}-2\Phi^{-\sigma}_{y,A=1,\rho}\end{array},\label{appPfinter1}\end{gather}\begin{gather}\begin{array}{*{20}l}\Phi^\sigma_{y\rho}+\Phi^\sigma_{y,d1}=&\Phi^\sigma_{y,A=0,\rho}+\Phi^\sigma_{y,A=1,\rho}+\Phi^\sigma_{y,n1}\\&-\Phi^\sigma_{y,n2}-\Phi^{-\sigma}_{y,A=0,\rho}\\\Phi^\sigma_{y\rho}-\Phi^\sigma_{y,d1}=&2\Phi^\sigma_{y,A=0,\rho}+\Phi^\sigma_{y,A=1,\rho}-\Phi^\sigma_{y,n1}\\&\Phi^\sigma_{y,n2}-2\Phi^{-\sigma}_{y,A=1,\rho}\\\Phi^\sigma_{y\rho}+\Phi^{-\sigma}_{y,d1}=&\Phi^\sigma_{y,A=0,\rho}+2\Phi^\sigma_{y,A=1,\rho}-\Phi^{-\sigma}_{y,A=0,\rho}\\&-\Phi^{-\sigma}_{y,A=1,\rho}+\Phi^{-\sigma}_{y,n1}-\Phi^{-\sigma}_{y,n2}\\\Phi^\sigma_{y\rho}-\Phi^{-\sigma}_{y,d1}=&2\Phi^\sigma_{y,A=0,\rho}-\Phi^{-\sigma}_{y,A=1,\rho}-\Phi^{-\sigma}_{y,n1}+\Phi^{-\sigma}_{y,n2}\end{array}.\label{appPfinter2}\end{gather}

{}




{}


\begin{thebibliography}{72}%
\makeatletter
\providecommand \@ifxundefined [1]{%
 \@ifx{#1\undefined}
}%
\providecommand \@ifnum [1]{%
 \ifnum #1\expandafter \@firstoftwo
 \else \expandafter \@secondoftwo
 \fi
}%
\providecommand \@ifx [1]{%
 \ifx #1\expandafter \@firstoftwo
 \else \expandafter \@secondoftwo
 \fi
}%
\providecommand \natexlab [1]{#1}%
\providecommand \enquote  [1]{``#1''}%
\providecommand \bibnamefont  [1]{#1}%
\providecommand \bibfnamefont [1]{#1}%
\providecommand \citenamefont [1]{#1}%
\providecommand \href@noop [0]{\@secondoftwo}%
\providecommand \href [0]{\begingroup \@sanitize@url \@href}%
\providecommand \@href[1]{\@@startlink{#1}\@@href}%
\providecommand \@@href[1]{\endgroup#1\@@endlink}%
\providecommand \@sanitize@url [0]{\catcode `\\12\catcode `\$12\catcode
  `\&12\catcode `\#12\catcode `\^12\catcode `\_12\catcode `\%12\relax}%
\providecommand \@@startlink[1]{}%
\providecommand \@@endlink[0]{}%
\providecommand \url  [0]{\begingroup\@sanitize@url \@url }%
\providecommand \@url [1]{\endgroup\@href {#1}{\urlprefix }}%
\providecommand \urlprefix  [0]{URL }%
\providecommand \Eprint [0]{\href }%
\providecommand \doibase [0]{http://dx.doi.org/}%
\providecommand \selectlanguage [0]{\@gobble}%
\providecommand \bibinfo  [0]{\@secondoftwo}%
\providecommand \bibfield  [0]{\@secondoftwo}%
\providecommand \translation [1]{[#1]}%
\providecommand \BibitemOpen [0]{}%
\providecommand \bibitemStop [0]{}%
\providecommand \bibitemNoStop [0]{.\EOS\space}%
\providecommand \EOS [0]{\spacefactor3000\relax}%
\providecommand \BibitemShut  [1]{\csname bibitem#1\endcsname}%
\let\auto@bib@innerbib\@empty
\bibitem [{\citenamefont {Wen}(2017)}]{WenRMP2017}%
  \BibitemOpen
  \bibfield  {author} {\bibinfo {author} {\bibfnamefont {X.-G.}\ \bibnamefont
  {Wen}},\ }\href {\doibase 10.1103/RevModPhys.89.041004} {\bibfield  {journal}
  {\bibinfo  {journal} {Rev. Mod. Phys.}\ }\textbf {\bibinfo {volume} {89}},\
  \bibinfo {pages} {041004} (\bibinfo {year} {2017})}\BibitemShut {NoStop}%
\bibitem [{\citenamefont {Laughlin}(1983)}]{Laughlin83}%
  \BibitemOpen
  \bibfield  {author} {\bibinfo {author} {\bibfnamefont {R.~B.}\ \bibnamefont
  {Laughlin}},\ }\href {\doibase 10.1103/PhysRevLett.50.1395} {\bibfield
  {journal} {\bibinfo  {journal} {Phys. Rev. Lett.}\ }\textbf {\bibinfo
  {volume} {50}},\ \bibinfo {pages} {1395} (\bibinfo {year}
  {1983})}\BibitemShut {NoStop}%
\bibitem [{\citenamefont {Wen}(1990)}]{Wentopologicalorder90}%
  \BibitemOpen
  \bibfield  {author} {\bibinfo {author} {\bibfnamefont {X.-G.}\ \bibnamefont
  {Wen}},\ }\href {\doibase 10.1142/S0217979290000139} {\bibfield  {journal}
  {\bibinfo  {journal} {Int. J. Mod. Phys. B}\ }\textbf {\bibinfo {volume}
  {04}},\ \bibinfo {pages} {239} (\bibinfo {year} {1990})}\BibitemShut
  {NoStop}%
\bibitem [{\citenamefont {Arovas}\ \emph {et~al.}(1984)\citenamefont {Arovas},
  \citenamefont {Schrieffer},\ and\ \citenamefont
  {Wilczek}}]{ArovasSchriefferWilczek84}%
  \BibitemOpen
  \bibfield  {author} {\bibinfo {author} {\bibfnamefont {D.}~\bibnamefont
  {Arovas}}, \bibinfo {author} {\bibfnamefont {J.~R.}\ \bibnamefont
  {Schrieffer}}, \ and\ \bibinfo {author} {\bibfnamefont {F.}~\bibnamefont
  {Wilczek}},\ }\href {\doibase 10.1103/PhysRevLett.53.722} {\bibfield
  {journal} {\bibinfo  {journal} {Phys. Rev. Lett.}\ }\textbf {\bibinfo
  {volume} {53}},\ \bibinfo {pages} {722} (\bibinfo {year} {1984})}\BibitemShut
  {NoStop}%
\bibitem [{\citenamefont {Wilczek}(1990)}]{Wilczekbook}%
  \BibitemOpen
  \bibfield  {author} {\bibinfo {author} {\bibfnamefont {F.}~\bibnamefont
  {Wilczek}},\ }\href@noop {} {\emph {\bibinfo {title} {Fractional Statistics
  and Anyon Superconductivity}}}\ (\bibinfo  {publisher} {World Scientific},\
  \bibinfo {year} {1990})\BibitemShut {NoStop}%
\bibitem [{\citenamefont {Cage}\ \emph {et~al.}(2012)\citenamefont {Cage},
  \citenamefont {Klitzing}, \citenamefont {Chang}, \citenamefont {Duncan},
  \citenamefont {Haldane}, \citenamefont {Laughlin}, \citenamefont {Pruisken},
  \citenamefont {Thouless}, \citenamefont {Prange},\ and\ \citenamefont
  {Girvin}}]{FQHE_Review}%
  \BibitemOpen
  \bibfield  {author} {\bibinfo {author} {\bibfnamefont {M.~E.}\ \bibnamefont
  {Cage}}, \bibinfo {author} {\bibfnamefont {K.}~\bibnamefont {Klitzing}},
  \bibinfo {author} {\bibfnamefont {A.}~\bibnamefont {Chang}}, \bibinfo
  {author} {\bibfnamefont {F.}~\bibnamefont {Duncan}}, \bibinfo {author}
  {\bibfnamefont {M.}~\bibnamefont {Haldane}}, \bibinfo {author} {\bibfnamefont
  {R.}~\bibnamefont {Laughlin}}, \bibinfo {author} {\bibfnamefont
  {A.}~\bibnamefont {Pruisken}}, \bibinfo {author} {\bibfnamefont
  {D.}~\bibnamefont {Thouless}}, \bibinfo {author} {\bibfnamefont {R.~E.}\
  \bibnamefont {Prange}}, \ and\ \bibinfo {author} {\bibfnamefont {S.~M.}\
  \bibnamefont {Girvin}},\ }\href@noop {} {\emph {\bibinfo {title} {The Quantum
  Hall Effect}}}\ (\bibinfo  {publisher} {Springer Science \& Business Media,
  Berlin},\ \bibinfo {year} {2012})\BibitemShut {NoStop}%
\bibitem [{\citenamefont {Hasan}\ and\ \citenamefont
  {Kane}(2010)}]{HasanKane10}%
  \BibitemOpen
  \bibfield  {author} {\bibinfo {author} {\bibfnamefont {M.~Z.}\ \bibnamefont
  {Hasan}}\ and\ \bibinfo {author} {\bibfnamefont {C.~L.}\ \bibnamefont
  {Kane}},\ }\href {\doibase 10.1103/RevModPhys.82.3045} {\bibfield  {journal}
  {\bibinfo  {journal} {Rev. Mod. Phys.}\ }\textbf {\bibinfo {volume} {82}},\
  \bibinfo {pages} {3045} (\bibinfo {year} {2010})}\BibitemShut {NoStop}%
\bibitem [{\citenamefont {Qi}\ and\ \citenamefont
  {Zhang}(2011)}]{QiZhangreview11}%
  \BibitemOpen
  \bibfield  {author} {\bibinfo {author} {\bibfnamefont {X.-L.}\ \bibnamefont
  {Qi}}\ and\ \bibinfo {author} {\bibfnamefont {S.-C.}\ \bibnamefont {Zhang}},\
  }\href {\doibase 10.1103/RevModPhys.83.1057} {\bibfield  {journal} {\bibinfo
  {journal} {Rev. Mod. Phys.}\ }\textbf {\bibinfo {volume} {83}},\ \bibinfo
  {pages} {1057} (\bibinfo {year} {2011})}\BibitemShut {NoStop}%
\bibitem [{\citenamefont {Chiu}\ \emph {et~al.}(2016)\citenamefont {Chiu},
  \citenamefont {Teo}, \citenamefont {Schnyder},\ and\ \citenamefont
  {Ryu}}]{RMP}%
  \BibitemOpen
  \bibfield  {author} {\bibinfo {author} {\bibfnamefont {C.-K.}\ \bibnamefont
  {Chiu}}, \bibinfo {author} {\bibfnamefont {J.~C.~Y.}\ \bibnamefont {Teo}},
  \bibinfo {author} {\bibfnamefont {A.~P.}\ \bibnamefont {Schnyder}}, \ and\
  \bibinfo {author} {\bibfnamefont {S.}~\bibnamefont {Ryu}},\ }\href {\doibase
  10.1103/RevModPhys.88.035005} {\bibfield  {journal} {\bibinfo  {journal}
  {Rev. Mod. Phys.}\ }\textbf {\bibinfo {volume} {88}},\ \bibinfo {pages}
  {035005} (\bibinfo {year} {2016})}\BibitemShut {NoStop}%
\bibitem [{\citenamefont {Maciejko}\ \emph {et~al.}(2010)\citenamefont
  {Maciejko}, \citenamefont {Qi}, \citenamefont {Karch},\ and\ \citenamefont
  {Zhang}}]{MaciejkoQiKarchZhang10}%
  \BibitemOpen
  \bibfield  {author} {\bibinfo {author} {\bibfnamefont {J.}~\bibnamefont
  {Maciejko}}, \bibinfo {author} {\bibfnamefont {X.-L.}\ \bibnamefont {Qi}},
  \bibinfo {author} {\bibfnamefont {A.}~\bibnamefont {Karch}}, \ and\ \bibinfo
  {author} {\bibfnamefont {S.-C.}\ \bibnamefont {Zhang}},\ }\href {\doibase
  10.1103/PhysRevLett.105.246809} {\bibfield  {journal} {\bibinfo  {journal}
  {Phys. Rev. Lett.}\ }\textbf {\bibinfo {volume} {105}},\ \bibinfo {pages}
  {\href{http://link.aps.org/doi/10.1103/PhysRevLett.105.246809}{246809}}
  (\bibinfo {year} {2010})}\BibitemShut {NoStop}%
\bibitem [{\citenamefont {Swingle}\ \emph {et~al.}(2011)\citenamefont
  {Swingle}, \citenamefont {Barkeshli}, \citenamefont {McGreevy},\ and\
  \citenamefont {Senthil}}]{SwingleBarkeshliMcGreevySenthil11}%
  \BibitemOpen
  \bibfield  {author} {\bibinfo {author} {\bibfnamefont {B.}~\bibnamefont
  {Swingle}}, \bibinfo {author} {\bibfnamefont {M.}~\bibnamefont {Barkeshli}},
  \bibinfo {author} {\bibfnamefont {J.}~\bibnamefont {McGreevy}}, \ and\
  \bibinfo {author} {\bibfnamefont {T.}~\bibnamefont {Senthil}},\ }\href
  {\doibase 10.1103/PhysRevB.83.195139} {\bibfield  {journal} {\bibinfo
  {journal} {Phys. Rev. B}\ }\textbf {\bibinfo {volume} {83}},\ \bibinfo
  {pages} {\href{http://link.aps.org/doi/10.1103/PhysRevB.83.195139}{195139}}
  (\bibinfo {year} {2011})}\BibitemShut {NoStop}%
\bibitem [{\citenamefont {Levin}\ \emph {et~al.}(2011)\citenamefont {Levin},
  \citenamefont {Burnell}, \citenamefont {Koch-Janusz},\ and\ \citenamefont
  {Stern}}]{LevinBurnellKochStern11}%
  \BibitemOpen
  \bibfield  {author} {\bibinfo {author} {\bibfnamefont {M.}~\bibnamefont
  {Levin}}, \bibinfo {author} {\bibfnamefont {F.~J.}\ \bibnamefont {Burnell}},
  \bibinfo {author} {\bibfnamefont {M.}~\bibnamefont {Koch-Janusz}}, \ and\
  \bibinfo {author} {\bibfnamefont {A.}~\bibnamefont {Stern}},\ }\href
  {\doibase 10.1103/PhysRevB.84.235145} {\bibfield  {journal} {\bibinfo
  {journal} {Phys. Rev. B}\ }\textbf {\bibinfo {volume} {84}},\ \bibinfo
  {pages} {\href{http://link.aps.org/doi/10.1103/PhysRevB.84.235145}{235145}}
  (\bibinfo {year} {2011})}\BibitemShut {NoStop}%
\bibitem [{\citenamefont {Maciejko}\ \emph {et~al.}(2012)\citenamefont
  {Maciejko}, \citenamefont {Qi}, \citenamefont {Karch},\ and\ \citenamefont
  {Zhang}}]{maciejko2012models}%
  \BibitemOpen
  \bibfield  {author} {\bibinfo {author} {\bibfnamefont {J.}~\bibnamefont
  {Maciejko}}, \bibinfo {author} {\bibfnamefont {X.-L.}\ \bibnamefont {Qi}},
  \bibinfo {author} {\bibfnamefont {A.}~\bibnamefont {Karch}}, \ and\ \bibinfo
  {author} {\bibfnamefont {S.-C.}\ \bibnamefont {Zhang}},\ }\href {\doibase
  10.1103/PhysRevB.86.235128} {\bibfield  {journal} {\bibinfo  {journal} {Phys.
  Rev. B}\ }\textbf {\bibinfo {volume} {86}},\ \bibinfo {pages} {235128}
  (\bibinfo {year} {2012})}\BibitemShut {NoStop}%
\bibitem [{\citenamefont {Ye}\ \emph {et~al.}(2016)\citenamefont {Ye},
  \citenamefont {Hughes}, \citenamefont {Maciejko},\ and\ \citenamefont
  {Fradkin}}]{ye2016composite}%
  \BibitemOpen
  \bibfield  {author} {\bibinfo {author} {\bibfnamefont {P.}~\bibnamefont
  {Ye}}, \bibinfo {author} {\bibfnamefont {T.~L.}\ \bibnamefont {Hughes}},
  \bibinfo {author} {\bibfnamefont {J.}~\bibnamefont {Maciejko}}, \ and\
  \bibinfo {author} {\bibfnamefont {E.}~\bibnamefont {Fradkin}},\ }\href
  {\doibase 10.1103/PhysRevB.94.115104} {\bibfield  {journal} {\bibinfo
  {journal} {Phys. Rev. B}\ }\textbf {\bibinfo {volume} {94}},\ \bibinfo
  {pages} {115104} (\bibinfo {year} {2016})}\BibitemShut {NoStop}%
\bibitem [{\citenamefont {Maciejko}\ and\ \citenamefont
  {Fiete}(2015)}]{maciejko2015fractionalized}%
  \BibitemOpen
  \bibfield  {author} {\bibinfo {author} {\bibfnamefont {J.}~\bibnamefont
  {Maciejko}}\ and\ \bibinfo {author} {\bibfnamefont {G.~A.}\ \bibnamefont
  {Fiete}},\ }\href {\doibase 10.1038/nphys3311} {\bibfield  {journal}
  {\bibinfo  {journal} {Nature Physics}\ }\textbf {\bibinfo {volume} {11}},\
  \bibinfo {pages} {385} (\bibinfo {year} {2015})}\BibitemShut {NoStop}%
\bibitem [{\citenamefont {Stern}(2016)}]{stern2016fractional}%
  \BibitemOpen
  \bibfield  {author} {\bibinfo {author} {\bibfnamefont {A.}~\bibnamefont
  {Stern}},\ }\href {\doibase 10.1146/annurev-conmatphys-031115-011559}
  {\bibfield  {journal} {\bibinfo  {journal} {Annual Review of Condensed Matter
  Physics}\ }\textbf {\bibinfo {volume} {7}},\ \bibinfo {pages} {349} (\bibinfo
  {year} {2016})}\BibitemShut {NoStop}%
\bibitem [{\citenamefont {Ye}\ \emph {et~al.}(2017)\citenamefont {Ye},
  \citenamefont {Cheng},\ and\ \citenamefont {Fradkin}}]{YeChengFradkin17}%
  \BibitemOpen
  \bibfield  {author} {\bibinfo {author} {\bibfnamefont {P.}~\bibnamefont
  {Ye}}, \bibinfo {author} {\bibfnamefont {M.}~\bibnamefont {Cheng}}, \ and\
  \bibinfo {author} {\bibfnamefont {E.}~\bibnamefont {Fradkin}},\ }\href
  {\doibase 10.1103/PhysRevB.96.085125} {\bibfield  {journal} {\bibinfo
  {journal} {Phys. Rev. B}\ }\textbf {\bibinfo {volume} {96}},\ \bibinfo
  {pages} {085125} (\bibinfo {year} {2017})}\BibitemShut {NoStop}%
\bibitem [{\citenamefont {Son}(2015)}]{Son15}%
  \BibitemOpen
  \bibfield  {author} {\bibinfo {author} {\bibfnamefont {D.~T.}\ \bibnamefont
  {Son}},\ }\href {\doibase 10.1103/PhysRevX.5.031027} {\bibfield  {journal}
  {\bibinfo  {journal} {Phys. Rev. X}\ }\textbf {\bibinfo {volume} {5}},\
  \bibinfo {pages} {031027} (\bibinfo {year} {2015})}\BibitemShut {NoStop}%
\bibitem [{\citenamefont {Wang}\ and\ \citenamefont
  {Senthil}(2015)}]{WangSenthil15}%
  \BibitemOpen
  \bibfield  {author} {\bibinfo {author} {\bibfnamefont {C.}~\bibnamefont
  {Wang}}\ and\ \bibinfo {author} {\bibfnamefont {T.}~\bibnamefont {Senthil}},\
  }\href {\doibase 10.1103/PhysRevX.5.041031} {\bibfield  {journal} {\bibinfo
  {journal} {Phys. Rev. X}\ }\textbf {\bibinfo {volume} {5}},\ \bibinfo {pages}
  {041031} (\bibinfo {year} {2015})}\BibitemShut {NoStop}%
\bibitem [{\citenamefont {Metlitski}\ and\ \citenamefont
  {Vishwanath}(2016)}]{MetlitskiVishwanath16}%
  \BibitemOpen
  \bibfield  {author} {\bibinfo {author} {\bibfnamefont {M.~A.}\ \bibnamefont
  {Metlitski}}\ and\ \bibinfo {author} {\bibfnamefont {A.}~\bibnamefont
  {Vishwanath}},\ }\href {\doibase 10.1103/PhysRevB.93.245151} {\bibfield
  {journal} {\bibinfo  {journal} {Phys. Rev. B}\ }\textbf {\bibinfo {volume}
  {93}},\ \bibinfo {pages} {245151} (\bibinfo {year} {2016})}\BibitemShut
  {NoStop}%
\bibitem [{\citenamefont {Kalmeyer}\ and\ \citenamefont
  {Zhang}(1992)}]{KalmeyerZhang92}%
  \BibitemOpen
  \bibfield  {author} {\bibinfo {author} {\bibfnamefont {V.}~\bibnamefont
  {Kalmeyer}}\ and\ \bibinfo {author} {\bibfnamefont {S.-C.}\ \bibnamefont
  {Zhang}},\ }\href {\doibase 10.1103/PhysRevB.46.9889} {\bibfield  {journal}
  {\bibinfo  {journal} {Phys. Rev. B}\ }\textbf {\bibinfo {volume} {46}},\
  \bibinfo {pages} {9889} (\bibinfo {year} {1992})}\BibitemShut {NoStop}%
\bibitem [{\citenamefont {Halperin}\ \emph {et~al.}(1993)\citenamefont
  {Halperin}, \citenamefont {Lee},\ and\ \citenamefont
  {Read}}]{HalperinLeeRead93}%
  \BibitemOpen
  \bibfield  {author} {\bibinfo {author} {\bibfnamefont {B.~I.}\ \bibnamefont
  {Halperin}}, \bibinfo {author} {\bibfnamefont {P.~A.}\ \bibnamefont {Lee}}, \
  and\ \bibinfo {author} {\bibfnamefont {N.}~\bibnamefont {Read}},\ }\href
  {\doibase 10.1103/PhysRevB.47.7312} {\bibfield  {journal} {\bibinfo
  {journal} {Phys. Rev. B}\ }\textbf {\bibinfo {volume} {47}},\ \bibinfo
  {pages} {7312} (\bibinfo {year} {1993})}\BibitemShut {NoStop}%
\bibitem [{\citenamefont {Girvin}(1984)}]{Girvin84}%
  \BibitemOpen
  \bibfield  {author} {\bibinfo {author} {\bibfnamefont {S.~M.}\ \bibnamefont
  {Girvin}},\ }\href {\doibase 10.1103/PhysRevB.29.6012} {\bibfield  {journal}
  {\bibinfo  {journal} {Phys. Rev. B}\ }\textbf {\bibinfo {volume} {29}},\
  \bibinfo {pages} {6012} (\bibinfo {year} {1984})}\BibitemShut {NoStop}%
\bibitem [{\citenamefont {Barkeshli}\ \emph {et~al.}(2015)\citenamefont
  {Barkeshli}, \citenamefont {Mulligan},\ and\ \citenamefont
  {Fisher}}]{BarkeshliMulliganFisher15}%
  \BibitemOpen
  \bibfield  {author} {\bibinfo {author} {\bibfnamefont {M.}~\bibnamefont
  {Barkeshli}}, \bibinfo {author} {\bibfnamefont {M.}~\bibnamefont {Mulligan}},
  \ and\ \bibinfo {author} {\bibfnamefont {M.~P.~A.}\ \bibnamefont {Fisher}},\
  }\href {\doibase 10.1103/PhysRevB.92.165125} {\bibfield  {journal} {\bibinfo
  {journal} {Phys. Rev. B}\ }\textbf {\bibinfo {volume} {92}},\ \bibinfo
  {pages} {165125} (\bibinfo {year} {2015})}\BibitemShut {NoStop}%
\bibitem [{\citenamefont {Wang}\ and\ \citenamefont
  {Senthil}(2016)}]{WangSenthil16}%
  \BibitemOpen
  \bibfield  {author} {\bibinfo {author} {\bibfnamefont {C.}~\bibnamefont
  {Wang}}\ and\ \bibinfo {author} {\bibfnamefont {T.}~\bibnamefont {Senthil}},\
  }\href {\doibase 10.1103/PhysRevB.93.085110} {\bibfield  {journal} {\bibinfo
  {journal} {Phys. Rev. B}\ }\textbf {\bibinfo {volume} {93}},\ \bibinfo
  {pages} {085110} (\bibinfo {year} {2016})}\BibitemShut {NoStop}%
\bibitem [{\citenamefont {Balram}\ and\ \citenamefont
  {Jain}(2017)}]{BalramJain17}%
  \BibitemOpen
  \bibfield  {author} {\bibinfo {author} {\bibfnamefont {A.~C.}\ \bibnamefont
  {Balram}}\ and\ \bibinfo {author} {\bibfnamefont {J.~K.}\ \bibnamefont
  {Jain}},\ }\href {\doibase 10.1103/PhysRevB.96.245142} {\bibfield  {journal}
  {\bibinfo  {journal} {Phys. Rev. B}\ }\textbf {\bibinfo {volume} {96}},\
  \bibinfo {pages} {245142} (\bibinfo {year} {2017})}\BibitemShut {NoStop}%
\bibitem [{\citenamefont {Nguyen}\ \emph {et~al.}(2018)\citenamefont {Nguyen},
  \citenamefont {Golkar}, \citenamefont {Roberts},\ and\ \citenamefont
  {Son}}]{NguyenGolkarRobertsSon18}%
  \BibitemOpen
  \bibfield  {author} {\bibinfo {author} {\bibfnamefont {D.~X.}\ \bibnamefont
  {Nguyen}}, \bibinfo {author} {\bibfnamefont {S.}~\bibnamefont {Golkar}},
  \bibinfo {author} {\bibfnamefont {M.~M.}\ \bibnamefont {Roberts}}, \ and\
  \bibinfo {author} {\bibfnamefont {D.~T.}\ \bibnamefont {Son}},\ }\href
  {\doibase 10.1103/PhysRevB.97.195314} {\bibfield  {journal} {\bibinfo
  {journal} {Phys. Rev. B}\ }\textbf {\bibinfo {volume} {97}},\ \bibinfo
  {pages} {195314} (\bibinfo {year} {2018})}\BibitemShut {NoStop}%
\bibitem [{\citenamefont {Liu}\ \emph {et~al.}(2016)\citenamefont {Liu},
  \citenamefont {Zhang},\ and\ \citenamefont {Qi}}]{liu2016quantum}%
  \BibitemOpen
  \bibfield  {author} {\bibinfo {author} {\bibfnamefont {C.-X.}\ \bibnamefont
  {Liu}}, \bibinfo {author} {\bibfnamefont {S.-C.}\ \bibnamefont {Zhang}}, \
  and\ \bibinfo {author} {\bibfnamefont {X.-L.}\ \bibnamefont {Qi}},\ }\href
  {\doibase 10.1146/annurev-conmatphys-031115-011417} {\bibfield  {journal}
  {\bibinfo  {journal} {Annual Review of Condensed Matter Physics}\ }\textbf
  {\bibinfo {volume} {7}},\ \bibinfo {pages} {301} (\bibinfo {year}
  {2016})}\BibitemShut {NoStop}%
\bibitem [{\citenamefont {Thouless}\ \emph {et~al.}(1982)\citenamefont
  {Thouless}, \citenamefont {Kohmoto}, \citenamefont {Nightingale},\ and\
  \citenamefont {den Nijs}}]{TKNN}%
  \BibitemOpen
  \bibfield  {author} {\bibinfo {author} {\bibfnamefont {D.~J.}\ \bibnamefont
  {Thouless}}, \bibinfo {author} {\bibfnamefont {M.}~\bibnamefont {Kohmoto}},
  \bibinfo {author} {\bibfnamefont {M.~P.}\ \bibnamefont {Nightingale}}, \ and\
  \bibinfo {author} {\bibfnamefont {M.}~\bibnamefont {den Nijs}},\ }\href
  {\doibase 10.1103/PhysRevLett.49.405} {\bibfield  {journal} {\bibinfo
  {journal} {Phys. Rev. Lett.}\ }\textbf {\bibinfo {volume} {49}},\ \bibinfo
  {pages} {405} (\bibinfo {year} {1982})}\BibitemShut {NoStop}%
\bibitem [{\citenamefont {Wang}\ \emph {et~al.}(2013)\citenamefont {Wang},
  \citenamefont {Potter},\ and\ \citenamefont
  {Senthil}}]{WangPotterSenthilgapTI13}%
  \BibitemOpen
  \bibfield  {author} {\bibinfo {author} {\bibfnamefont {C.}~\bibnamefont
  {Wang}}, \bibinfo {author} {\bibfnamefont {A.~C.}\ \bibnamefont {Potter}}, \
  and\ \bibinfo {author} {\bibfnamefont {T.}~\bibnamefont {Senthil}},\ }\href
  {\doibase 10.1103/PhysRevB.88.115137} {\bibfield  {journal} {\bibinfo
  {journal} {Phys. Rev. B}\ }\textbf {\bibinfo {volume} {88}},\ \bibinfo
  {pages} {115137} (\bibinfo {year} {2013})}\BibitemShut {NoStop}%
\bibitem [{\citenamefont {Metlitski}\ \emph {et~al.}(2015)\citenamefont
  {Metlitski}, \citenamefont {Kane},\ and\ \citenamefont
  {Fisher}}]{MetlitskiKaneFisher13b}%
  \BibitemOpen
  \bibfield  {author} {\bibinfo {author} {\bibfnamefont {M.~A.}\ \bibnamefont
  {Metlitski}}, \bibinfo {author} {\bibfnamefont {C.~L.}\ \bibnamefont {Kane}},
  \ and\ \bibinfo {author} {\bibfnamefont {M.~P.~A.}\ \bibnamefont {Fisher}},\
  }\href {\doibase 10.1103/PhysRevB.92.125111} {\bibfield  {journal} {\bibinfo
  {journal} {Phys. Rev. B}\ }\textbf {\bibinfo {volume} {92}},\ \bibinfo
  {pages} {125111} (\bibinfo {year} {2015})}\BibitemShut {NoStop}%
\bibitem [{\citenamefont {Chen}\ \emph {et~al.}(2014)\citenamefont {Chen},
  \citenamefont {Fidkowski},\ and\ \citenamefont
  {Vishwanath}}]{ChenFidkowskiVishwanath14}%
  \BibitemOpen
  \bibfield  {author} {\bibinfo {author} {\bibfnamefont {X.}~\bibnamefont
  {Chen}}, \bibinfo {author} {\bibfnamefont {L.}~\bibnamefont {Fidkowski}}, \
  and\ \bibinfo {author} {\bibfnamefont {A.}~\bibnamefont {Vishwanath}},\
  }\href {\doibase 10.1103/PhysRevB.89.165132} {\bibfield  {journal} {\bibinfo
  {journal} {Phys. Rev. B}\ }\textbf {\bibinfo {volume} {89}},\ \bibinfo
  {pages} {165132} (\bibinfo {year} {2014})}\BibitemShut {NoStop}%
\bibitem [{\citenamefont {Bonderson}\ \emph {et~al.}(2013)\citenamefont
  {Bonderson}, \citenamefont {Nayak},\ and\ \citenamefont
  {Qi}}]{BondersonNayakQi13}%
  \BibitemOpen
  \bibfield  {author} {\bibinfo {author} {\bibfnamefont {P.}~\bibnamefont
  {Bonderson}}, \bibinfo {author} {\bibfnamefont {C.}~\bibnamefont {Nayak}}, \
  and\ \bibinfo {author} {\bibfnamefont {X.-L.}\ \bibnamefont {Qi}},\ }\href
  {http://stacks.iop.org/1742-5468/2013/i=09/a=P09016} {\bibfield  {journal}
  {\bibinfo  {journal} {Journal of Statistical Mechanics: Theory and
  Experiment}\ }\textbf {\bibinfo {volume} {2013}},\ \bibinfo {pages} {P09016}
  (\bibinfo {year} {2013})}\BibitemShut {NoStop}%
\bibitem [{\citenamefont {Read}\ and\ \citenamefont {Moore}(1991)}]{MooreRead}%
  \BibitemOpen
  \bibfield  {author} {\bibinfo {author} {\bibfnamefont {N.}~\bibnamefont
  {Read}}\ and\ \bibinfo {author} {\bibfnamefont {G.}~\bibnamefont {Moore}},\
  }\href {\doibase 10.1016/0550-3213(91)90407-O} {\bibfield  {journal}
  {\bibinfo  {journal} {Nucl. Phys. B}\ }\textbf {\bibinfo {volume} {360}},\
  \bibinfo {pages} {362} (\bibinfo {year} {1991})}\BibitemShut {NoStop}%
\bibitem [{\citenamefont {Jain}(1989)}]{Jain89}%
  \BibitemOpen
  \bibfield  {author} {\bibinfo {author} {\bibfnamefont {J.~K.}\ \bibnamefont
  {Jain}},\ }\href {\doibase 10.1103/PhysRevB.40.8079} {\bibfield  {journal}
  {\bibinfo  {journal} {Phys. Rev. B}\ }\textbf {\bibinfo {volume} {40}},\
  \bibinfo {pages} {8079} (\bibinfo {year} {1989})}\BibitemShut {NoStop}%
\bibitem [{\citenamefont {WEN}(1992)}]{WenEdgeStatesQHE92}%
  \BibitemOpen
  \bibfield  {author} {\bibinfo {author} {\bibfnamefont {X.-G.}\ \bibnamefont
  {WEN}},\ }\href {\doibase 10.1142/S0217979292000840} {\bibfield  {journal}
  {\bibinfo  {journal} {International Journal of Modern Physics B}\ }\textbf
  {\bibinfo {volume} {06}},\ \bibinfo {pages} {1711} (\bibinfo {year}
  {1992})}\BibitemShut {NoStop}%
\bibitem [{\citenamefont {Sahoo}\ \emph {et~al.}(2017)\citenamefont {Sahoo},
  \citenamefont {Sirota}, \citenamefont {Cho},\ and\ \citenamefont
  {Teo}}]{SahooSirotaChoTeo17}%
  \BibitemOpen
  \bibfield  {author} {\bibinfo {author} {\bibfnamefont {S.}~\bibnamefont
  {Sahoo}}, \bibinfo {author} {\bibfnamefont {A.}~\bibnamefont {Sirota}},
  \bibinfo {author} {\bibfnamefont {G.~Y.}\ \bibnamefont {Cho}}, \ and\
  \bibinfo {author} {\bibfnamefont {J.~C.~Y.}\ \bibnamefont {Teo}},\ }\href
  {\doibase 10.1103/PhysRevB.96.161108} {\bibfield  {journal} {\bibinfo
  {journal} {Phys. Rev. B}\ }\textbf {\bibinfo {volume} {96}},\ \bibinfo
  {pages} {161108} (\bibinfo {year} {2017})}\BibitemShut {NoStop}%
\bibitem [{\citenamefont {Cho}\ \emph {et~al.}(2017)\citenamefont {Cho},
  \citenamefont {Teo},\ and\ \citenamefont {Fradkin}}]{ChoTeoFradkin17}%
  \BibitemOpen
  \bibfield  {author} {\bibinfo {author} {\bibfnamefont {G.~Y.}\ \bibnamefont
  {Cho}}, \bibinfo {author} {\bibfnamefont {J.~C.~Y.}\ \bibnamefont {Teo}}, \
  and\ \bibinfo {author} {\bibfnamefont {E.}~\bibnamefont {Fradkin}},\ }\href
  {\doibase 10.1103/PhysRevB.96.161109} {\bibfield  {journal} {\bibinfo
  {journal} {Phys. Rev. B}\ }\textbf {\bibinfo {volume} {96}},\ \bibinfo
  {pages} {161109} (\bibinfo {year} {2017})}\BibitemShut {NoStop}%
\bibitem [{\citenamefont {Kane}\ and\ \citenamefont
  {Fisher}(1997)}]{KaneFisher97}%
  \BibitemOpen
  \bibfield  {author} {\bibinfo {author} {\bibfnamefont {C.~L.}\ \bibnamefont
  {Kane}}\ and\ \bibinfo {author} {\bibfnamefont {M.~P.~A.}\ \bibnamefont
  {Fisher}},\ }\href {\doibase 10.1103/PhysRevB.55.15832} {\bibfield  {journal}
  {\bibinfo  {journal} {Phys. Rev. B}\ }\textbf {\bibinfo {volume} {55}},\
  \bibinfo {pages} {15832} (\bibinfo {year} {1997})}\BibitemShut {NoStop}%
\bibitem [{\citenamefont {Cappelli}\ \emph {et~al.}(2002)\citenamefont
  {Cappelli}, \citenamefont {Huerta},\ and\ \citenamefont
  {Zemba}}]{Cappelli01}%
  \BibitemOpen
  \bibfield  {author} {\bibinfo {author} {\bibfnamefont {A.}~\bibnamefont
  {Cappelli}}, \bibinfo {author} {\bibfnamefont {M.}~\bibnamefont {Huerta}}, \
  and\ \bibinfo {author} {\bibfnamefont {G.~R.}\ \bibnamefont {Zemba}},\ }\href
  {\doibase http://dx.doi.org/10.1016/S0550-3213(02)00340-1} {\bibfield
  {journal} {\bibinfo  {journal} {Nuclear Physics B}\ }\textbf {\bibinfo
  {volume} {636}},\ \bibinfo {pages} {568 } (\bibinfo {year}
  {2002})}\BibitemShut {NoStop}%
\bibitem [{\citenamefont {Kitaev}(2006)}]{Kitaev06}%
  \BibitemOpen
  \bibfield  {author} {\bibinfo {author} {\bibfnamefont {A.}~\bibnamefont
  {Kitaev}},\ }\href {\doibase https://doi.org/10.1016/j.aop.2005.10.005}
  {\bibfield  {journal} {\bibinfo  {journal} {Annals of Physics}\ }\textbf
  {\bibinfo {volume} {321}},\ \bibinfo {pages} {2 } (\bibinfo {year} {2006})},\
  \bibinfo {note} {january Special Issue}\BibitemShut {NoStop}%
\bibitem [{\citenamefont {O'Hern}\ \emph {et~al.}(1999)\citenamefont {O'Hern},
  \citenamefont {Lubensky},\ and\ \citenamefont
  {Toner}}]{OHernLubenskyToner99}%
  \BibitemOpen
  \bibfield  {author} {\bibinfo {author} {\bibfnamefont {C.~S.}\ \bibnamefont
  {O'Hern}}, \bibinfo {author} {\bibfnamefont {T.~C.}\ \bibnamefont
  {Lubensky}}, \ and\ \bibinfo {author} {\bibfnamefont {J.}~\bibnamefont
  {Toner}},\ }\href {\doibase 10.1103/PhysRevLett.83.2745} {\bibfield
  {journal} {\bibinfo  {journal} {Phys. Rev. Lett.}\ }\textbf {\bibinfo
  {volume} {83}},\ \bibinfo {pages} {2745} (\bibinfo {year}
  {1999})}\BibitemShut {NoStop}%
\bibitem [{\citenamefont {Emery}\ \emph {et~al.}(2000)\citenamefont {Emery},
  \citenamefont {Fradkin}, \citenamefont {Kivelson},\ and\ \citenamefont
  {Lubensky}}]{EmeryFradkinKivelsonLubensky00}%
  \BibitemOpen
  \bibfield  {author} {\bibinfo {author} {\bibfnamefont {V.~J.}\ \bibnamefont
  {Emery}}, \bibinfo {author} {\bibfnamefont {E.}~\bibnamefont {Fradkin}},
  \bibinfo {author} {\bibfnamefont {S.~A.}\ \bibnamefont {Kivelson}}, \ and\
  \bibinfo {author} {\bibfnamefont {T.~C.}\ \bibnamefont {Lubensky}},\ }\href
  {\doibase 10.1103/PhysRevLett.85.2160} {\bibfield  {journal} {\bibinfo
  {journal} {Phys. Rev. Lett.}\ }\textbf {\bibinfo {volume} {85}},\ \bibinfo
  {pages} {2160} (\bibinfo {year} {2000})}\BibitemShut {NoStop}%
\bibitem [{\citenamefont {Vishwanath}\ and\ \citenamefont
  {Carpentier}(2001)}]{VishwanathCarpentier01}%
  \BibitemOpen
  \bibfield  {author} {\bibinfo {author} {\bibfnamefont {A.}~\bibnamefont
  {Vishwanath}}\ and\ \bibinfo {author} {\bibfnamefont {D.}~\bibnamefont
  {Carpentier}},\ }\href {\doibase 10.1103/PhysRevLett.86.676} {\bibfield
  {journal} {\bibinfo  {journal} {Phys. Rev. Lett.}\ }\textbf {\bibinfo
  {volume} {86}},\ \bibinfo {pages} {676} (\bibinfo {year} {2001})}\BibitemShut
  {NoStop}%
\bibitem [{\citenamefont {Sondhi}\ and\ \citenamefont
  {Yang}(2001)}]{SondhiYang01}%
  \BibitemOpen
  \bibfield  {author} {\bibinfo {author} {\bibfnamefont {S.~L.}\ \bibnamefont
  {Sondhi}}\ and\ \bibinfo {author} {\bibfnamefont {K.}~\bibnamefont {Yang}},\
  }\href {\doibase 10.1103/PhysRevB.63.054430} {\bibfield  {journal} {\bibinfo
  {journal} {Phys. Rev. B}\ }\textbf {\bibinfo {volume} {63}},\ \bibinfo
  {pages} {054430} (\bibinfo {year} {2001})}\BibitemShut {NoStop}%
\bibitem [{\citenamefont {Mukhopadhyay}\ \emph {et~al.}(2001)\citenamefont
  {Mukhopadhyay}, \citenamefont {Kane},\ and\ \citenamefont
  {Lubensky}}]{MukhopadhyayKaneLubensky01}%
  \BibitemOpen
  \bibfield  {author} {\bibinfo {author} {\bibfnamefont {R.}~\bibnamefont
  {Mukhopadhyay}}, \bibinfo {author} {\bibfnamefont {C.~L.}\ \bibnamefont
  {Kane}}, \ and\ \bibinfo {author} {\bibfnamefont {T.~C.}\ \bibnamefont
  {Lubensky}},\ }\href {\doibase 10.1103/PhysRevB.63.081103} {\bibfield
  {journal} {\bibinfo  {journal} {Phys. Rev. B}\ }\textbf {\bibinfo {volume}
  {63}},\ \bibinfo {pages} {081103} (\bibinfo {year} {2001})}\BibitemShut
  {NoStop}%
\bibitem [{\citenamefont {Kane}\ \emph {et~al.}(2002)\citenamefont {Kane},
  \citenamefont {Mukhopadhyay},\ and\ \citenamefont
  {Lubensky}}]{KaneMukhopadhyayLubensky02}%
  \BibitemOpen
  \bibfield  {author} {\bibinfo {author} {\bibfnamefont {C.~L.}\ \bibnamefont
  {Kane}}, \bibinfo {author} {\bibfnamefont {R.}~\bibnamefont {Mukhopadhyay}},
  \ and\ \bibinfo {author} {\bibfnamefont {T.~C.}\ \bibnamefont {Lubensky}},\
  }\href {\doibase 10.1103/PhysRevLett.88.036401} {\bibfield  {journal}
  {\bibinfo  {journal} {Phys. Rev. Lett.}\ }\textbf {\bibinfo {volume} {88}},\
  \bibinfo {pages} {036401} (\bibinfo {year} {2002})}\BibitemShut {NoStop}%
\bibitem [{\citenamefont {Haldane}(1983)}]{Haldane83}%
  \BibitemOpen
  \bibfield  {author} {\bibinfo {author} {\bibfnamefont {F.~D.~M.}\
  \bibnamefont {Haldane}},\ }\href {\doibase 10.1103/PhysRevLett.51.605}
  {\bibfield  {journal} {\bibinfo  {journal} {Phys. Rev. Lett.}\ }\textbf
  {\bibinfo {volume} {51}},\ \bibinfo {pages}
  {\href{http://link.aps.org/doi/10.1103/PhysRevLett.51.605}{605}} (\bibinfo
  {year} {1983})}\BibitemShut {NoStop}%
\bibitem [{\citenamefont {Halperin}(1984)}]{Halperin84}%
  \BibitemOpen
  \bibfield  {author} {\bibinfo {author} {\bibfnamefont {B.~I.}\ \bibnamefont
  {Halperin}},\ }\href {\doibase 10.1103/PhysRevLett.52.1583} {\bibfield
  {journal} {\bibinfo  {journal} {Phys. Rev. Lett.}\ }\textbf {\bibinfo
  {volume} {52}},\ \bibinfo {pages}
  {\href{http://link.aps.org/doi/10.1103/PhysRevLett.52.1583}{1583}} (\bibinfo
  {year} {1984})}\BibitemShut {NoStop}%
\bibitem [{\citenamefont {Teo}\ and\ \citenamefont
  {Kane}(2014)}]{TeoKaneCouplewires}%
  \BibitemOpen
  \bibfield  {author} {\bibinfo {author} {\bibfnamefont {J.~C.~Y.}\
  \bibnamefont {Teo}}\ and\ \bibinfo {author} {\bibfnamefont {C.~L.}\
  \bibnamefont {Kane}},\ }\href {\doibase 10.1103/PhysRevB.89.085101}
  {\bibfield  {journal} {\bibinfo  {journal} {Phys. Rev. B}\ }\textbf {\bibinfo
  {volume} {89}},\ \bibinfo {pages} {085101} (\bibinfo {year}
  {2014})}\BibitemShut {NoStop}%
\bibitem [{\citenamefont {Klinovaja}\ and\ \citenamefont
  {Loss}(2014)}]{KlinovajaLoss14}%
  \BibitemOpen
  \bibfield  {author} {\bibinfo {author} {\bibfnamefont {J.}~\bibnamefont
  {Klinovaja}}\ and\ \bibinfo {author} {\bibfnamefont {D.}~\bibnamefont
  {Loss}},\ }\href {\doibase 10.1140/epjb/e2014-50395-6} {\bibfield  {journal}
  {\bibinfo  {journal} {The European Physical Journal B}\ }\textbf {\bibinfo
  {volume} {87}},\ \bibinfo {pages} {171} (\bibinfo {year} {2014})}\BibitemShut
  {NoStop}%
\bibitem [{\citenamefont {Meng}\ \emph {et~al.}(2014)\citenamefont {Meng},
  \citenamefont {Stano}, \citenamefont {Klinovaja},\ and\ \citenamefont
  {Loss}}]{MengStanoKlinovajaLoss14}%
  \BibitemOpen
  \bibfield  {author} {\bibinfo {author} {\bibfnamefont {T.}~\bibnamefont
  {Meng}}, \bibinfo {author} {\bibfnamefont {P.}~\bibnamefont {Stano}},
  \bibinfo {author} {\bibfnamefont {J.}~\bibnamefont {Klinovaja}}, \ and\
  \bibinfo {author} {\bibfnamefont {D.}~\bibnamefont {Loss}},\ }\href {\doibase
  10.1140/epjb/e2014-50445-1} {\bibfield  {journal} {\bibinfo  {journal} {The
  European Physical Journal B}\ }\textbf {\bibinfo {volume} {87}},\ \bibinfo
  {pages} {203} (\bibinfo {year} {2014})}\BibitemShut {NoStop}%
\bibitem [{\citenamefont {Sagi}\ \emph {et~al.}(2015)\citenamefont {Sagi},
  \citenamefont {Oreg}, \citenamefont {Stern},\ and\ \citenamefont
  {Halperin}}]{SagiOregSternHalperin15}%
  \BibitemOpen
  \bibfield  {author} {\bibinfo {author} {\bibfnamefont {E.}~\bibnamefont
  {Sagi}}, \bibinfo {author} {\bibfnamefont {Y.}~\bibnamefont {Oreg}}, \bibinfo
  {author} {\bibfnamefont {A.}~\bibnamefont {Stern}}, \ and\ \bibinfo {author}
  {\bibfnamefont {B.~I.}\ \bibnamefont {Halperin}},\ }\href {\doibase
  10.1103/PhysRevB.91.245144} {\bibfield  {journal} {\bibinfo  {journal} {Phys.
  Rev. B}\ }\textbf {\bibinfo {volume} {91}},\ \bibinfo {pages} {245144}
  (\bibinfo {year} {2015})}\BibitemShut {NoStop}%
\bibitem [{\citenamefont {Kane}\ \emph {et~al.}(2017)\citenamefont {Kane},
  \citenamefont {Stern},\ and\ \citenamefont {Halperin}}]{KaneSternHalperin17}%
  \BibitemOpen
  \bibfield  {author} {\bibinfo {author} {\bibfnamefont {C.~L.}\ \bibnamefont
  {Kane}}, \bibinfo {author} {\bibfnamefont {A.}~\bibnamefont {Stern}}, \ and\
  \bibinfo {author} {\bibfnamefont {B.~I.}\ \bibnamefont {Halperin}},\ }\href
  {\doibase 10.1103/PhysRevX.7.031009} {\bibfield  {journal} {\bibinfo
  {journal} {Phys. Rev. X}\ }\textbf {\bibinfo {volume} {7}},\ \bibinfo {pages}
  {031009} (\bibinfo {year} {2017})}\BibitemShut {NoStop}%
\bibitem [{\citenamefont {Read}\ and\ \citenamefont {Moore}(1992)}]{ReadMoore}%
  \BibitemOpen
  \bibfield  {author} {\bibinfo {author} {\bibfnamefont {N.}~\bibnamefont
  {Read}}\ and\ \bibinfo {author} {\bibfnamefont {G.}~\bibnamefont {Moore}},\
  }\href {\doibase 10.1143/PTPS.107.157} {\bibfield  {journal} {\bibinfo
  {journal} {Progress of Theoretical Physics Supplement}\ }\textbf {\bibinfo
  {volume} {107}},\ \bibinfo {pages} {157} (\bibinfo {year}
  {1992})}\BibitemShut {NoStop}%
\bibitem [{\citenamefont {Read}\ and\ \citenamefont
  {Rezayi}(1999)}]{ReadRezayi}%
  \BibitemOpen
  \bibfield  {author} {\bibinfo {author} {\bibfnamefont {N.}~\bibnamefont
  {Read}}\ and\ \bibinfo {author} {\bibfnamefont {E.}~\bibnamefont {Rezayi}},\
  }\href {\doibase 10.1103/PhysRevB.59.8084} {\bibfield  {journal} {\bibinfo
  {journal} {Phys. Rev. B.}\ }\textbf {\bibinfo {volume} {59}},\ \bibinfo
  {pages} {8084} (\bibinfo {year} {1999})}\BibitemShut {NoStop}%
\bibitem [{\citenamefont {Di~Francesco}\ \emph {et~al.}(1999)\citenamefont
  {Di~Francesco}, \citenamefont {Mathieu},\ and\ \citenamefont
  {Senechal}}]{bigyellowbook}%
  \BibitemOpen
  \bibfield  {author} {\bibinfo {author} {\bibfnamefont {P.}~\bibnamefont
  {Di~Francesco}}, \bibinfo {author} {\bibfnamefont {P.}~\bibnamefont
  {Mathieu}}, \ and\ \bibinfo {author} {\bibfnamefont {D.}~\bibnamefont
  {Senechal}},\ }\href@noop {} {\emph {\bibinfo {title} {Conformal Field
  Theory}}}\ (\bibinfo  {publisher} {Springer, New York},\ \bibinfo {year}
  {1999})\BibitemShut {NoStop}%
\bibitem [{\citenamefont {Bais}\ and\ \citenamefont
  {Slingerland}(2009)}]{BaisSlingerlandCondensation}%
  \BibitemOpen
  \bibfield  {author} {\bibinfo {author} {\bibfnamefont {F.~A.}\ \bibnamefont
  {Bais}}\ and\ \bibinfo {author} {\bibfnamefont {J.~K.}\ \bibnamefont
  {Slingerland}},\ }\href {\doibase 10.1103/PhysRevB.79.045316} {\bibfield
  {journal} {\bibinfo  {journal} {Phys. Rev. B}\ }\textbf {\bibinfo {volume}
  {79}},\ \bibinfo {pages} {045316} (\bibinfo {year} {2009})}\BibitemShut
  {NoStop}%
\bibitem [{\citenamefont {Haldane}(1995)}]{Haldane95}%
  \BibitemOpen
  \bibfield  {author} {\bibinfo {author} {\bibfnamefont {F.~D.~M.}\
  \bibnamefont {Haldane}},\ }\href {\doibase 10.1103/PhysRevLett.74.2090}
  {\bibfield  {journal} {\bibinfo  {journal} {Phys. Rev. Lett.}\ }\textbf
  {\bibinfo {volume} {74}},\ \bibinfo {pages} {2090} (\bibinfo {year}
  {1995})}\BibitemShut {NoStop}%
\bibitem [{\citenamefont {Jordan}(1927)}]{Jordan27}%
  \BibitemOpen
  \bibfield  {author} {\bibinfo {author} {\bibfnamefont {P.}~\bibnamefont
  {Jordan}},\ }\href {\doibase 10.1007/BF01397395} {\bibfield  {journal}
  {\bibinfo  {journal} {Zeitschrift f{\"u}r Physik}\ }\textbf {\bibinfo
  {volume} {44}},\ \bibinfo {pages} {473} (\bibinfo {year} {1927})}\BibitemShut
  {NoStop}%
\bibitem [{\citenamefont {Jordan}\ and\ \citenamefont
  {Wigner}(1928)}]{JordanWigner28}%
  \BibitemOpen
  \bibfield  {author} {\bibinfo {author} {\bibfnamefont {P.}~\bibnamefont
  {Jordan}}\ and\ \bibinfo {author} {\bibfnamefont {E.}~\bibnamefont
  {Wigner}},\ }\href {\doibase 10.1007/BF01331938} {\bibfield  {journal}
  {\bibinfo  {journal} {Zeitschrift f{\"u}r Physik}\ }\textbf {\bibinfo
  {volume} {47}},\ \bibinfo {pages} {631} (\bibinfo {year} {1928})}\BibitemShut
  {NoStop}%
\bibitem [{\citenamefont {Mross}\ \emph {et~al.}(2015)\citenamefont {Mross},
  \citenamefont {Essin},\ and\ \citenamefont {Alicea}}]{MrossEssinAlicea15}%
  \BibitemOpen
  \bibfield  {author} {\bibinfo {author} {\bibfnamefont {D.~F.}\ \bibnamefont
  {Mross}}, \bibinfo {author} {\bibfnamefont {A.}~\bibnamefont {Essin}}, \ and\
  \bibinfo {author} {\bibfnamefont {J.}~\bibnamefont {Alicea}},\ }\href
  {\doibase 10.1103/PhysRevX.5.011011} {\bibfield  {journal} {\bibinfo
  {journal} {Phys. Rev. X}\ }\textbf {\bibinfo {volume} {5}},\ \bibinfo {pages}
  {011011} (\bibinfo {year} {2015})}\BibitemShut {NoStop}%
\bibitem [{\citenamefont {Jain}(1990)}]{Jain90}%
  \BibitemOpen
  \bibfield  {author} {\bibinfo {author} {\bibfnamefont {J.~K.}\ \bibnamefont
  {Jain}},\ }\href {\doibase 10.1103/PhysRevB.41.7653} {\bibfield  {journal}
  {\bibinfo  {journal} {Phys. Rev. B}\ }\textbf {\bibinfo {volume} {41}},\
  \bibinfo {pages} {7653} (\bibinfo {year} {1990})}\BibitemShut {NoStop}%
\bibitem [{\citenamefont {Levin}\ \emph {et~al.}(2007)\citenamefont {Levin},
  \citenamefont {Halperin},\ and\ \citenamefont
  {Rosenow}}]{LevinHalperinRosenow07}%
  \BibitemOpen
  \bibfield  {author} {\bibinfo {author} {\bibfnamefont {M.}~\bibnamefont
  {Levin}}, \bibinfo {author} {\bibfnamefont {B.~I.}\ \bibnamefont {Halperin}},
  \ and\ \bibinfo {author} {\bibfnamefont {B.}~\bibnamefont {Rosenow}},\ }\href
  {\doibase 10.1103/PhysRevLett.99.236806} {\bibfield  {journal} {\bibinfo
  {journal} {Phys. Rev. Lett.}\ }\textbf {\bibinfo {volume} {99}},\ \bibinfo
  {pages} {236806} (\bibinfo {year} {2007})}\BibitemShut {NoStop}%
\bibitem [{\citenamefont {Lee}\ \emph {et~al.}(2007)\citenamefont {Lee},
  \citenamefont {Ryu}, \citenamefont {Nayak},\ and\ \citenamefont
  {Fisher}}]{LeeRyuNayakFisher07}%
  \BibitemOpen
  \bibfield  {author} {\bibinfo {author} {\bibfnamefont {S.-S.}\ \bibnamefont
  {Lee}}, \bibinfo {author} {\bibfnamefont {S.}~\bibnamefont {Ryu}}, \bibinfo
  {author} {\bibfnamefont {C.}~\bibnamefont {Nayak}}, \ and\ \bibinfo {author}
  {\bibfnamefont {M.~P.~A.}\ \bibnamefont {Fisher}},\ }\href {\doibase
  10.1103/PhysRevLett.99.236807} {\bibfield  {journal} {\bibinfo  {journal}
  {Phys. Rev. Lett.}\ }\textbf {\bibinfo {volume} {99}},\ \bibinfo {pages}
  {236807} (\bibinfo {year} {2007})}\BibitemShut {NoStop}%
\bibitem [{\citenamefont {{Fuji}}\ and\ \citenamefont
  {{Furusaki}}(2018)}]{FujiFurusaki18}%
  \BibitemOpen
  \bibfield  {author} {\bibinfo {author} {\bibfnamefont {Y.}~\bibnamefont
  {{Fuji}}}\ and\ \bibinfo {author} {\bibfnamefont {A.}~\bibnamefont
  {{Furusaki}}},\ }\href@noop {} {\bibfield  {journal} {\bibinfo  {journal}
  {ArXiv e-prints}\ } (\bibinfo {year} {2018})},\ \Eprint
  {http://arxiv.org/abs/1808.07648} {arXiv:1808.07648 [cond-mat.str-el]}
  \BibitemShut {NoStop}%
\bibitem [{\citenamefont {Mross}\ \emph
  {et~al.}(2016{\natexlab{a}})\citenamefont {Mross}, \citenamefont {Alicea},\
  and\ \citenamefont {Motrunich}}]{MrossAliceaMotrunich16}%
  \BibitemOpen
  \bibfield  {author} {\bibinfo {author} {\bibfnamefont {D.~F.}\ \bibnamefont
  {Mross}}, \bibinfo {author} {\bibfnamefont {J.}~\bibnamefont {Alicea}}, \
  and\ \bibinfo {author} {\bibfnamefont {O.~I.}\ \bibnamefont {Motrunich}},\
  }\href {\doibase 10.1103/PhysRevLett.117.016802} {\bibfield  {journal}
  {\bibinfo  {journal} {Phys. Rev. Lett.}\ }\textbf {\bibinfo {volume} {117}},\
  \bibinfo {pages} {016802} (\bibinfo {year} {2016}{\natexlab{a}})}\BibitemShut
  {NoStop}%
\bibitem [{\citenamefont {Mross}\ \emph
  {et~al.}(2016{\natexlab{b}})\citenamefont {Mross}, \citenamefont {Alicea},\
  and\ \citenamefont {Motrunich}}]{MrossAliceaMotrunich16PRL}%
  \BibitemOpen
  \bibfield  {author} {\bibinfo {author} {\bibfnamefont {D.~F.}\ \bibnamefont
  {Mross}}, \bibinfo {author} {\bibfnamefont {J.}~\bibnamefont {Alicea}}, \
  and\ \bibinfo {author} {\bibfnamefont {O.~I.}\ \bibnamefont {Motrunich}},\
  }\href {\doibase 10.1103/PhysRevLett.117.136802} {\bibfield  {journal}
  {\bibinfo  {journal} {Phys. Rev. Lett.}\ }\textbf {\bibinfo {volume} {117}},\
  \bibinfo {pages} {136802} (\bibinfo {year} {2016}{\natexlab{b}})}\BibitemShut
  {NoStop}%
\bibitem [{\citenamefont {Mross}\ \emph {et~al.}(2017)\citenamefont {Mross},
  \citenamefont {Alicea},\ and\ \citenamefont
  {Motrunich}}]{MrossAliceaMotrunich17}%
  \BibitemOpen
  \bibfield  {author} {\bibinfo {author} {\bibfnamefont {D.~F.}\ \bibnamefont
  {Mross}}, \bibinfo {author} {\bibfnamefont {J.}~\bibnamefont {Alicea}}, \
  and\ \bibinfo {author} {\bibfnamefont {O.~I.}\ \bibnamefont {Motrunich}},\
  }\href {\doibase 10.1103/PhysRevX.7.041016} {\bibfield  {journal} {\bibinfo
  {journal} {Phys. Rev. X}\ }\textbf {\bibinfo {volume} {7}},\ \bibinfo {pages}
  {041016} (\bibinfo {year} {2017})}\BibitemShut {NoStop}%
\bibitem [{\citenamefont {Mross}\ \emph {et~al.}(2018)\citenamefont {Mross},
  \citenamefont {Oreg}, \citenamefont {Stern}, \citenamefont {Margalit},\ and\
  \citenamefont {Heiblum}}]{MrossOregSternMargalitHeiblum18}%
  \BibitemOpen
  \bibfield  {author} {\bibinfo {author} {\bibfnamefont {D.~F.}\ \bibnamefont
  {Mross}}, \bibinfo {author} {\bibfnamefont {Y.}~\bibnamefont {Oreg}},
  \bibinfo {author} {\bibfnamefont {A.}~\bibnamefont {Stern}}, \bibinfo
  {author} {\bibfnamefont {G.}~\bibnamefont {Margalit}}, \ and\ \bibinfo
  {author} {\bibfnamefont {M.}~\bibnamefont {Heiblum}},\ }\href {\doibase
  10.1103/PhysRevLett.121.026801} {\bibfield  {journal} {\bibinfo  {journal}
  {Phys. Rev. Lett.}\ }\textbf {\bibinfo {volume} {121}},\ \bibinfo {pages}
  {026801} (\bibinfo {year} {2018})}\BibitemShut {NoStop}%
\bibitem [{\citenamefont {{Banerjee}}\ \emph {et~al.}(2017)\citenamefont
  {{Banerjee}}, \citenamefont {{Heiblum}}, \citenamefont {{Rosenblatt}},
  \citenamefont {{Oreg}}, \citenamefont {{Feldman}}, \citenamefont {{Stern}},\
  and\ \citenamefont
  {{Umansky}}}]{BanerjeeHeiblumRosenblattOregFeldmanSternUmansky17}%
  \BibitemOpen
  \bibfield  {author} {\bibinfo {author} {\bibfnamefont {M.}~\bibnamefont
  {{Banerjee}}}, \bibinfo {author} {\bibfnamefont {M.}~\bibnamefont
  {{Heiblum}}}, \bibinfo {author} {\bibfnamefont {A.}~\bibnamefont
  {{Rosenblatt}}}, \bibinfo {author} {\bibfnamefont {Y.}~\bibnamefont
  {{Oreg}}}, \bibinfo {author} {\bibfnamefont {D.~E.}\ \bibnamefont
  {{Feldman}}}, \bibinfo {author} {\bibfnamefont {A.}~\bibnamefont {{Stern}}},
  \ and\ \bibinfo {author} {\bibfnamefont {V.}~\bibnamefont {{Umansky}}},\
  }\href {\doibase 10.1038/nature22052} {\bibfield  {journal} {\bibinfo
  {journal} {Nature}\ }\textbf {\bibinfo {volume} {545}},\ \bibinfo {pages}
  {75} (\bibinfo {year} {2017})}\BibitemShut {NoStop}%
\bibitem [{\citenamefont {{Banerjee}}\ \emph {et~al.}(2018)\citenamefont
  {{Banerjee}}, \citenamefont {{Heiblum}}, \citenamefont {{Umansky}},
  \citenamefont {{Feldman}}, \citenamefont {{Oreg}},\ and\ \citenamefont
  {{Stern}}}]{BanerjeeHeiblumUmanskyFeldmanOregStern18}%
  \BibitemOpen
  \bibfield  {author} {\bibinfo {author} {\bibfnamefont {M.}~\bibnamefont
  {{Banerjee}}}, \bibinfo {author} {\bibfnamefont {M.}~\bibnamefont
  {{Heiblum}}}, \bibinfo {author} {\bibfnamefont {V.}~\bibnamefont
  {{Umansky}}}, \bibinfo {author} {\bibfnamefont {D.~E.}\ \bibnamefont
  {{Feldman}}}, \bibinfo {author} {\bibfnamefont {Y.}~\bibnamefont {{Oreg}}}, \
  and\ \bibinfo {author} {\bibfnamefont {A.}~\bibnamefont {{Stern}}},\ }\href
  {\doibase 10.1038/s41586-018-0184-1} {\bibfield  {journal} {\bibinfo
  {journal} {Nature}\ }\textbf {\bibinfo {volume} {559}},\ \bibinfo {pages}
  {205} (\bibinfo {year} {2018})}\BibitemShut {NoStop}%
\end{thebibliography}

%

\end{document}